\documentclass[a4paper,10pt]{article}
\usepackage[italian,english]{babel}
\usepackage{float}
\usepackage[utf8]{inputenc}
\usepackage{graphicx}
\usepackage{amsmath}
\usepackage{caption}
\usepackage{setspace}
\usepackage{textcomp}
\usepackage{longtable}
\usepackage{relsize}
\usepackage{subfigure}

\title{}
\author{Ester Ricci}
\date{}

\begin{document}
 \begin{titlepage}
  \thispagestyle{empty}
\setlength{\unitlength}{1cm}
\begin{figure}[h]
\begin{center}
\includegraphics[height=3cm]{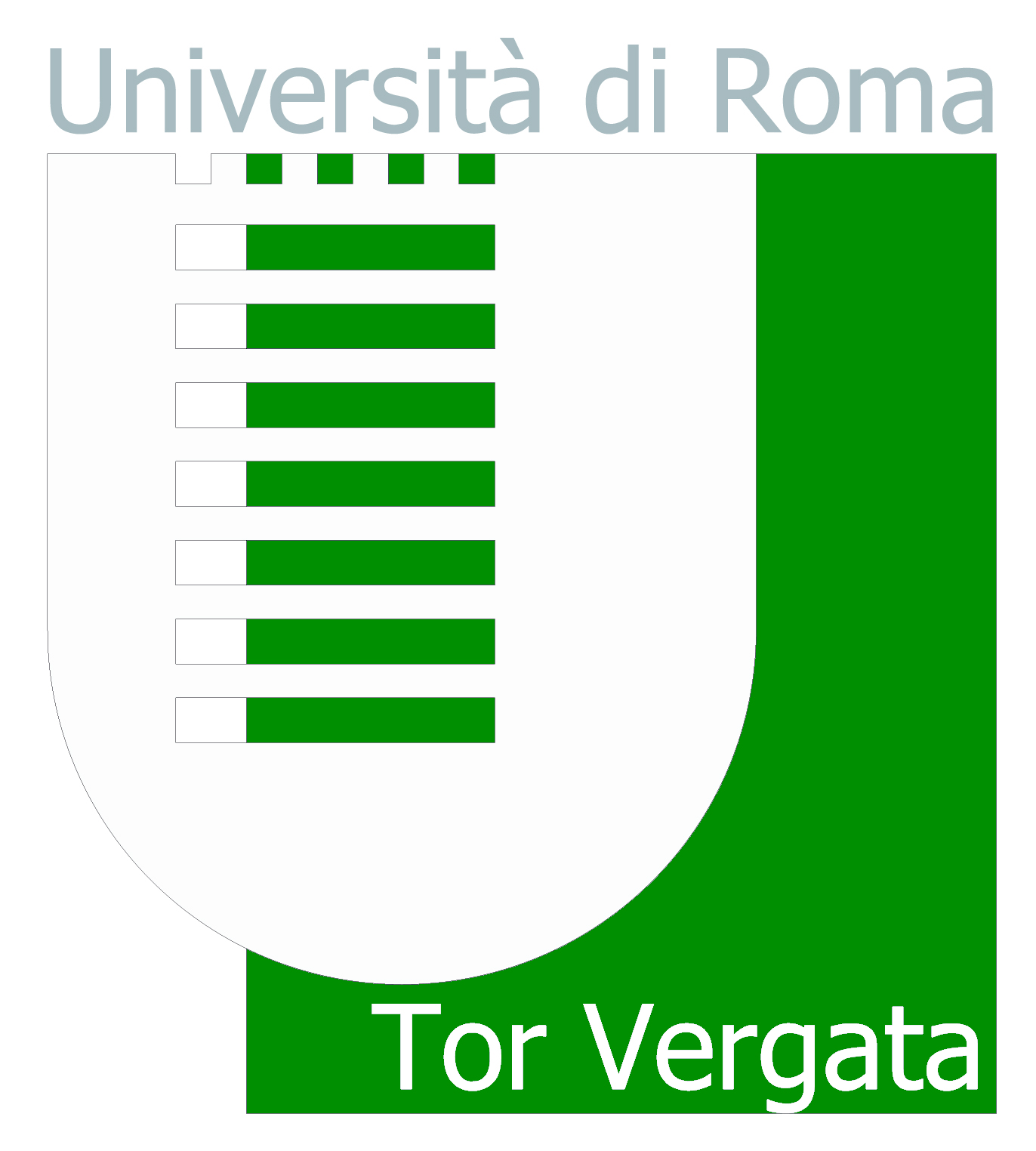}
\end{center}
\end{figure}
\begin{center} \Large{UNIVERSIT\`{A} DEGLI STUDI DI ROMA \\ ``TOR VERGATA''}
\makebox[\textwidth][c]{\hrulefill}
\vspace{1 cm} \\ \emph{\normalsize{Facolt\`{a} di Scienze Matematiche, Fisiche e Naturali}} \\ \normalsize{Corso di Laurea Magistrale in Fisica}
\end{center}
\vfill
\begin{center}
Tesi di Laurea
\end{center}
\vfill
\begin{center}
\Large{\textbf{Search for double beta decay of $^{106}$Cd} \\ Monte Carlo simulation and cosmogenic activation of the $^{106}$CdWO$_4$ detectors}
\end{center}
\vfill
\begin{center}
\Large{\textbf{Ricerca del doppio decadimento beta del $^{106}$Cd} \\ Simulazione Monte Carlo e valutazione dell'attivazione cosmogenica dei rivelatori di $^{106}$CdWO$_4$}
\end{center}
\vfill \vfill
Relatori: \hfill Candidata:\\
Dott. Fedor A. Danevich\hfill Ester Ricci\\
Prof.ssa Rita Bernabei\\
\begin{center}
\makebox[\textwidth][c]{\hrulefill}
\end{center}
\begin{center}
Anno Accademico 2015-2016
\end{center}
 \end{titlepage}
 \newpage
 \begin{flushright}
 
\emph{To my parents, and to Andrea.}
\end{flushright}
\vspace{\stretch{2}}\null
\newpage
\thispagestyle{empty}
 \newpage
 
\thispagestyle{empty}
 \newpage
\selectlanguage{italian}%
\begin{abstract}
  \thispagestyle{empty}

  Il $^{106}$Cd è un isotopo candidato a decadere mediante doppio decadimento $\beta$ positivo, un evento che non è ancora mai stato osservato sperimentalmente. È infatti più complesso da indagare del doppio decadimento $\beta$ negativo, in quanto può avvenire attraverso tre differenti canali: tramite doppia cattura elettronica, tramite l'emissione di due positroni o tramite l'emissione di un positrone e una cattura elettronica. Presenta inoltre una vita media più lunga, e la presenza di vari canali di decadimento rende la funzione di distribuzione di energia molto complessa.

 Il lavoro di tesi si propone di realizzare la simulazione Monte Carlo di un esperimento il cui fine è lo studio del doppio decadimento $\beta$ del $^{106}$Cd. L'esperimento utilizza un approccio calorimetrico; la sorgente del decadimento è infatti un cristallo scintillatore di CdWO$_4$ arricchito al 66\% in $^{106}$Cd.
 
 Il cristallo arricchito è completamente circondato da due rivelatori a scintillazione, anch'essi di CdWO$_4$ realizzati a partire da cadmio naturale. La presenza dei due rivelatori ausiliari permette di selezionare eventi in coincidenza o anticoincidenza fra i tre rivelatori, al fine di distinguere gli eventi di background radioattivo dagli eventi interessanti. La riduzione del fondo è cruciale in questo tipo di esperimenti, in quanto il numero di eventi attesi è estremamente basso. 
 
 La simulazione è stata realizzata utilizzando il tool GEANT4, e riproduce i cristalli rivelatori, le guide di luce e il primo strato della schermatura in rame dell'esperimento. Permette di utilizzare come sorgente delle particelle primarie sia il generatore di eventi di GEANT4 che una sorgente esterna, il software DECAY0. L'utilizzo di una sorgente esterna è necessario in quanto il generatore di eventi non prevede il doppio decadimento $\beta$, oggetto di questo studio.
 
 La simulazione è stata utilizzata per stimare i contributi al fondo radioattivo degli isotopi prodotti dall'attivazione cosmogenica nei cristalli scintillatori. Per stabilire quali siano i possibili contaminanti presenti nei cristalli, naturali e arricchito, abbiamo utilizzato due software, COSMO1 e Activia, che calcolano le sezioni d'urto per i processi di spallazione dei raggi cosmici secondari con i materiali e restituiscono un elenco di possibili contaminati e l'attività prevista per ciascuno di essi in base al tempo di esposizione all'attivazione e di permanenza sotto terra, impostati dall'utente.
 
 Per quanto riguarda il cristallo scintillatore arricchito abbiamo selezionato tra i possibili contaminanti quelli che decadono per emissione $\beta^+$ o $\beta^-$ con un Q-valore della reazione superiore a 1000 keV, un'attività superiore a 0.001 conteggi kg$^{-1}$ day$^{-1}$ e una vita media superiore a 100 giorni.
 Per i due cristalli arricchiti abbiamo preso in considerazione i nuclei con attività superiore a 1 conteggi kg$^{-1}$ day$^{-1}$, il cui Q-valore fosse superiore a 500 keV e la cui vita media fosse maggiore di 100 giorni.
 Questa selezione permette di ridurre a sette il numero di contaminati studiati per il cristallo arricchito e a cinque quelli per i cristalli non arricchiti. Gli isotopi simulati sono $^{102}$Rh, $^{148}$Re, $^{182}$Ta, $^{65}$Zn, $^{108}$Ag, $^{110m}$Ag, $^{174}$Lu per il cristallo arricchito e $^{172}$Hf, $^{182}$Ta, $^{184}$Re, $^{110m}$Ag e $^{173}$Lu per i cristalli naturali.
 
 I risultati delle simulazioni del decadimento di questi isotopi sono stati confrontati con la funzione di distribuzione di energia prodotta dal canale $2\nu\epsilon\beta^+$ del doppio decadimento $\beta$ del $^{106}$Cd, le cui condizioni iniziali sono state calcolate utilizzando il software DECAY0. Questo canale è stato selezionato per la sua vita media, stimata intorno a $\sim 10^{22}$ yr, inferiore a quella del canale $2\nu2\beta^+$ ($\sim 10^{26}$ anni). Il canale $2\nu2\epsilon$ presenta una vita media attesa ancora più breve, ma una contaminazione di $^{113}$Cd del cristallo arricchito rende molto difficoltoso lo studio della regione a bassa energia in cui andrebbero cercate le tracce di un simile processo. 
 
 A partire dal valore atteso per la vita media e dai dati di composizione del cristallo abbiamo ricavato per il canale $2\nu\epsilon\beta^+$ un numero di conteggi attesi all'interno del rivelatore arricchito pari a circa 17 per un'esposizione di un anno.
 
 A partire dal più alto valore di attività fornito dai softwares COSMO1 e Activia abbiamo calcolato il numero di conteggi attesi per ciascun contaminate all'interno del cristallo arricchito. Abbiamo quindi utilizzato i valori ottenuti per normalizzare i vari contributi al fondo radioattivo dell'esperimento, che abbiamo poi sommato insieme per ottenere il background totale. Anche in questo caso abbiamo considerato un anno di esposizione. 
 
 Abbiamo infine effettuato un test dell'efficienza dell'acquisizione in coincidenza tra i rivelatori nella riduzione del backgraund selezionando gli eventi che depositano energia all'interno di un range stabilito attorno a 511 keV nei cristalli naturali. Si osserva che questo tipo di selezione permette di ridurre il rapporto tra il contributo dovuto alle contaminazioni e quello del canale di decadimento doppio beta che stiamo valutando.

   \thispagestyle{empty}
\end{abstract}
\selectlanguage{english}
\newpage
\tableofcontents

\newpage

\section{Introduction: Double beta decay}

The double beta (double-$\beta$) decay is a second order process in the Standard Model (SM). This process was at first proposed by Maria Goeppert-Mayer in 1935 \cite{GM-theory}. 
Two years later, in 1937, Ettore Majorana proposed his theory of the neutrino nature \cite{majorana}. One of the main aspects of this new theory is that, in Majorana's picture, the neutrino and antineutrino are the same particle. The use of double-$\beta$ decay to test this theory was suggested in the same year by Gulio Racah \cite{racah}. In Majorana's theory the neutrino can give rise to a neutrinoless double-$\beta$ decay, as we are going to explain.



Double-$\beta$ decay is a process that can provide us information about two  different open questions on neutrino's nature. The first is whether it is a Dirac or a Majorana particle, and we can at the same time use the information to evaluate neutrino absolute mass \cite{theo1,theo2,theo3,theo4,giunti,exp1,exp2,exp3,exp4}.

There are two other experimental approaches that are used for the search of neutrino mass: the study of the end-point of $\beta$ decay in experiments like KATRIN \cite{katrin}, and a cosmological approach, that uses data from the cosmic microwave background collected by spatial experiments like Planck \cite{plack_neutrini,plack_neutrini2}. 
The nature of the particle, on the other hand, can be distinguished only with the observation of a physical phenomenon that violates the leptonic number conservation, and right now the only theorized process that could have this property is neutrinoless double-$\beta$ decay.

In the Dirac picture, neutrinos and antineutrinos have a different property called leptonic number, usually labeled as L. Neutrinos have L=1, and antineutrinos L=-1. In this picture, neutrinoless double-$\beta$ decay is not possible. However, if the neutrino is a Majorana particle it is massive, and this condition implies that it is not in a leptonic number eigenstate. The particle would be in a superimposition among the L=1 and L=-1 states and the mixing would allow the neutrinoless double-$\beta$ decay \cite{giunti}. 


We notice that a first effect that violates the SM prescriptions for neutrinos has already been observed, and it is the neutrino oscillation. Nowadays, almost all the experiments that study solar neutrinos have registered a lack of events with respect to the expected rate, that can be calculated in particular for neutrinos from p-p reaction. This quantity is in fact related to the solar luminosity $L_{\odot}$ \cite{particle_astro}.

This lack of events is a feature of those detectors that are only sensitive to electronic neutrinos ($\nu_e$), and it disappears in the experiments that are sensitive also to the other neutrino flavors.
The existence of these fluctuations implies that the neutrinos must have a nonzero mass, although very small \cite{particle_astro}. 

Unfortunately, the oscillation experiments can not offer a solution to the question whether the neutrino is a Dirac or Majorana particle or about the absolute scale of the neutrino mass. From the oscillation experiments in fact we can only extract information about the difference squared between the mass eigenstates \cite{giunti}.

The cases of double-$\beta$ decay allowed by the SM are always  accompanied by the emission of two neutrinos or antineutrinos. If the energetic conditions of the nucleus allow a $\beta^-$  two antineutrinos are emitted, whereas, if $\beta^+$ decay is allowed, we deal with two neutrinos.  When double-$\beta^+$ decay is allowed, there are two other channels in competition: double electronic capture (2$\nu$2$\epsilon$) and a mixed mode where an electronic capture and a $\beta^+$ emission happen simultaneously ($2\nu\epsilon\beta^+$) \cite{double_beta_review}.

An isotope candidate to be a double-$\beta^-$ emitter is always a even-even nuclide which pairing forces make more bound than its (Z+1,A) neighbor, but less so than the (Z+2,A) nuclide, as it is shown in figure \ref{fig_2beta_energy}. For double-$\beta^+$ emitter the same conditions are applied to (Z-1,A) and (Z-2,A) nuclei. We will expose later the other conditions that have to be fulfilled by a nucleus to be a candidate for this decay.
As we previously said, if the neutrino is a Majorana particle, all the double-$\beta$ emitters could decay without neutrino emission.

\begin{figure}
 \includegraphics[width=0.5\textwidth]{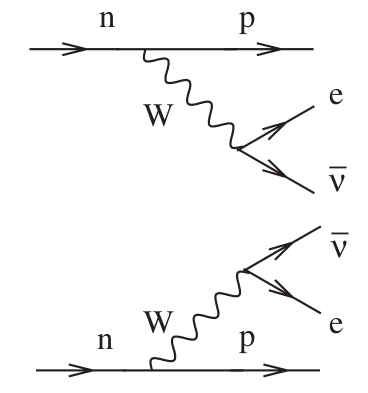}
 \includegraphics[width=0.5\textwidth]{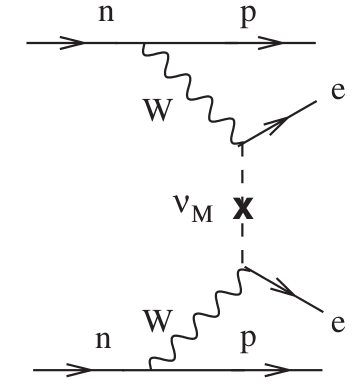}
 \caption{On the left side, the Feynman diagram for $2\nu2\beta$ decay is reported. On the right side the Feynman diagram for $0\nu2\beta$ decay is shown. \cite{double_beta_review}}\label{fig_feynman}
\end{figure}

There are also more exotic processes that involve the emission of hypothetical particles called Majorons, that we won't treat in this work. We only report that the half-lives expected for these processes is even longer than that calculated for the channels we are going to take into account \cite{double_beta_review}. 

The expected rate for double-$\beta$ is proportional to $G_F^4$, where $G_F$ is the Fermi constant \cite{particle_astro}. These decays are then strongly suppressed with respect to the ordinary decay modes, such as $\beta$ or $\alpha$ decay, and we expect the half-lives to be longer than for ordinary decays. 
The neutrinoless double-$\beta$ processes are further suppressed by the proportionality of the transition amplitude to the effective Majorana mass of the neutrino, that can be expressed as:

$$\sum_i U_{ei}^2m_i,$$

in which we label with $m_i$ the mass eigenstates of the neutrino and with  $U_{ei}$ the elements of the neutrino mixing matrix. The extremely small value of this factor gives rise to the rarest process we have ever searched for.

Two-neutrino double-$\beta$ decay has already been observed for some nuclei. Some of them are reported in table \ref{tab_double_beta_active}.
Nowadays, the experimental efforts focus on neutrinoless double-$\beta$ decay search, since it is the key to demonstrate that the neutrino is a Majorana particle and, at the same time, it can give us a value for the neutrino mass. The mass of the particle in fact could be evaluated since it is related to the half-life of the process.

\subsection{Two-neutrino double-$\beta^-$ decay}

The two-neutrino double-$\beta^-$ decay  (2$\nu$2$\beta^-$) conserves lepton number, and does not discriminate between Majorana and Dirac neutrinos. At the same time, it does not depend significantly on the mass of the neutrino \cite{double_beta_review}.
The reaction can be written as:

$$(A,Z)\longrightarrow (A,Z+2)+2 e^-+ 2\overline{\nu}_e,$$

where we use A to indicate the mass number of the isotope, Z for the atomic number, $e$ for the electrons and $\overline{\nu}_e$ for the electron antineutrinos. 

\begin{figure}
 \centering
 \includegraphics[width=0.7\textwidth]{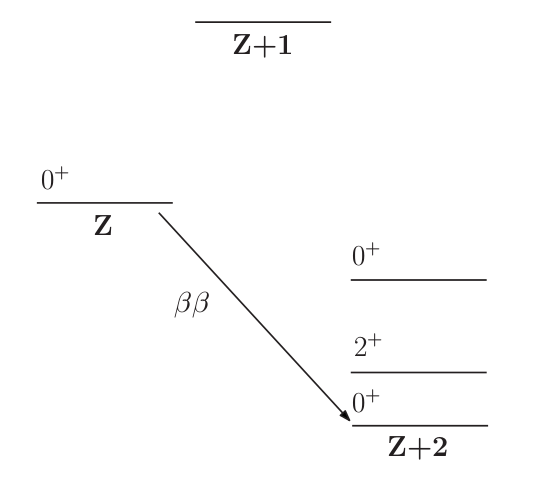}
 \caption{Energetical condition for double-$\beta$ decay. \cite{double_beta_review}}\label{fig_2beta_energy}
\end{figure}

A nuclide for which the sum of the masses of the daughter nucleus (A,Z+2) and the masses of two electrons is lighter than the mass of the father nucleus (A,Z), is a good candidate for double-$\beta^-$ decay. This condition can be written as $M(A,Z\pm2)+2m_e<M(A,Z)$ and it is graphically reproduced in figure \ref{fig_2beta_energy}.


The half-life $T_{1/2}^{2\nu}$ can be theoretically deduced by the relation \cite{double_beta_review}:

$$ (T_{1/2}^{2\nu})^{-1}=G^{2\nu}(Q,Z)|M^{2\nu}|^2, $$

in which we labeled with $G^{2\nu}(Q,Z)$ the four-particle phase-space factor, and with $|M^{2\nu}|$ the nuclear matrix element for the second-order process. 

From this relation it is clear that  the measurement of the half life of double-$\beta$ decay can improve our knowledge on the nuclear matrix elements for the double-$\beta$ emitter, for which there are several different theoretical calculations, that are not in agreement with each other. 
As we will see, a good knowledge of the nuclear matrix elements is necessary if we want to extract information on the neutrino mass absolute value from possible future observations of neutrinoless double-$\beta$ decay. The nuclear matrix elements for the two processes are largely different. Nevertheless, theorists can use the improved knowledge of the nuclear structures achieved by studying two-neutrino mode to improve the calculations for neutrinoless double-$\beta$ decay matrix elements \cite{double_beta_review}.

There are many other reasons to investigate such a rare process, whose experimental search is quite difficult. One  is that the energy spectrum gives us information about the actual decay mode. Several channels have been proposed by theoreticians, and only the measure of the energy spectrum and the evaluation of its shape could enlighten at how this process actually happens.

We want also to remark that 2$\nu$2$\beta^-$ decay has been found for several nuclides by different experiments, and is nowadays the rarest process ever observed by men. A list of measured half-lives of some isotopes is presented in table \ref{tab_double_beta_active}.

\begin{table}
\centering
 \begin{tabular}{lc}
 \hline
 Nuclide & Half-life [yr]\\
 \hline
 $^{48}$Ca &4.4$^{+0.6}_{-0.5}\times 10^{19}$\\
 $^{76}$Ge &$(1.5\pm0.1)\times 10^{21}$\\
 $^{82}$Se &$(0.92\pm0.07)\times10^{20}$\\
 $^{96}$Zr &$(2.3\pm0.2)\times10^{19}$\\
 $^{100}$Mo&$(7.1\pm0.4)\times10^{18}$\\
 $^{116}$Cd&$(2.8\pm0.2)\times10^{19}$\\
 $^{128}$Te&$(1.9\pm0.4)\times10^{24}$\\
 $^{130}$Te&$6.8^{+1.2}_{-1.1}\times10^{20}$\\
 $^{150}$Nd&$(8.2\pm0.9)\times10^{18}$\\
 \hline
 \end{tabular}
\caption{Some of the nuclides for which double-$\beta^-$ decay has been observed. All the half-lives are taken from \cite{double_beta_hl}.}\label{tab_double_beta_active}
\end{table}

\subsection{Two-neutrino double-$\beta^+$, double-$\epsilon$ and $\epsilon\beta^+$  decays}


Although 2$\nu$2$\beta^-$ decay has already been observed for many isotopes (see table \ref{tab_double_beta_active}), double-$\beta^+$ decay is still unobserved. Only lower limits for the half-lives have been set at the level of $10^{18} - 10^{22}$ yr. 

There are three ways for this decay to happen, since electronic capture always acts as a competitive process to $\beta^+$ emission. The three possible reactions are \cite{2beta_plus}:

$$ 2\nu2\beta^+: \ \ \ \ \ (Z,A)\longrightarrow (Z-2,A)+2 e^+ + 2\nu_e, $$
$$ 2\nu\epsilon\beta^+ : \ \ \ \ e^- + (Z,A)\longrightarrow (Z-2,A)+ e^- +2\nu_e,$$
$$ 2\nu2\epsilon: \ \ \ \ \ 2e^- +(Z,A)\longrightarrow (Z-2,A) + 2\nu_e.$$

It is important to notice that the three processes have different Q-values, that can be written as \cite{2beta_plus}:

$$Q_{2\beta^+} = M(A,Z)-M(A,Z-2)-4m_ec^2, $$
$$Q_{\epsilon\beta^+} = M(A,Z)-M(A,Z-2)-2m_ec^2, $$
$$Q_{2\epsilon} = M(A,Z)-M(A,Z-2).$$

The energetic conditions that rule these processes are close to those indicated for double-$\beta^-$ decay. 
A nucleus (Z,A) is a double-$\beta^+$ candidate if the energy of the (Z-1,A) nuclide is higher than for the father nucleus, and the (Z-2,A) nucleus has a lower energy. The difference have to be at least equal to the Q-value for double-$\beta^+$ processes we have indicated \cite{double_beta_review}.


The research of these decays has not been taken into account as has the one about double-$\beta^-$ decay for several reasons. The first we mention is that there are only a few candidates for this process, reported in table \ref{tab_2b_plus}. 


Another reason for this lack of interest is that the expected half-life is higher for this processes than for double-$\beta^-$ decay, due to the fact that the phase space factor is smaller. A theoretical calculation of expected half-lives for double-$\beta^+$ emitters is reported in table \ref{tab_2b_plus}.
Moreover, the competition between the two processes, $\beta^+$ emission and electronic capture, causes an increment of the expected half-life.

Despite all these problems, an attractive aspect of these transitions is the possible use of coincidence signals from four annihilation $\gamma$-ray produced by the two positrons in the $2 \beta^+$ case, or two in the $\epsilon\beta^+$ case, to select interesting events from the background. In order to investigate $2\epsilon$ events, anticoincience techniques can be used.

The Q-value expected for these processes is quite low. This characteristic is another factor to take into account, since the energy spectrum at low energies is usually harder to study because the large number of possible radioactive contaminations that release energy in the range.
This further difficulty makes the possibility to select events by means of coincidence analysis results particularly interesting.

Among the three channels, the Q-value for double-$\epsilon$ is expected to be higher, but the signal from this decay is harder to find because only X-rays are emitted from the nucleus that undergoes such kind of process and there is not positron emission. This channel is then really difficult to detect, making the observation of the process still harder. 

\begin{table}[h]
\centering
 \begin{tabular}{lccc}
 \hline
 Nuclide & T$_{1/2}^{2\nu2\beta^+}$ [yr]& T$_{1/2}^{2\nu\epsilon\beta^+}$ [yr]&T$_{1/2}^{2\nu2\epsilon}$ [yr]\\
 \hline
 $^{78}$Kr &$2.3\times10^{26}$ & $5.3\times10^{22}$ &$3.7\times10^{21}$\\
 $^{96}$Ru &$5.8\times10^{26}$ & $1.2\times10^{22}$ &$2.1\times10^{21}$\\
 $^{106}$Cd&$4.2\times10^{26}$ & $4.1\times10^{21}$ &$8.7\times10^{20}$\\
 $^{124}$Xe&$1.4\times10^{27}$ & $3.0\times10^{22}$ &$2.9\times10^{21}$\\
 $^{130}$Ba&$1.7\times10^{29}$ & $1.0\times10^{23}$ &$4.2\times10^{21}$\\
 $^{136}$Ce&$5.2\times10^{31}$ & $9.2\times10^{23}$ &$1.7\times10^{22}$\\
 \hline
 \end{tabular}
\caption{List of possible double-$\beta^+$ emitters, with a theoretical prediction of half-lives. \cite{2b_plus_list}}\label{tab_2b_plus}
\end{table}

\subsection{Neutrinoless double-$\beta$ decay}



We now introduce a simplified model of the neutrinoless double-$\beta$ decay. 
The final reaction we observe is \cite{particle_astro}:

$$(A,Z)\longrightarrow (A,Z+2)+2 e^-. $$

We present double-$\beta^-$ decay because it is the one most studied, both experimentally and theoretically. The other modes can be easily deduced from this first example. 
For a better understanding of the whole process, we divide it into two virtual intermediate stages \cite{particle_astro}.
 
During the first stage, an antineutrino is produced in the first single $\beta$ decay.

$$ (A,Z)\longrightarrow (A,Z+1)+ e^- +\overline{\nu}_e. $$

If we take into account Majorana's picture, the antineutrino and the neutrino are the same particles, so the antineutrino emitted in the first $\beta$ decay can be absorbed by the daughter nucleus according to the equation:

$$(A,Z+1)+\nu_e\longrightarrow (A,Z+2)+ e^-. $$

Actually, the two processes happen simultaneously, so the effect we see is the one presented in the equation for neutrinoless double-$\beta$ decay.

The existence of this decay also implies that the neutrinos are massive particles. This assumption is a consequence of the fact that helicity is defined to be left-handed for neutrinos and right-handed for antineutrinos. Massless particles are helicity eigenstates, whereas massive particle are in a state that is mixed. To have a mixture of both states in the particles, that is a condition implied in the neutrino and antineutrino equality, the particle must be massive.

\begin{figure}
 \centering
 \includegraphics[width=\textwidth]{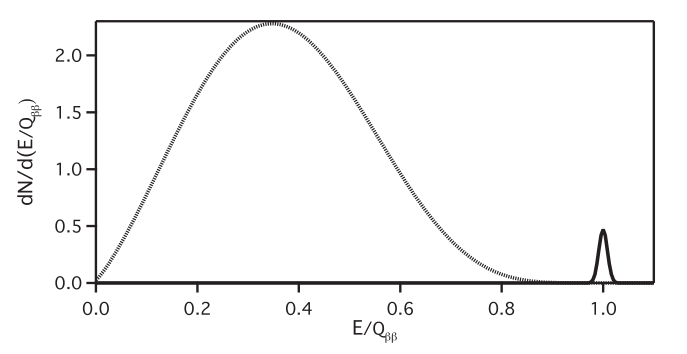}
 \caption{Scheme of the energy spectrum due to $2\nu2\beta$ and $0\nu2\beta$ decay. The case reported in figure refers to a detector with a high energy resolution. In the case of low energetic resolution, the peak from $0\nu$ effect can be superimposed and covered by $2\nu$ decays. \cite{double_beta_review}}\label{fig_2beta_spectrum}
\end{figure}

The experimental search of 0$\nu$2$\beta$ decay is really difficult, but in the case of 0$\nu$2$\beta^-$ help comes from the fact that the signal we expect has some clear characteristics we can use to identify it. In particular, without the neutrino emission almost all the energy released in the reaction is shared between the two electrons, that are easily stopped inside a material. This leads to a complete deposition of the energy inside the detector. So, the experimental signature of a 0$\nu$2$\beta$ decay is a peak at the Q-value of the reaction, as we can see in figure \ref{fig_2beta_spectrum}.
Despite the simple shape we expect, it is not easy to reveal this kind of signal, mainly because of the extremely low rate we expect for this decay.

So, a deeply radiopure setup is needed to perform this kind of measurement. And we always have to take into account the continuous spectrum from 2$\nu$2$\beta$ decay, that is a background source superimposed on the peak. We are not able to avoid this background because the two processes are in competition for the same nuclei. The detectors used in the experiments must then have a sufficiently high energy resolution to allow the separation of the peak from the continuous background.

We previously said that the investigation of 0$\nu$2$\beta$ decay can also give us information about the neutrino mass absolute scale. This is possible if we take into account the half-life of the process. In fact, the half-life of this decay can be related to three factors, as we can see in the formula \cite{particle_astro}: 

$$(T_{1/2}^{0\nu})^{-1}=|m_{\beta\beta}|^2 |M^{0\nu}|^2 G^{0\nu}(Q,Z), $$

where $T_{1/2}^{0\nu}$ is the half-life, $|m_{\beta\beta}|$ is the absolute value neutrino mass, $|M^{0\nu}|$ is the modulus of the nuclear matrix element and $ G^{0\nu}(Q,Z)$ is a phase space factor.

Since the half-life is related to the absolute value of neutrino mass, its measure can be used to evaluate the neutrino mass absolute scale, if we would be able to evaluate independently all the other two parameters.
The second factor is the modulus squared of the nuclear matrix element, that has to be calculated on the basis of our knowledge on nuclear physics. This factor represents the main problem in the evaluation of neutrino mass, because we only have approximate models that describe the many-body interaction in a nucleus.

Several models have been developed, taking into account different aspects of nuclear physics, and different numerical approximations of the many-body interaction have been realized in the last years, each of them leading to different values of the nuclear matrix element. 
The study of $2\nu2\beta$ spectra gives a great help in the determination of these quantities, but the problem of nuclear matrix elements is far from being solved.

The last factor is a phase space factor, that has to be calculated too. But in this case the evaluation is easier, and it has been calculated with a good accuracy for the most important double-$\beta$ emitter candidates. The mathematical procedure for this calculation can be found in \cite{2beta_general}. 


We finally report that Klapdor et al. have presented some analyses suggesting a possible observation of the $0\nu2\beta^-$ decay of  $^{76}$Ge  \cite{hm,hm2} using the Heidelberg-Moskow data; as usual for such relevant arguments, a wide debate has been raised. At present, the GERDA (GErmanium Detector Array) experiment, also at Laboratori Nazionali del Gran Sasso, is further investigating the process, but with different assembling \cite{gerda}.

\newpage

\section{Status of $^{106}$Cd double-$\beta$ experiments}
\subsection{Early experiments}
$^{106}$Cd is one of the most widely studied double-$\beta^+$ candidate, mainly for two reasons. The first is the large energy released in the reaction (Q$_{\beta\beta}=2775.39(10)$ keV \cite{106cd_2012}). This energy sets the expected emission spectrum in a region in which the  background from natural radioactive contaminants is lower than in the low energy range. The second reason for this large use in research is the quite large natural abundance ($1.25\pm 0.06$\% of the natural composition of Cd \cite{106cd_2012}) of the isotope, with respect to other candidates, that are listed in table \ref{tab_2b_plus}.

Moreover, this isotope has some interesting properties from a theoretical point of view. The estimated half-life for the two-neutrino mode of the $2\epsilon$ (double electronic capture) and $\epsilon\beta^+$ (one electronic capture and one positron emission) is in fact at the level of $T_{1/2}\sim10^{20}- 10^{22}$ yr, in a range reachable with the present low counting techniques, whose sensibilities are at the level of $10^{20}-10^{21}$ yr. This predicted half-life is widely shorter than the results of the theoretical calculation for the other double-$\beta^+$ emitters, indicated in table \ref{tab_2b_plus}.



\begin{figure}[hb]
 \centering
 \includegraphics[width=0.8\textwidth]{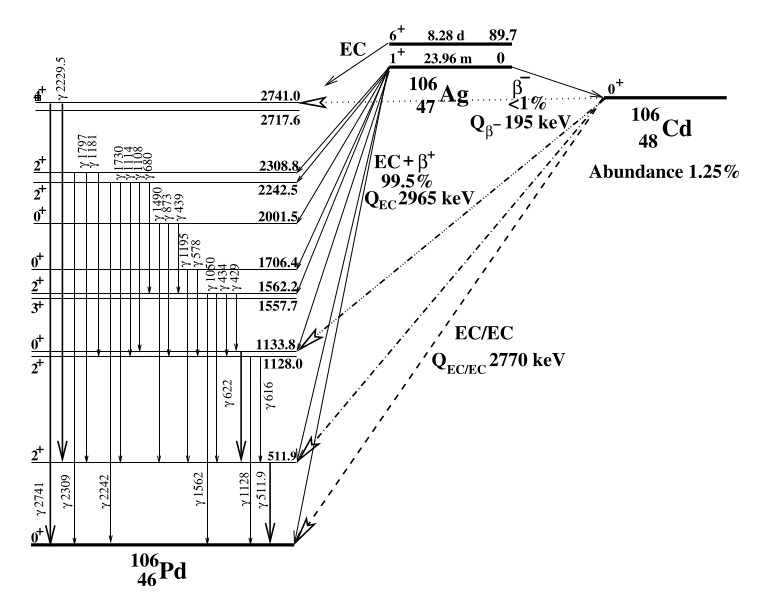}
 \caption{$^{106}$Cd decay scheme. \cite{tgv}}\label{fig_excited}
\end{figure}

Another  interesting property is that in the case of $0\nu$ capture of two electrons, the energy released from the K shell is  2727 keV. This value is close to the energy of an excited level of the daughter nucleus $^{106}$Pd, as we can appreciate in figure \ref{fig_excited}. This particular condition could give rise to a resonant enhancement of the 0$\nu$2$\epsilon$ capture.

A long series of experiments have chosen to investigate the double-$\beta$ decay of this interesting isotope. A summary of the obtained results related to $^{106}$Cd double-$\beta$ decay search are summarized in table \ref{tab_experiments}.

All these measurements can be divided in two main groups. Some experiments use Cd samples, sometimes enriched in $^{106}$Cd, and external detectors, others experiments use detectors containing cadmium. This last method is called ``calorimetric''. The calorimetric approach has two main advantages: the efficiency is higher, and it is possible to resolve easily between two-neutrino and neutrinoless mode.
For the search of double-$\beta$ decay of cadmium by means of this second approach, several detectors have been developed: CdTe or CdZnTe \cite{cobra}, and CdWO$_4$ crystal scintillators \cite{detector_development}.


\begin{table}
{\relsize{-3}
 \begin{tabular}{lcc}
 \hline
 Description & T$_{1/2}$ limit [yr] &Year [Ref. No.]\\
 \hline
 Cd samples between photographic emulsions & $\sim 10^{15}$ ($0\nu2\beta^+$, $0\nu\epsilon\beta^+$) & 1952 \cite{exp_1}\\
 Cd foil in a Wilson cloud chamber & $6 \times 10^{16}$ ($0\nu2\beta^+$) & 1955 \cite{exp_2}\\
 Cd sample between two NaI(Tl) scintillators in coincidence & $(2.2-2.6)\times 10^{17}$ ($2\beta^+$ & 1984 \cite{exp_3}\\
 & $(4.9-5.7)\times 10^{17}$ $(\epsilon\beta^+)$ & \\
 & $1.5\times 10^{17}$ ($2\nu2\epsilon$)& \\
 $^{116}$CdWO$_4$ crystal scintillator& $(0.5-1.4)\times 10^{18}$ $(0\nu2\beta^+)$ & 1995 \cite{exp_4}\\
 & $(0.3-1.1)\times 10^{19}$ ($0\nu\epsilon\beta^+$)& \\
 & $5.8\times 10^{17}$ ($2\nu2\epsilon$)& \\
 CdWO$_4$ crystal scintillator & $2.2 \times 10^{19}$ ($0\nu2\beta^+$) & 1996 \cite{exp_5}\\
 &$9.2\times 10^{17}$ ($2\nu2\beta^+$) & \\
 &$5.5\times 10^{19}$ ($0\nu\epsilon\beta^+$& \\
 &$2.6\times 10^{17}$ ($2\nu\epsilon\beta$)& \\
 Cd sample measured by HPGe detector & $1.0\times 10^{19}$ ($2\beta^+$)& 1996 \cite{exp_6}\\
 &$(6.6-8.1)\times10^{18}$ ($\epsilon\beta^+$) &\\
 &$(3.5-6.2)\times10^{18}$ ($2\epsilon$) &\\
 CdTe cryogenic bolometer & $1.4\times 10^{16}$ ($0\nu\epsilon\beta^+$) & 1997 \cite{exp_7}\\
 $^{106}$Cd sample between two NaI(Tl) scintillators in coincidence & $(1.6-2.4)\times 10^{20}$ ($2\beta^+$) & 1999 \cite{exp_8}\\
 & $(1.1-4.1)\times 10^{20}$ ($\epsilon\beta^+$)&\\
 & $(3.0-7.3)\times10^{19}$ ($2\epsilon$)&\\
 $^{116}$CdWO$_4$ crystal scintillators&$(0.5-1.4)\times 10^{19}$ ($2\beta^+$)& 2003 \cite{exp_9}\\
 &$(0.1-7.0)\times10^{19}$ ($\epsilon\beta^+$)&\\
 & $(0.6-8.0)\times 10^{18}$ ($2\epsilon$)&\\
 \hline
 \end{tabular}
 }
 \caption{Previous experiments about $^{106}$Cd double-$\beta$ decay \cite{106cd_2012}.}\label{tab_experiments}

\end{table}

\subsection{Other ongoing experiments}

Among all the experiments performed with $^{106}$Cd, we focus on the two that are still ongoing. We deal with an experiment that uses a calorimetric approach, COBRA \cite{cobra}, and an experiment that uses an external source coupled with semiconductor detectors, TGV-II \cite{tgv}.

\subsubsection{COBRA}

The COBRA experiment uses CdTe and CdZnTe semiconductor crystals, that contain nine double-$\beta$ emitter candidates. We expect double-$\beta^-$ emission from $^{116}$Cd, $^{130}$Te, $^{90}$Zn, $^{128}$Te and $^{114}$Cd, and double-$\epsilon$, $\epsilon\beta^+$ or double-$\beta^+$ emission from $^{108}$Cd, $^{64}$Zn, $^{120}$Te, $^{106}$Cd. 
The most promising isotope for 0$\nu$2$\beta$ decay in this setup is $^{116}$Cd, because with its high Q-value (2809 keV for 0$\nu$2$\beta^-$ channel  \cite{cobra}), it is above the gamma lines from U and Th natural background \cite{tgv}.

The setup installation started in 2006. The experiment consists in a layer of $ 4\times4$ detectors, each of them is $11 \times 11 \times 11$ mm$^3$. The total mass of the detectors is 103.9 g \cite{cobra}. The array is housed in an inner copper shield 5 cm thick, and it is surrounded by 20 cm of lead. The setup is then surrounded by a Faraday cage and by a neutron shield, made up of a 7 cm thick boron-loaded polyethylene plates, and finally there are 20 cm of paraffin wax. COBRA is located at the Laboratori Nazionali del Gran Sasso, that provides a 3500 m w. e. shielding from muons \cite{cobra}.

The setup is calibrated using external sources of $^{22}$Na, $^{57}$Co and $^{228}$Th, and with the photopeaks due to internal contaminations in the low background spectrum. A linear relation between the energy deposition and the peak position in channel is found.

We only focus on the results concerning the isotope we are interested in, $^{106}$Cd. The experiment have to take into account that the 0$\nu$2$\beta^+$ spectrum is more complex than the single line we expect for 0$\nu$2$\beta^-$ decay. We expect several different lines, due for example to the escape of an annihilation gamma photon at 511 keV. We can also expect, in addition to the full energy peak and the mentioned escape peak, other satellite peaks from additional X-ray escape \cite{cobra}.
In order to evaluate all these contributions, COBRA collaboration realized a Monte Carlo simulation based on GEANT4 libraries to reconstruct all the features of the spectrum, and then used this model to fit the data from the measurements, that were previously selected by omitting runs that resulted too populated.

This experiment studies a large number of double-$\beta $ emitters, and $^{106}$Cd is not one of the most favorable. So the main part of released data are related to other emitters. In \cite{cobra} we find a table of upper limits set for some decay modes of our isotope, that range between $1.6 \times 10^{17}$ yr ($0\nu2\beta^+$ mode to ground state) and $4.6 \times 10^{18}$ yr ($0\nu2\beta^+$ mode to 512 keV state). 
All these results are orders of magnitude below the limit set by other experiments, as we can see if we compare them with table \ref{tab_experiments}.

\subsubsection{TGV-II}
 
The TGV-II experiment is operating at the Modane Underground Laboratory (4800 m w. e. \cite{tgv}). 
The experimental setup is a low background spectrometer, made up by 32 HPGe planar detectors, each of them with a sensitive volume of $20.4$ cm$^2 \times 0.6$ cm. The basic cell of the detector is a sandwich-like structure, with two face-to-face detectors and a thin foil of cadmium placed between them.
The 16 couples are mounted one over the other in a common cryostat.
The total mass of the setup is about 3 kg of germanium, and the sensitive volume is about 400 cm$^3$. The energy resolution ranges between 3.0 and 4.0 keV at 1332 keV \cite{tgv}.
The efficiency of the setup is estimated to be $50 \div 70 \%$ \cite{tgv}, depending on the energy threshold.

A passive shield is also present to reduce contaminations from the environment. At first, near the detector tower there is a copper layer copper ($\geq 20$ cm thick), then a lead layer ($\geq 10$ cm thick) and a box where a purified gas is flushed to avoid radon contamination. At last a neutron shielding of borated polyethylene ($16 cm$) closes the setup. 
A further selection is performed in the data analysis.

Two runs were performed with $^{106}$Cd samples. In the first, 12 foils of enriched material ($75\%$ of $^{106}$Cd) and four of natural Cd were used. The total number of $^{106}$Cd atoms was $4.249(2)\times 10^{22}$\cite{tgv}. The samples were exposed for 8687 hr.
The data analysis focused its attention on 2$\nu$2$\epsilon$ decay channel, in particular on the transition to the ground state of the daughter nucleus $^{106}$Pd. Results on this measure were largely limited by background, whose sources were located on the central part of the end cap of the cryostat. The replacement of the part reduced the background by a factor 4.

The second run used only enriched foils, so the total number of $^{106}$Cd rose to $\sim 5.787(2)\time10^{22}$ \cite{tgv}. The attention focused on the search for resonant 0$\nu$2$\epsilon$ decay in addition to the 2$\nu$2$\epsilon$ channel. In order to detect this signal, whose energy is expected to be about 2741 keV, the energy range was extended up to $\sim 3$MeV in one of the spectroscopy channels for each detector. The other was kept at the same range of phase I (up to 800 keV), preserving the high sensitivity for low X-ray emission of double electronic capture.
Unfortunately, inside the four additional cadmium foils contaminations by $^{241}$Am were found, so the data that came from them have been excluded, with a consequent reduction of the exposed mass.

Analysis of the two run data has been performed, in particular by searching for double coincidences between the two characteristic palladium KX-rays in neighboring face-to-face detectors. Data obtained do not present any evidence of a significant signal for 2$\nu$2$\epsilon$ events. Only a lower limit can be set using the formula

$$T_{1/2}=\frac{\ln 2 \cdot N_0 \cdot \eta \cdot t}{N_{EXCL}}, $$

where $N_0$ is the number of examined nuclei, $t$ is the exposition time, $\eta$ is the detector efficiency and $N_{EXCL}$ is the number of excluded signal events.

From the data and the indicated equation, the following limits have been set for 2$\nu$2$\epsilon$ decay of $^{106}$Cd, for phase I and II respectively: \cite{tgv}

$$T_{1/2}(2\nu2\epsilon,g.s.)\geq 3.0 \times 10^{20} y,$$
$$T_{1/2}(2\nu2\epsilon,g.s.)\geq 3.6 \times 10^{20} y, $$

As we said, during phase II data have been analyzed to search also for resonant 0$\nu$2$\epsilon$ decay. For this analysis single particle energy spectra and spectra released in events with high multiplicity have been taken into account, with a special care for the cases in which KX-rays of palladium were registered. The best result was found in the second case, searching the 2741-2762 keV doublet in the total energy spectra of neighboring detectors in double coincidence events.

All the 16 foils were examined, since the $^{241}$Am contamination does not affect this channel. From simulation, an efficiency of $0.67(3)\%$ \cite{tgv} was estimated. No events were observed in the region of interest, so only an upper limit can be set:

$$T_{1/2}(0\nu2\epsilon,2741 keV)> 1.1\times 10^{20} y.$$

The TGV collaboration plans to continue the studies on $^{106}$Cd with larger masses and higher isotopic enrichment.

\subsection{Search  for double-$\beta$ decay of $^{106}$Cd with enriched $^{106}$CdWO$_4$ detector}

\begin{table}[b]
\centering
\begin{tabular}{ lc}
 \hline
 Density [g/cm$^3$]&7.9\\
 Melting point [°C]&1271\\
 Structural type & Wolframite\\
 Hardness [Mohs] & 4 - 4.5\\
 Wavelength of emission maximum [nm]&480\\
 Refractive index & 2.2 - 2.3\\
 Decay Time & 1.1 $\mu$s (40\%),14.5 $\mu$s(60\%)\\
 \hline
\end{tabular}
\caption{Main characteristics of CdWO$_4$ scintillators. \cite{cdwo4_study}}\label{tab_cdwo}
\end{table}

\subsubsection{Enriched  $^{106}$CdWO$_4$ crystal scintillator }

The core of the experiment is a CdWO$_4$ crystal scintillator, enriched in $^{106}$Cd. The main properties of this kind of detector is reported in table \ref{tab_cdwo}. In addition to the properties reported in the table, we want also to remark that these detectors have good energy resolution. For a 1 cm$^3$ detector it can reach a value of 6.7\% \cite{knoll}, and it is around 10\%-12\% in bigger crystals \cite{knoll}. A limit to the resolution is the high refractive index, 2.3 \cite{knoll}, that makes the optical coupling with light-guides and photomultiplier tubes (PMTs) difficult.
The light yield is about 40\% of NaI(Tl).

The CdWO$_4$ crystal scintillators have already been used for double-$\beta$ decay studies of $^{116}$Cd in the Solotvina Underground Laboratory and at Laboratori Nazionali del Gran Sasso, and the two-neutrino mode was observed with a half life of $T_{1/2}=(2.9^{+0.4}_{-0.3}\times 10^{19})$ yr \cite{detector_development}. The measurements were performed with a detector isotopically enriched in $^{116}$Cd. 
These experiments demonstrated that CdWO$_4$ crystals possess several important properties that allow the enhancement of the sensitivity in double-$\beta$ experiments. In particular, it is possible to grow crystals whose intrinsic radioactivity is extremely low, and with good scintillation characteristics.  Moreover, these detectors offer the possibility to perform pulse-shape discrimination, are relatively cheap and have a good stability in operation conditions.

The possibility to realize detectors that contain the double-$\beta$ emitter maximizes the detection efficiency with respect to the cases in which the source of the radiation is external to the detector. On the other hand the experiments that use external sources allow us to obtain more detailed information about the decay, like the tracks of emitted $\beta$ particles, their individual energies and angle of emission, but the efficiency of the detection is widely poorer. 
We prefer the calorimetric approach, in which the source and the detector coincide, because of the extremely low rate of the processes we are studying.

The production of the enriched crystals requires a certain number of steps, that we briefly enumerate.
The first step is the purification of the enriched materials, that is required at first to obtain good scintillation characteristics, and also to reduce the radioactive background. Then the CdWO$_4$ must be synthesized and crystal must be grown. 
The enriched cadmium used for the detector construction has the isotopic composition reported in table \ref{tab_106cd_composition}.

\begin{table}[b]
\centering
\begin{tabular}{ccc}
\hline
Cd isotope & Relative abundance [\%] & Natural abundance [\%]\\
\hline
106 & 66.40$\pm$0.05 &1.25$\pm$0.06\\
108 &0.658$\pm$0.004 &0.89$\pm$0.03\\
110&$5.06\pm0.01$&$12.49\pm0.18$\\
111&$4.83\pm0.01$&$12.80\pm0.012$\\
112&$8.85\pm0.03$&$24.13\pm0.21$\\
113&$3.935\pm0.003$&$12.22\pm 0.12$\\
114&$8.77\pm0.03$&$28.73\pm0.42$\\
116&$1.497\pm0.008$&$7.49\pm0.18$\\
\hline

\end{tabular}
\caption{Cadmium isotopical composition of the enriched detector. Natural cadmium values are given in the second column.\cite{detector_development}} \label{tab_106cd_composition}

\end{table}

\begin{figure}
 \centering
 \includegraphics[width=0.8\textwidth]{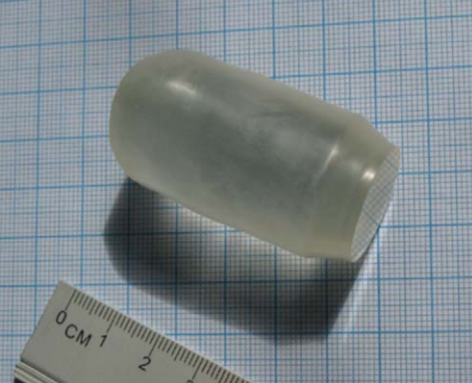}
 \caption{Enriched detector: the crystal has a total mass of about 215.8 g, and the size is $\oslash 27 \times 50$ mm$^2$. \cite{detector_development}}\label{fig_enr_detector}
\end{figure}

After the growth of the crystal, all the optical properties were evaluated. The most important for our aims are the transmittance and the scintillation properties.
The transmittance was measured in the spectral range between $300-700$ nm using a UV-vis spectrometer along the longer axis of the crystal. The results obtained are displayed in \cite{detector_development} and show the sample has good transmission properties in the range of the emission spectrum of scintillation light.
An explanation of the goodness of the optical properties is the deep purification of the initial materials \cite{detector_development}.

The scintillation properties were tested using several $\gamma$ sources. For the test, the detector was optically coupled to a photomultiplier tube and surrounded by a reflecting cup. 
An energy resolution 10.0\% (FWHM) was obtained for the 662 keV $\gamma$ line of $^{137}$Cs. It is quite good if we take into account the irregular shape of the crystal and if we compare the result with the data in literature \cite{knoll}. 

\subsubsection{First experiment with the enriched $^{106}$CdWO$_4$ crystal}

One of the first experiments performed in Laboratori Nazionali del Gran Sasso using the enriched crystal ended in 2012, after 6590 hours of exposition. 
The setup was installed inside the DAMA/R\&D barrack.
The crystal was set inside a cavity in the central part of a polystyrene light guide. The cavity was then filled using high-purity silicon oil. 
Two high purity quartz light-guides were optically coupled with the polystyrene at the opposite extremities. The scintillation light was collected using two low-radioactive PMTs.

The passive shield was made up by copper bricks, that completely surrounded the setup, and a copper box in which pure nitrogen was flushed enclosed this first shield.
The copper box was surrounded by other passive shields, namely a 10 cm thick layer of copper, a 15 cm thick low radioactive lead layer,  1.5 mm of cadmium and a variable (from 4 to 10 cm) layer of paraffin. 
While copper and lead are used to shield the setup from $\gamma$ quanta and charged particles in general, cadmium and paraffin have to reduce the neutron flux, that is one of the most dangerous background sources in underground experiments, since it is higher in underground spaces than in open spaces.
The paraffin reduces the neutron energies, and the cadmium foil absorbs them.
All the shield was contained in a plexiglass box, continuously flushed with high-purity nitrogen \cite{first_enriched}.

Data analysis allows us to investigate the enriched crystal background. Several techniques have been applied to determinate and analytically reduce the background.

\begin{figure}
 \centering
 \includegraphics[width=0.8\textwidth]{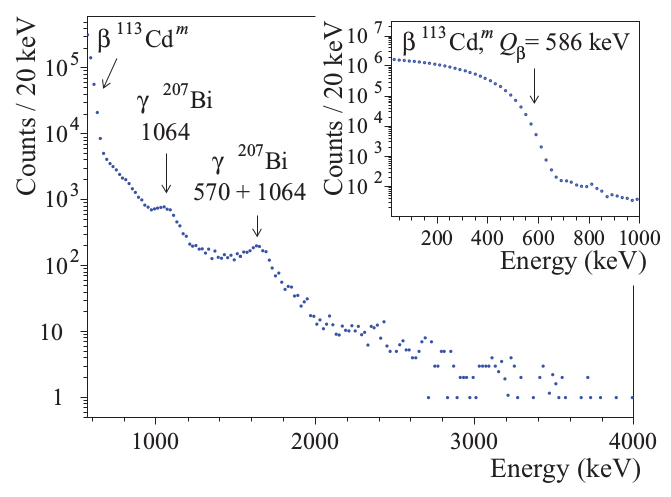}
 \caption{Energy spectrum measured with the $^{106}$CdWO$_4$ crystal scintillator. The inset shows the dominating $^{113}$Cd beta activity for energy lower than 0.65 MeV. The exposure for this figure is 283 hr. \cite{first_enriched} }\label{fig_113activity}
\end{figure}

At first we notice that the low energy region of the spectrum is dominated by $^{113m}$Cd $\beta$ emission, as we can see in figure \ref{fig_113activity}. The activity measured for this isotope is $116\pm4$ Bq/kg \cite{first_enriched}. The contamination due to this nuclide was found also in TGV experiments \cite{tgv}. The presence of this contamination excludes the possibility to investigate the $2\nu2\epsilon$ channel, since the energy range up to 600 keV is dominated by the $\beta^-$ emission from $^{113m}$Cd.

Contributions to the background above the energy of 0.6 MeV was analyzed and compared with the results of Monte Carlo simulations to identify the background sources. The comparison was performed only after the data from the experiment underwent some cuts, that select the interesting events from some components of the background. In particular, time-amplitude and pulse-shape discrimination techniques were applied to the collected events.

Since the techniques used for the data analysis in this first measurement are used also for the subsequent experiments performed using the same enriched crystal, we will spend some words about them, starting from the pulse-shape discrimination method.
To each signal, an appropriate transformation is applied. The optimal one was proposed by Gatti and De Martini in 1962 \cite{gatti}. A numerical indicator, called shape indicator (SI) is calculated \cite{first_enriched}:

$$SI = \frac{\sum f(t_k) \times P(t_k)}{\sum f(t_k)},  $$

in which the  sum runs over the time channels, starting from the origin of signal up to 50 $\mu$s. $f(t_k)$ is the digitized amplitude of the signal in the time channel $t_k$. The weight function $P(t)$ is defined as  \cite{first_enriched}:

$$P(t)=\frac{f_{\alpha}(t)-f_{\gamma}(t)}{f_{\alpha}(t)+f_{\gamma}(t)}, $$

where $f_{\alpha}(t)$ and $f_{\gamma}(t)$ are the reference pulse-shapes for $\alpha $ particles and $\gamma$ quanta, measured for the crystal.

\begin{figure}
 \centering
 \includegraphics[width=0.8\textwidth]{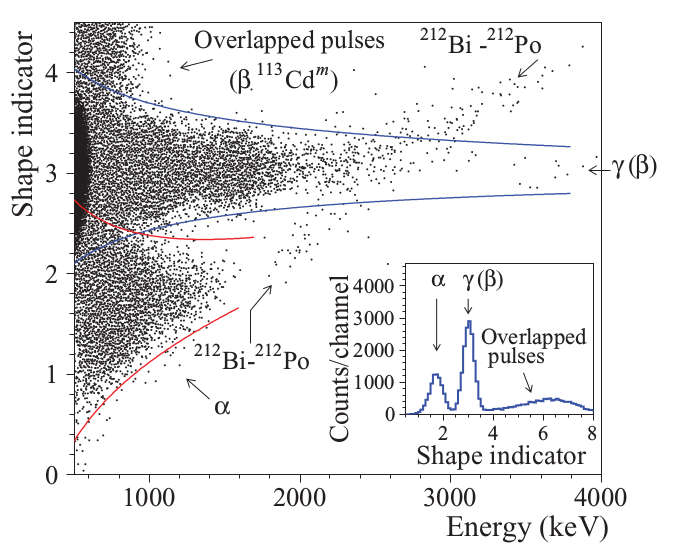}
 \caption{Pulse shape discrimination for the signal from the enriched crystal. Also several other features are indicated. \cite{first_enriched} }\label{fig_pulse_shape}
\end{figure}

In figure \ref{fig_pulse_shape}, the scatter plot of the shape indicator as function of the energy deposited inside the detector is shown. We can clearly see that the events are disposed in two separate structures that can be associated to $\alpha$ and electromagnetic events.
This index allows us to reject the background produced from $\alpha$ events.

From pulse-shape analysis it is also possible to identify the fast chain produced by the decay of $^{212}$Bi$\longrightarrow^{212}$Po, that always produces double pulses, whose shape indicator is clearly identifiable in the scatter plot in figure \ref{fig_pulse_shape}. The index for these events is slightly different from $\alpha$ or single electromagnetic events.

The electromagnetic background has been then reconstructed using Monte Carlo simulations, after the spectrum was cleaned from $\alpha$ events.
After the background identification and reconstruction, no peculiarities from double-$\beta$ decay are visible. So, only a lower limit can be set for a long list of possible channels. All the results are presented in \cite{first_enriched}. In general, the limits found were of the order of $10^{19}-10^{21}$ yr for the different channels at 90\% confidence level.

\subsubsection{$^{106}$CdWO$_4$ inside GeMulti setup}

\begin{figure}[h]
 \centering
 \includegraphics[width=0.8\textwidth]{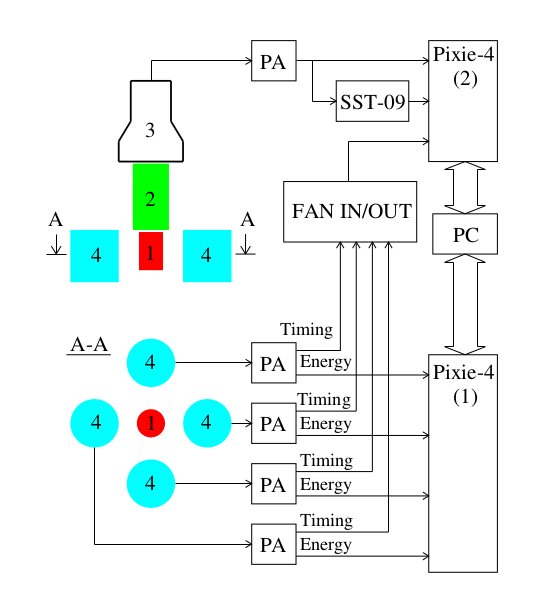}
 \caption{GeMulti setup with $^{106}$CdWO$_4$ scintillator inside, lateral and top-down view. The electronic chain is shown on the right. \cite{gemulti} }\label{fig_gemulti}
\end{figure}

Figure \ref{fig_gemulti} shows the experimental setup of the second experiment performed in Laboratori Nazionali del Gran Sasso by the DAMA group with the enriched $^{106}$CdWO$_4$ crystal. The setup was set inside the STELLA facility of the laboratory, in the cryostat of the GeMulti spectrometer. GeMulti is composed of four ultra-low background HPGe detectors, that share the same cryostat. The HPGe detectors are labeled with the number 4 in figure \ref{fig_gemulti}. Each of them has a volume of approximately 225 cm$^3$.

The light from the enriched $^{106}$CdWO$_4$ crystal scintillator, labeled with the number 1 in figure \ref{fig_gemulti}, is viewed through a lead tungstate (PbWO$_4$) crystal light-guide by a low radioactive photomultiplier tube (PMT), labeled with the number 2. The light-guide was developed from deeply purified archaeological lead. It is the same light-guide we use in the present setup of the experiment, that we are going to describe.

The use of coincidence and anticoincidence between scintillator and germanium detector signals increases the experimental sensitivity to the double-$\beta$ processes with emission of $\gamma$ quanta.

The setup has an event-by-event acquisition system, based on two four channel digital spectrometers (Pixie-4). One of them is used to collect the data for the HPGe detectors (Pixie(1) in figure \ref{fig_gemulti}), whereas the other is connected to the PMT that collects the light from $^{106}$CdWO$_4$ scintillator (Pixie(2) in figure \ref{fig_gemulti}). The second Pixie unit also receives a trigger pulse when the amplitude of the $^{106}$CdWO$_4$ signal exceeds $\sim 0.6 $ MeV, to avoid acquisition of large amounts of events caused by the decay of $^{113m}$Cd ($Q_{\beta}=586$ keV \cite{gemulti}). This Pixie module also receives a signal that is the sum of the timing signal from the four HPGe detectors, in order to be able to select coincidence between the scintillator and the germanium detectors. 

After 13085 hours of exposition, data were released \cite{gemulti}.
Contributions from possible radioactive contaminants have been simulated using EGS4 code. The list includes the possible contaminants of the crystal, the light-guide, and mainly the PMT. Moreover, the background from double-$\beta$ decay of $^{116}$Cd has been taken into account.
Response to the several decay modes of $^{106}$Cd has also been simulated using the EGS4 model and the software DECAY0, that allows simulations of rare processes not implemented in ordinary Monte Carlo simulators as EGS and GEANT.

The collaboration does not find any peculiarities that could be ascribed to double-$\beta$ processes from $^{106}$Cd. Then, only an upper limit can be given, according to the formula \cite{gemulti2}:

$$\lim T_{1/2}=\frac{\ln 2 N \eta t}{\lim S}, $$

where $N$ is the number of $^{106}$Cd nuclei in the enriched crystal, $\eta$ is the detection efficiency, $t$ is the time of measurement and $\lim S$  is the number of events of the effect searched that can be excluded at a given confidence level.

Different analysis have been performed for the different decay channels to estimate the limits on half-lives. For example, the best fit to investigate the channel $0\nu\epsilon\beta^+$ has been achieved in the interval 1000-3200 keV, and the estimated number of counts is $S=27\pm49$, that means no evidence of signal \cite{gemulti2}.
According to Feldman-Cousins procedure, this value corresponds to $\lim S=109$ at 90\% confidence level \cite{gemulti2}.

The data analysis must also take into account the efficiency of the setup, that can be evaluated by Monte Carlo simulations, and it is quantified to be 69.3\% \cite{gemulti2}, and the efficiency in pulse shape discrimination to select $\beta$ and $\gamma$ events, that results to be 95.5\% \cite{gemulti2}.
From these data, the limit that can be set is:

$$T_{1/2}\geq 1.1 \times 10^{21} yr, $$

that is very close to some theoretical predictions, that fix the half-life for this channel at $T_{1/2}=(1.4 - 1.6)\times 10^{21} yr$. \cite{gemulti2}

Limits set for other channels can be found in \cite{gemulti}. The comparison between the data obtained with this coincidence method and previous results regarding the same isotope suggests that the improving of the efficiency in the $\gamma$ quanta detection can allow a further improvement in the sensitivity of the experiment, that could be able to investigate ranges of half-lives very close to the ones obtained from theoretical calculations. 
The experiment that is the subject of this work has exactly this aim, as we will see.


\newpage

\section{A new stage of experiment with $^{106}$CdWO$_4$ detector}
\subsection{$^{106}$CdWO$_4$ setup in coincidence with large volume CdWO$_4$ detectors}

 The experimental setup around which this work has been performed is located inside the DAMA/CRYS structure inside the Laboratori Nazionali del Gran Sasso. 

\begin{figure}[H]
 \centering
 \includegraphics[width=0.8\textwidth]{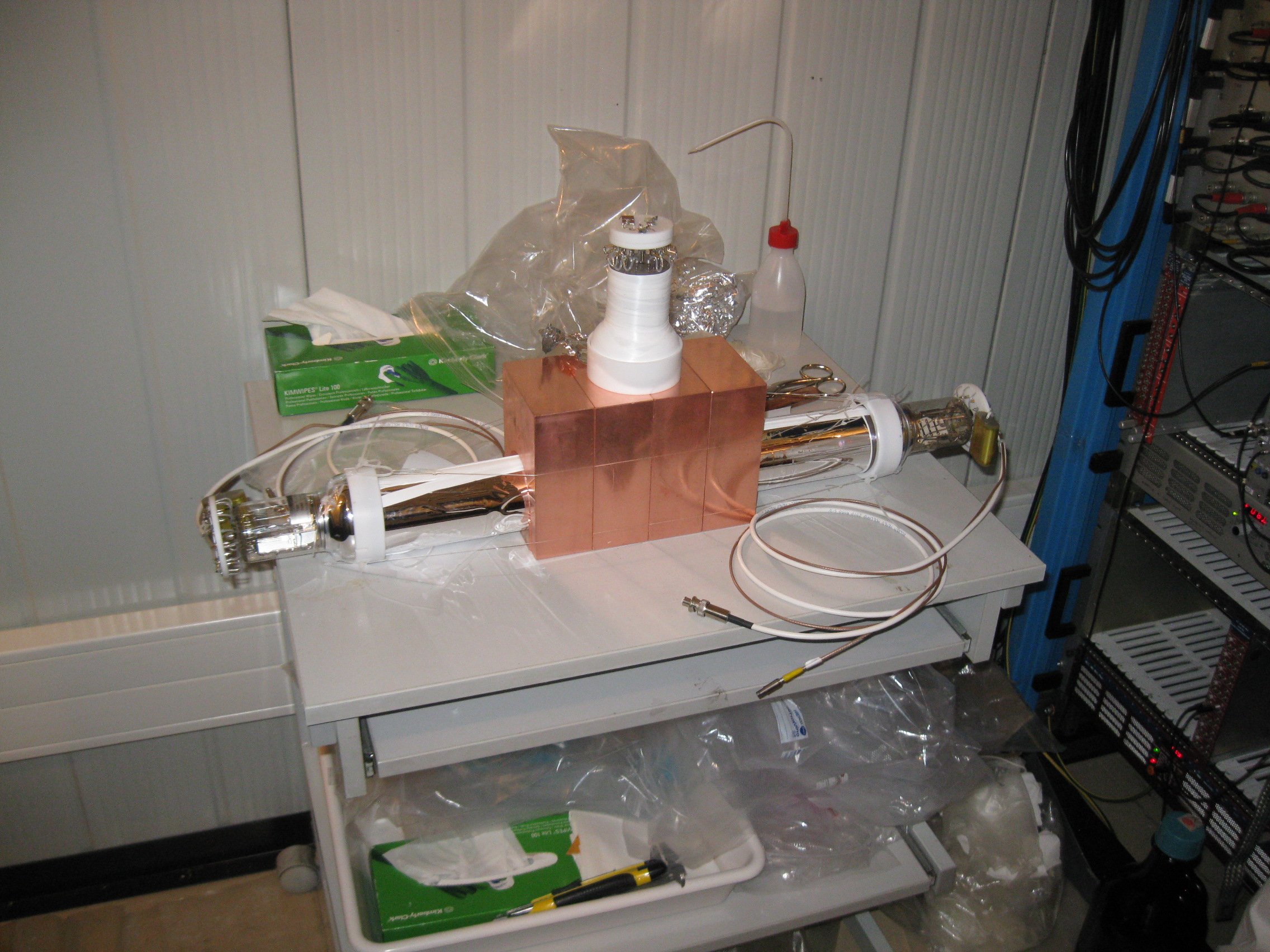}
 \caption{Preliminary assemble of the full setup.}\label{fig_full_setup}
\end{figure}

In figure \ref{fig_full_setup} we can see the experiment after the preliminary assembly of the setup. 
The core of the experiment is the enriched $^{106}$CdWO$_4$ detector we described in the previous section. It is at the same time one of the sensitive elements and the source of the radiation we want to study.
This means that our experiment is based on a calorimetric approach, whose advantages have already been explained in previous sections.


The enriched detector is surrounded by two CdWO$_4$ scintillation detectors, made up using natural cadmium. These two detectors can be used to acquire data in a coincidence mode. The coincidence mode is really important for this measurement because we expect $\beta^+$ emission. The positrons emitted in the reaction annihilate with the electrons of the surrounding matter and emit characteristic back-to-back 511 keV $\gamma$ quanta, that are stopped with a very good efficiency by the scintillators. Anticoincidence mode can be used to select the events of double electronic capture.

This new experiment has the same theoretical conception of the previous one, realized inside GeMulti setup, but uses CdWO$_4$ detectors instead of low-background HPGe detectors. 
The new setup improves both the geometrical and the absolute detection efficiency. The geometrical efficiency improves because the enriched detector is completely surrounded by the natural crystals, and the absolute efficiency because the high atomic number Z of the CdWO$_4$, in particular of tungsten, enhances the cross-section for $\gamma$ quanta absorption.



\begin{figure}
 \centering
 \includegraphics[width=0.8\textwidth]{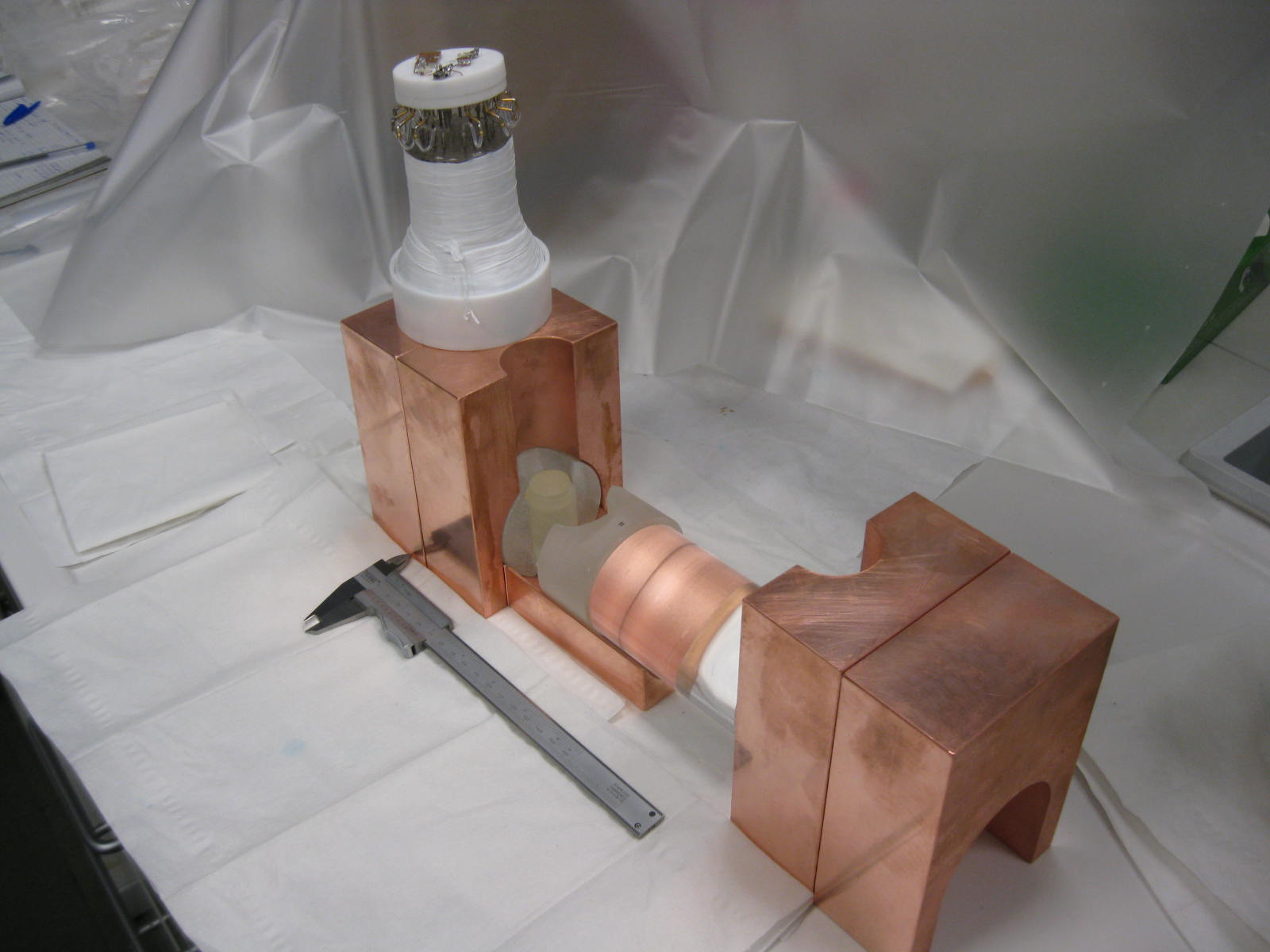}
 \caption{View of the detectors inside the primary copper shield. }\label{fig_detectors}
\end{figure}

In figure \ref{fig_detectors} we can see the enriched detector and the two  $^{nat}$CdWO$_4$ scintillators. In this picture we can also appreciate the shape of the two scintillators, that have been built to encompass the enriched one.
The three detectors are covered by a material which, together with the opacity of the surfaces of the detectors that can be appreciated in figure \ref{fig_detectors} and in figure \ref{fig_enr_detector}, diffuse the scintillation light and make the detector response more uniform. This artifice reduces the dependency of the output light from the event position. 

Above this material, a layer of mylar covers the detector, to avoid scintillation light loss. These materials are effective on the visible scintillation light, but do not affect the $\gamma$ quanta diffusion, and in particular the 511 keV emission we search from $\beta^+$ annihilation, that can reach the two detectors we use for coincidence measure.

Each detector is connected to a photomultiplier tube by a light-guide. The enriched detector is optically coupled to the same PbWO$_4$ light-guide used inside the GeMULTI setup in the previous configuration of the experiment. The PbWO$_4$ crystal is highly photosensitive, so during the preparation of the setup we had to take care to not allow contact with direct light. Its properties could be ruined by the exposition to the light.

The light-guide that is optically coupled to the natural detector is made up of two different pieces. The first one is a quartz cylinder, the other is a plexiglass cylinder. The diameter and the length of the two pieces are the same. 
The use of this double light-guide, whose elements have been carefully optically coupled using a silicon gel, is due to the fact that we want to introduce a considerable space between the active part of the experiments and the PMTs, that are the main sources of radioactive contaminations in  the whole setup. 
The light-guides are wrapped in a reflective material. 


Optical coupling between the detectors, the light-guides and the PMTs is a process that must be performed with the maximum care. Silicon gel is used, and it must be enough to allow the coupling, but at the same time the gel must be limited only to the areas in which it is necessary. After the coupling, a careful cleaning of the surfaces is performed.

The crystals, detectors and light-guides, are posed on a copper base, that has a cylindrical shape. The detectors and the PbWO$_4$ light-guide are then completely surrounded by copper, as we show in figure \ref{fig_full_setup} and in figure \ref{fig_detectors}. The copper pieces are shaped to completely surround the crystals, both the detectors and the PbWO$_4$ light-guide.

To keep the enriched detector in the most favorable position between the two natural CdWO$_4$ detectors, a base made of teflon is used. Teflon is also used to keep the PMTs in the correct position with respect to the light-guides. In fact, PMT diameter is always bigger than the light-guides one. The teflon rings give more stability to the setup.
PMTs connected to natural CdWO$_4$ detectors are also connected to each other by teflon wires, to enhance the stability of the whole setup. 

\subsection{DAMA/CRYS passive shield and data acquisition}

\begin{figure}
 \centering
 \includegraphics[width=\textwidth]{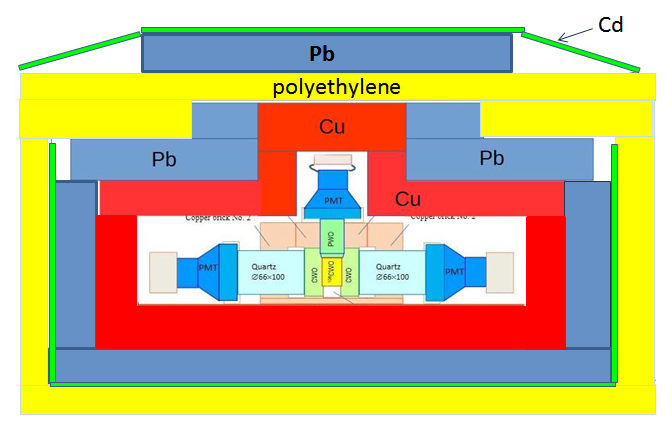}
 \caption{Scheme of the passive shield of DAMA/CRYS setup.}\label{fig_shield}
\end{figure}

DAMA/CRYS is a low background facility realized for low counting experiments that uses scintillating detectors.
Its passive shield is made of a first layer of high-purity copper 11 cm thick, followed by a lead layer 10 cm thick. These two layers reduce the $\gamma$ activity from external sources.
A further shield is made up of cadmium, 2 mm thick, and polyethylene 10 cm thick. These layers are introduced to avoid neutron contaminations \cite{shield}.

The whole setup is sealed and flushed with high purity nitrogen gas to avoid contact of the detector with environmental air, that contains a large amount of radioactive radon.
The inner volume available to install the detectors is (82$\times$16$\times$14) cm$^3$\cite{shield}.
A scheme of the passive shield is reported in figure \ref{fig_shield}.

\begin{figure}
 \centering
 \includegraphics[width=\textwidth]{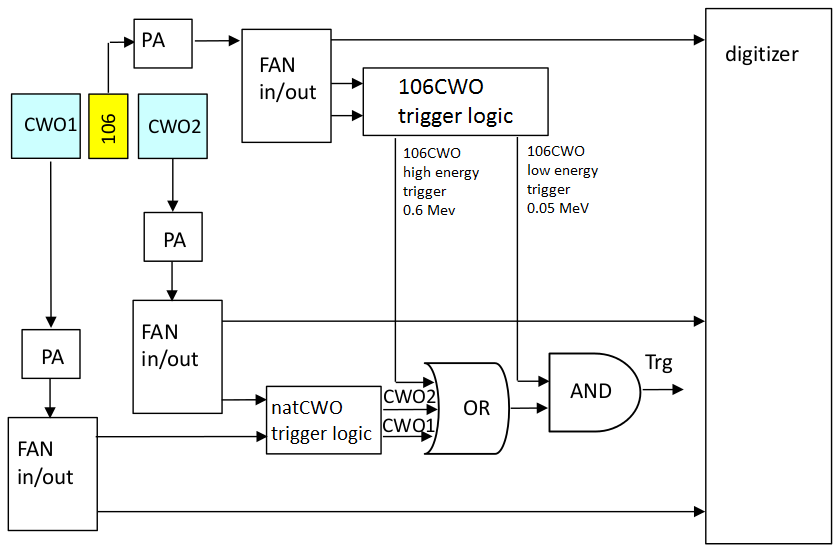}
 \caption{Electronic readout of the setup. 106 labels the enriched detector, CWO1 and CWO2 are the two natural detectors.}\label{fig_electronics}
\end{figure}

The readout of the setup is reported in figure \ref{fig_electronics}. The signal from the three detectors is sent to a preamplifier module (PA). Then, it is doubled by a FAN in/out module. A first copy of the signal is sent to the digitizer module, the second copy to an especially developed module that produce the logical signal we use as trigger. 

The module that treats the output of $^{106}$CdWO$_4$ detector has two different output signals, defined as ``high'' and ``low'' energy trigger. The ``high'' gate become true when the impulse collected from the detector crosses the value of $\sim$0.6 MeV, whereas the ``low'' gate refers to a threshold of $\sim$0.05 MeV.
A single signal is produced from the signal collected by the two $^{nat}$CdWO$_4$.

The logic gate ``AND'' becomes true if it is reached by a signal from the ``low'' gate of the enriched detector trigger module and from a positive signal from the ``OR'' logic. The ``OR'' takes as input the two signals from the natural crystals and the signal from the ``high'' gate of the enriched crystal trigger. Its output is positive if at least one of these signal is positive.

This choice of trigger allows the setup to work in both coincidence and anticoincidence mode. The double output used to select the events that involve only the enriched crystal is used to avoid the collection of the energy deposited below $\sim$0.6 MeV. Under this threshold the rate of counts is too high to investigate the double-$\beta$ decay of $^{106}$Cd, since the crystal is contaminated by $^{113}$Cd, as we reported before (see figure \ref{fig_113activity}). 

%

 
\subsection{Calibration measurements}

\begin{figure}
\centering%
\subfigure[]{
 \includegraphics[width=0.45\textwidth]{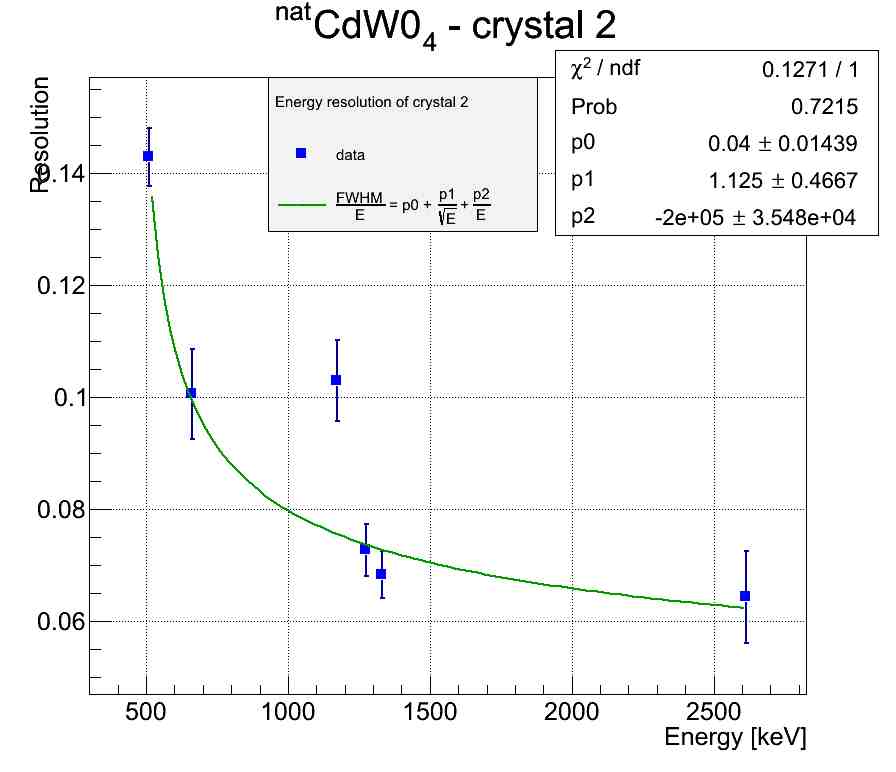}}\qquad
 \subfigure[]{
 \includegraphics[width=0.45\textwidth]{./figures/res_nat1.jpg}}\qquad
 \subfigure[]{
 \includegraphics[width=0.45\textwidth]{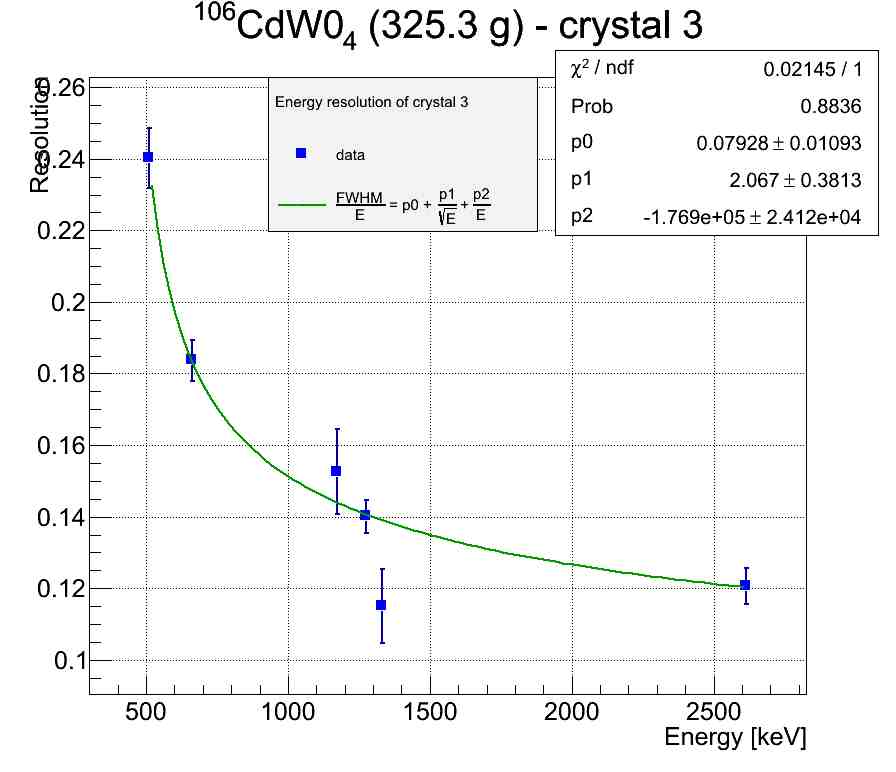}}
\caption{Calibration of the crystals inside the setup.}\label{fig_calibration}
 \end{figure}

The energy resolution in a crystal scintillator depends on the energy deposited inside the detector. The numerical values that rule this dependency are obtained from the calibration of the crystals with $\gamma$ sources. 
The calibration has been performed using different sources that emit inside the energy interval we are interested in studying, from some hundreds of keV to $\sim$3 MeV.
Calibration curves are reported in figure \ref{fig_calibration}.

The data are fitted using the formula:

$$\frac{FWHM}{E} = p_0 + \frac{p_1}{\sqrt{E}} +\frac{p_2}{E}. $$

For the $^{106}$CdWO$_4$ detector the values obtained for the parameters are:

$$p_0= 0.079\pm 0.011, $$
$$p_1 = 2.1\pm 0.4 \ keV^{1/2}, $$
$$p_2=(-1.8\pm0.2)\times 10^{5} \ keV. $$

For the $^{nat}$CdWO$_4$ we report an average value between the results obtained for the two detectors, that are comparable:

$$p_0= 0.053\pm 0.010, $$
$$p_1 = 0.9\pm 0.3 \ keV^{1/2}, $$
$$p_2=(-2.2\pm0.2)\times 10^{5} \ keV. $$


\newpage
\section{Monte Carlo simulation of the experiment}

\subsection{GEANT4 software}

The main part of this work is the realization of a simulation of the experiment using the software GEANT4, developed by CERN laboratories in Geneva. The program aim is to simulate the interactions between radiation and matter. Developed at first for high energy application, it is now extremely versatile, and with the correct implementation in the code, it gives good results also in low energy range, in which we are interested in the present work.

GEANT4 versatility comes from the structure of the program. Almost all the code, C++ based, has to be written by the user, because the source only provides a huge number of virtual classes that the program expects to find to correctly simulate the characteristics of the experiment the user is interested in. This choice is really interesting, because it leaves to single users the full control of the processes involved in the simulation. 

GEANT4 is a very complicated tool, and a full description of all its capabilities is beyond the aims of this work. We only give some hints to allow the reader to understand the complexity of the work performed. For more complete description it is possible to refer to the documentation provided by CERN at \cite{geant_support}.

We briefly introduce the steps that are necessary to build a simulation.
The first step in the simulation construction is to build the structure of the experiment whose answer we want to simulate. We have to fix geometry, material and position in the space of each component of the experimental setup. These three characteristics are given to the object in three passages, that must be repeated for each component of the simulation. The program is also able to produce pieces equal to the ones already built, to simplify the process.

We define the geometry by building the volume shape. GEANT4 offers a large number of classes of solids that can be used alone or in combination. The structure and the properties of the classes are thought to have a great versatility. Sum, subtraction and intersection are the operations allowed by the program for the solids. For each operation, reciprocal position and rotation between the volumes can be set. If the shapes are particularly complex, an implementation from programs for computer-aided design (CAD) is possible.

The geometrical volume is then used to define a ``logical'' volume. This step assigns a material to the volume. GEANT4 has a defined  database that contains a large number of materials among the most used in nuclear physics. If some of the materials we are dealing with are not in this database, we can construct them by indicating the density and the chemical composition. In our simulation, quartz and PbWO$_4$ have been defined by the user, whereas all the other materials have been selected from the internal material database.

Then, the logical object must be set inside the surrounding environment. The simulation structure expects to find a maximum logical volume, called ``world'', to be returned from the detector construction class, inside which all the processes we are describing take place. All the other pieces must be located inside this volume, or inside other volumes already located inside the ``world''. This is the last step of geometry construction. The logical volume is rotated and posed in the right place with respect to the other components of the setup. At this step we can use the function that allows the replication of a single solid.
This last process returns a ``Physical volume''.

A further step is required for the volumes from whom we want to extract information. They have to be marked as ``touchable''. This assignment labels the volume with a unique identifier, used by the software to correctly reconstruct the events inside the setup.
We use this label to associate to these volumes a user defined class that inherits from ``G4VSensitiveDetector'' virtual class. These functions allow us to extract both the energy deposited in the volume by radiation and the length of the tracks of the particles. This second information is useless for our purpose, because we are building the simulation of a calorimeter, and we will not include it in our class. It is instead the most important quantity to study when we are studying the answer of a tracker detector.

Once the geometry of the setup is built, we have to establish the physical processes we expect in the radiation-matter interaction. GEANT4 offers some of these ``PhysicsLists'', but they are not completely reliable in the low energy range, under 1 GeV. So for our aim we wrote a personalized class, based on the virtual class ``Penelope'', that is optimized for the range we are dealing with. 
After these classes have been written and compiled, the simulation is ready to work. 

Now, we focus on the real simulation mechanism.
The simulation is divided into single events. An event contains at least one primary particle. In the case of radioactive decays, the particles can be even more than one. 
When a simulation is running, the software follows every particle, starting from the primaries, and for each it builds a series of ``steps''. In each step, the mean free path of the particle in the material is calculated, and the particle keeps the initial direction, that is defined  by the user or calculated from the radioactive decay for the first step. If the path does not hit a boundary between different volumes, the particle moves for a length equal to the mean free path. At the end of the path the interaction probability is calculated. If the particle meets a boundary, it stops and the interaction probability is calculated. 

If the particle survives the interaction, its motion continues according to the new characteristics produced by the interaction itself. Interactions can change the direction or the energy of the particle, or both.

If the interaction produces other particles, the program keeps them in memory, but it follows the primary until it finishes its motion or escapes from the ``world'' volume defined by the geometry.
Then, the program starts to follow the first generated particle according to the same rules exposed for the primary, and so on until all the tracks have been calculated.

At the end of one event, the program is able to store the information about the energy deposited in the ``touchable volumes'', or the track length if we are studying a tracker, in a output file, then it starts to manage the following event.


GEANT4 is able to produce different kinds of output files. Our analysis is performed by using the software ROOT, also developed at CERN for data analysis in particle physics. So, we set the simulation to obtain ROOT files as output. Inside our file we have different objects. In particular, we have a TTree object, that works as a database, in which we insert the energy deposited in each of the three detectors for every single event.

We also have three histograms, that report the spectrum of the energy deposited in each detector. Obviously, we can also obtain the same spectra from the data inside the TTree structure. 

We are going to use mainly the data inside the TTree structure for our analysis, since the histograms can not provide any correlation between the data. And we are interested in selection of coincidence events between the detectors under some conditions that we will explain in the data analysis section.

\subsection{Simulation of the experiment}

To perform the study of cosmogenical contamination inside the crystals, four different versions of the simulation have been prepared.
The simulations use the same geometry, physics list and data collection structures, but are distinguished by two features. Two of them have the enriched crystal as the source of the decay, whereas the other two have the natural crystals. For each source, we use two simulations because one is prepared to use the GEANT4 internal particle generator, whereas the second one works with external sources of primary decay. 
It is obviously possible to produce a single simulation with all these features but, as a starting project, we think this solution is easier to realize and at the same time provides the results we need. The unification of the four simulations can be performed as a future development of this work.

The use of an external source for primary decay is mandatory, since the GEANT4 decay generator is not able to simulate the  double-$\beta$ decay we are interested in. 
We will explain later the software we used to produce the primary decay, named DECAY0. We want at first to focus on our simulation features.

\begin{figure}
 \centering
 \includegraphics[width=0.8\textwidth]{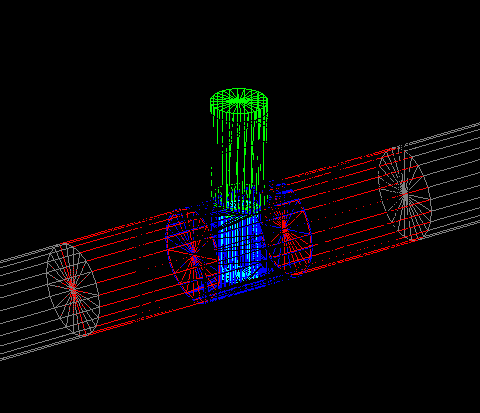}
 \centering
 \includegraphics[width=0.8\textwidth]{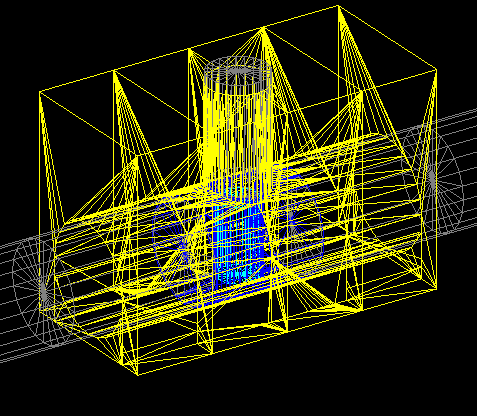}
 \caption{Simulation geometrical structure. In the first picture detectors and light-guides are evidenced. The cyan cylinder is the enriched crystal, the natural crystals are in blue. The PbWO$_4$ light-guide is in green and the quartz light-guides are in red. In the second picture the yellow components represent the copper shield around the core of the setup.}\label{fig_simulation_geom}
\end{figure}

In figure \ref{fig_simulation_geom} we can appreciate the geometry constructed for the simulation. It can be easily compared with figures \ref{fig_full_setup} and \ref{fig_detectors}. In the first picture inside figure \ref{fig_simulation_geom} the colors indicate the scintillating crystals and the light guides, whereas in the second one the first layer of the copper shield is shown in yellow. The geometrical shapes are correctly modeled and located inside the space. 

As we said, one of the main differences between the simulations is the place in which the events take place. To produce the decays exactly inside the selected regions, we produce at first a decay in a bigger volume using a random number generator. 
The generator produces a random number between 0 and 1. For each events, we produce three of these values that are then modified in order to indicate a point inside the volume that is the source of the decays.

Then we check, using the GEANT4 tools, that the point where the decay happens is actually inside the selected volume. This check is particularly significant in the case of natural crystals, because their shape is more complicated than the one of the enriched crystal. 
In fact, we produce the points for these decays without taking care of the empty space that contains the enriched crystal. The points generated inside that volume are refused by the GEANT4 check, that is based on the name we assigned to the physical volumes in the construction class. For those events that are rejected, a new point is created, until it is inside the right volume.

A second distinction between the simulations is the possibility they give to use external sources for primary particle generation.
The use of these two different sources is binding, because the code we use to produce double-$\beta$ decay, DECAY0, is able to produce the initial conditions of the decay only for a limited number of isotopes among the most common inside this kind of experiment, whereas from cosmogenical activation we could find isotopes whose initial conditions are not provided by DECAY0. 

The event generator provided by GEANT4 is instead able to produce also these decays, once we provide it the atomic and mass number of the isotope.

An important characteristic we have to insert inside the simulation code is the detector resolution. GEANT4 software in fact gives as output the exact value of the energy deposited inside each volume. The resolution of the system has been introduced taking into account the values obtained of the calibration of the setup using $\gamma$ sources, previously exposed.




\subsection{DECAY0 code description}

The DECAY0 code was developed to generate the initial conditions of events in the low energy nuclear and particle physics, in particular for double-$\beta$ decays, that are the main interest in this development, and radioactive nuclide decays \cite{decay0}.
The program is able to generate initial conditions for forty double-$\beta$ emitters, selected to be the most interesting in research applications, for the most used calibration sources and for the most dangerous and frequent contaminants that are usually found in low-background nuclear physics experiments.

The program is divided in two main sections. The first, named INIT, searches and collects all the parameters of the selected nuclide and its decay. The data are acquired from several libraries. One example is NuDat \cite{nudat}.
The collected information regards decay modes, their probabilities and energy release, half-lives. Also parameters of nuclear levels and  particles emitted in de-excitation processes are registered by the program \cite{decay0}.

The second process, named GENDEC, generates the particles produced from the decays using a Monte Carlo code. The code generates the energy, the time of emission, and the direction and polarization for each emitted particle. The code is able to produce electrons and positrons for single and double-$\beta $ decay, $\alpha$ particles, protons and neutrons, as well as $\gamma$ and X photons from nuclear de-excitation processes.

\begin{figure}
 \centering
 \includegraphics[width=\textwidth]{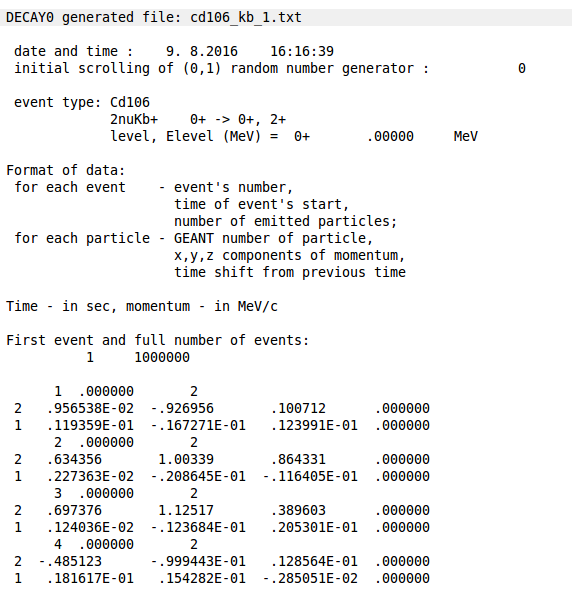}
 \caption{First lines of DECAY0 produced file. We see the header information, regarding the time in which the file has been written, the selected isotope and decay mode. An explanation of the data organization follows. We also see the lines related to the first events. }\label{fig_decay0}
\end{figure}

The output file produced by the DECAY0 code has a header in which the input parameters are summarized. Then, the decays are reported in the following way: a first line contains the number of the event inside the simulation, the absolute time of decay and the number of particles involved in the process. The file header is shown in figure \ref{fig_decay0}.

Then, for each particle produced a line reports the kind of particle, using the same code that GEANT4 uses to identify particles, the three components of impulse vector and the delay between the emission of the single particle from the absolute time given for the events.

In figure \ref{fig_decay0} we can notice that all the particles in a event are emitted simultaneously; their delay time is always 0, as we expect for double-$\beta$ decay.

We use this software only for the simulation of double-$\beta$ decay, since the contaminations we are going to take into account are not provided, so we now focus on the features the software offers to produce such kind of data. 

DECAY0 provides a large number of possible channels for double-$\beta $ processes, and each of them can be separately simulated. At first, it is able to produce data about double-$\beta^-$, double-$\beta^+$, double-$\epsilon$ and $\beta^+\epsilon$ decays, in both two-neutrino and neutrinoless modes.
Each transition can reach ground or excited level, and in the latter case, also de-excitation emissions are listed inside the simulation.
For all the listed modes, several channels are taken into account. Each of them produces a different shape in the spectrum deposited inside the detector, in particular inside the enriched one, in which we expect to have the main effect due to positrons. Some of the proposed channels involve Majoron emission, or more exotic decays. All the information about the channels the software offers are listed in \cite{decay0}.

In our simulation we decide to focus on $2\nu\epsilon\beta^+$ decay. Its expected half life is short enough to produce a signal in a year of measurements, and the signal is also easy to identify in coincidence mode, because the emitted positron annihilates and produces the characteristic $\gamma$ quanta emission at 511 keV, that can be registered by the natural CdWO$_4$ detectors.

So we produce a file from DECAY0 containing 1$\times 10^6$ decay events, that we used inside the GEANT4 simulation.

\newpage

\section{Cosmogenic activation of the detectors}

Since double-$\beta$ decay is a very rare process, an ultra-low radioactive background is required inside the experimental setup and in particular inside the detectors.

The direct contribution from cosmic rays can be usually neglected, because the experiments are always located deep underground, and the contribution from environmental radioactivity can be shielded using both passive and active shielding techniques.
The principal contribution to the background is then due to the intrinsic radioactive impurities inside the detector and the surrounding materials.
The main difficulty for studying quantitatively the cosmogenic activation is the lack of accurate information on the isotope production cross sections for the targets of interest in double-$\beta$ experiments \cite{cosmogenic_general}.

Cosmogenic radioactivity is mostly produced in the atmosphere, but it is also relevant at ground level. At the earth's surface, the isotopic production is dominated by interactions with  neutrons ($\sim$95\%), whereas protons are a minor contribution ($\sim$5\%). \cite{cosmogenic_general}
High energy neutrons interact mainly by spallation processes. Activation due to primary cosmic rays interaction is negligible since most of these particles are absorbed by the atmosphere before they reach the ground level.
The production rate R of an isotope with a decay constant $\lambda$ by exposure to cosmic rays can be calculated from the formula

$$R=\int \sigma(E) \phi(E)dE, $$

where we labeled with $\sigma(E)$ the cross section of the process, that depends on the energy of the impinging particle, and with $\phi(E)$ the flux of neutrons with a given energy $E$.

So, the activity induced after a time of exposure $t_{exp}$ can be calculated using the relation

$$A= R(1-\exp(-\lambda/t_{exp})) .$$

To calculate the cosmogenical activation in a target material, we need three main ingredients. The first is the history of the material, namely the time it has spent at sea level and the time it has spent underground. Then, we need the cosmic ray flux spectrum, and finally the cross section for the spallation reactions between neutrons and target materials, namely all the materials used to build the experiment. 
We have to focus particularly on the detector materials and the immediately surrounding pieces.

Cross sections are without doubt the elements that introduce the main uncertainties on the activation studies, so the main effort has to be dedicated to them. In the following exposition, we will analyze the different ways in which the programs we used obtain the information about cross sections.

The only way to avoid cosmic ray activation is keeping materials inside a proper shielding against the hadronic components of the cosmic rays, but this is not always possible, since the materials must be manufactured in order to realize the different elements of the experimental setup \cite{cosmogenic_general}. 
Also, flying must be avoided, since at higher altitudes the neutron flux can be up to two orders of magnitude greater that at sea level. 
At last, the total time the materials spend outside the deep underground sites must be minimized.

Since direct activation measurements are long and expensive, and the facilities that allow their performance are not numerous, the use of a nuclear reaction code is an appealing solution \cite{cosmogenic_general}.

These codes can use two different strategies. The first is the use of semi-empirical formulae to calculate the cross sections of isotope production, and the second is the use of Monte Carlo simulation of the hadronic interaction of impinging nucleons with target nuclei.

Both the programs we used in this work follow the first approach, whose main advantage is the shorter time it requires for the calculations. The most famous Monte Carlo codes like GEANT4 and FLUKA can use the second approach, but discussion of this kind of approach is outside the aim of this work. More information about it can be found in \cite{cosmogenic_general}.

%

We will now spend some words on the description of the two tools we used, named Activia and COSMO1, but at first we will report the cross-section semi-empirical calculation method proposed by Sildeberg and Tsao used by both the softwares \cite{sildeberg_rao}.

\subsection{Cross-section calculation}
The cross-sections for isotopic production due to a proton or a neutron beam impinging the target material are calculated using the set of semi-empirical formulae from Sildberg and Tsao \cite{sildeberg_rao}. 

The general formula that rules the production of a given result (Z,A) from a target (Z$_t$,A$_t$), in the case of an impinging nucleon with energy E, given in MeV, is \cite{activia_description}:

$$\sigma =\sigma_0 f(A)f(E) e^{-P\Delta A} e^{-R|Z-SA + TA^2 + UA^3|^{\nu}}\Omega \eta \xi.$$

In the previous formula we find several parameters that have to be explained. At first, the normalization factor $\sigma_0$ is the cross-section in millibarns at the energy $E_0\sim 3$ GeV, above which the cross sections are assumed to be independent from the energy of incident particle. The two functions $f(A)$ and $f(E)$ are correction factors usually applied in the cases in which A$_t>30$ and when the difference (A$_t$-A) $>10$. This difference is labeled $\Delta A$ in the formula.

The first exponential term, in which we find $\Delta A$, describes the reduction of the cross-section as the difference $\Delta A$ increases. The parameter $P$ is actually a function of A$_t$ and $E$.

The second exponential term describes the distribution of the cross section for several isotopes of the same element with atomic number $Z$. The parameters that appears, $R$, $S$, $T$, $U$, $\nu$, are a function of the target material and depend also on the values Z and A of the product as well as of the energy $E$.

The parameter $R$ represents the distribution of the cross-section values among the isotopes of the product material, while $S$ identifies the peaks around some mass numbers. Then $T$ describes the increase in the number of neutrons as A increases. The last parameter $U$ is equal to $3.0 \times 10^{-7}$ and it is inserted to correct the produced cross-section and improve the agreement with experimental data.
The exponent $\nu$ can vary from 1.3 to 2.0, depending on target and product isotopes.
Then we have some multiplying factors. $\Omega$ is related to nuclear structure, $\eta$ is a nuclear pairing factor and $\xi$ is an enhancement factor for light nuclei production.

Activia software uses this approach only in the cases in which no nuclear data are available, whereas COSMO1 always performs this kind of calculation to evaluate cross-sections. 

\subsection{Activia software description}

Activia is a C++ based package, that uses a combination of semi-empirical formulae and tables of data based on experimental results to calculate the cross-section, production rates and yields of radioactive isotopes from cosmic ray activation \cite{activia_description}.

The code takes advantage of objected-oriented design principles of C++, and can be split in two main sections. The first part performs the calculation of the cross section and production rates, and the second calculates the final yields of the radioactive isotope products \cite{activia_description}.

The input file has to contain some information. The first is the target isotopic composition, then we have to specify a list of allowed products, the input beam spectrum, the exposure time and the cooling time, i.e. the time the material has spent deep underground after the activation.

\begin{figure}
 \centering
 \includegraphics[width=\textwidth]{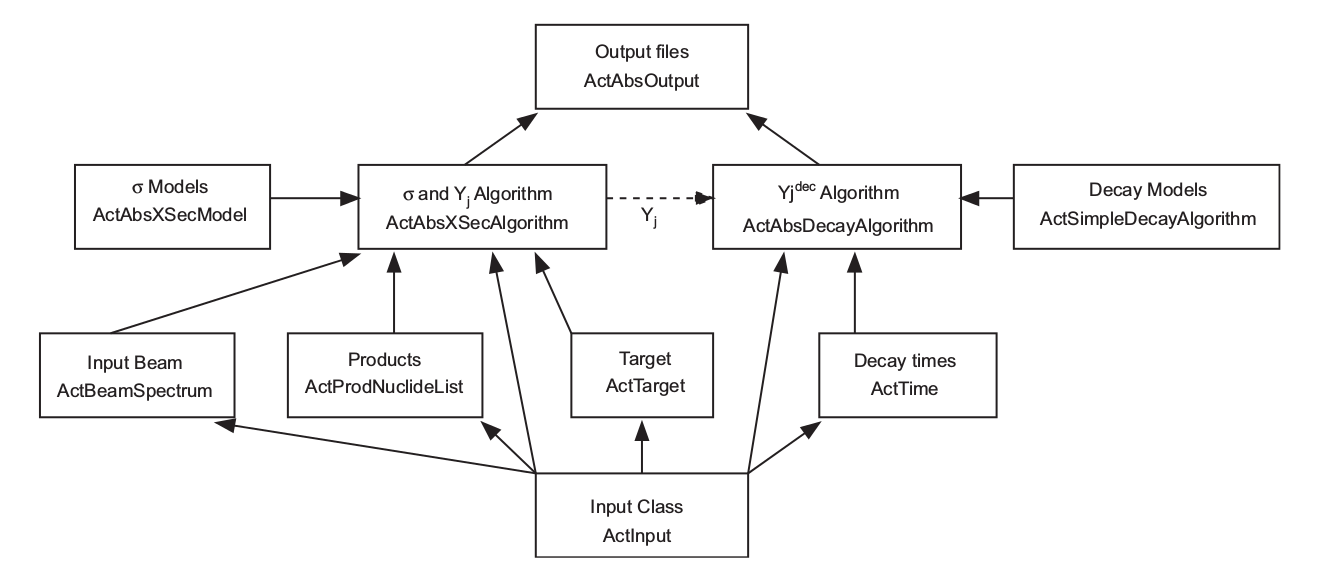}
 \caption{Simplified version of the overall Activia flow. In the scheme, $\sigma$ indicates the cross section for the spallation process, $Y^j$ is the total production rate for a nuclide $j$ and $Y_j^{dec}$ indicates the final yield for the $j$ isotope after the cooling time.  \cite{activia_description}}\label{fig_activia_flux}
\end{figure}

Figure \ref{fig_activia_flux} shows a simplified scheme of the program flow, and indicates the name of the classes that are called by the program when it is running.

From the ``ActInput'' class, that reads the input file, the data are sent to other classes, that contain all the functions and data that are useful to manage them. We see that target material is managed by ``ActTarget'' class, the products we are searching for are collected by the ``ActProdNuclideList'', the characteristics of the input beam, that in our case is the cosmic ray spectrum, is sent to ``ActBeamSpectrum''.

The final required input variables are the time of exposure to the flux, that can be labeled as $t_{exp}$ and the ``cooling'' time, that is the time elapsed since the target has be removed from the beam exposure. All the time measurements are given in days, and this information is stored inside a ``ActTime'' class \cite{activia_description}.

The ``ActIsotopeProduction'' class calculates the cross-section between all target and product isotope pairs, that uses the information stored from the input in the previously enumerated classes. 

The values of Z and A of the products must satisfy the requirements Z$\leq$Z$_t$+Z$_{beam}$ and A$\leq$A$_t$+A$_{beam}$+1, which ensure that there are at least two isotope products in each reaction.
The energy of the reaction must be larger than a threshold energy, typically of the order of 10 MeV, defined as the mass excess from the target-product reaction \cite{activia_description}.

After the evaluation of the cross-sections, that is performed according to the formulae previously given, the production rates of the isotopes are calculated by the ``ActIsotopeProduction'' class. 

As we said, whenever it is possible, Activia uses experimental data instead of the calculated cross-sections.
A large amount of dataset can be used to provide cross-section in function of energy for target-product isotope pairs. Activia uses this kind of data any time they exist. 
A linear interpolation between existent data is also used to calculate the  more probable cross-section for beam energies for which no experimental data are available.

It is important that the data have the correct units, and enough energy bins to be useful. In addition, they must have contiguous, increasing energy range.
If the available data are below a lower limit, set at 0.001 mb, the previously exposed formula is used to calculate the cross-section again.
%

The results of cross-section and yield calculations are stored in output files by the abstract class ``ActAbsOutput''. The class is able to provide ASCII text output as well as ROOT files, according to the requests we made in the input file.

The output classes are able to produce texts, tables and graphs based on the data produced by the calculations performed by the program.
Each output class produces two different output files. The first contains details from the cross section and production rate calculations, while the second one contains the yields from the isotope decay algorithm. \cite{activia_description}

A summary table of the yields of each produced isotope at the start and end of the decay period, weighted over all the target isotopes, is written to the decay output file. The table also reports the half life of the isotope.
For our purposes, we manually add to these data the modes of decay and their Q-value, that we used to select isotopes to simulate.

The code itself is structured to allow modification or extension of the components. In particular, cross-sections and yield calculation mechanisms can be improved leaving the main code practically intact.
Also, different input beams can be selected or constructed to analyze particular experimental conditions.
By default, the code only takes into account isotopes with a half-life bigger than one day. Shorter living isotopes can be manually added to the list provided in ``decayData.dat'' file.

Many other properties of the code can be modified if it is needed for the application we deal with, but in our case we used the basic code without any modification.
 
\subsection{COSMO1: software description}

COSMO1 is a FORTRAN77 based program. As Activia, it computes production cross-sections and activities for nuclides produced by nucleon-induced spallation on a specified target material. 

The Sildeberg and Tsao theory \cite{sildeberg_rao} describes cross-sections for nuclear spallation, evaporation and fission. Mainly they give a set of decision rules that allow to decide what is the most probable reaction for the production of a selected product from the target material. There is also a cross-section calculation formula associated to any production mode, that allows the calculation of the rate of production. \cite{cosmo_description}
The program is found to correctly predict which will be the most important radionuclides, and generally gives activities with an accuracy of the order of 50\% \cite{cosmo_description}.
We will now briefly describe the code structure, as we did for Activia. 

The code is divided into nine files, each of them containing a major program unit and an associated function. Some of these functions are called by different subroutines, and some subroutines call each other in the program economy. 
The input file has to contain every useful data, with the only exception of the table of product properties that is already loaded inside the source code.

The main subroutine is named COSMO, and takes the input values from bash or from the input file. The input file must contain the following data: at first, the time of exposure at ground level and the cooling time of the sample in days. Then, the program needs to know the target atomic number and the number of isotopes that are present inside the sample.
For each isotope, the mass number and the abundance must be specified.
Then, the number of stable isotopes of the element with atomic number Z+1 must be specified, and the mass number and the abundance of each isotope of it. The same must be provided for the element with atomic number Z-1.

After these data, we have to select the shape of the impinging beam. The cosmic ray beam is labeled with letter ``a'', but we can also evaluate activation due to single-energy beam, using other options that can be found in \cite{cosmo_description}.

From these input data, the subroutine ``SELECT'' chooses for each product the main process that causes the reaction and then the cross-sections are calculated.
Once cross-sections are known, the program calculates the yield of the reaction and print them on the bash. At first, it produces the yields for each target isotope, then a summary table is created, where the various result are added to each other. 

We finally have a table that contains the products coupled with their half-life, the produced activity and the activity we observe after the cooling time set in the input file. \cite{cosmo_description}

COSMO1 code is easier to describe than Activia, but at the same time it does not have the same versatility we find in the previous example. 

\subsection{Calculation and selection of contaminants}

The calculation of cosmogenic activation inside a composite material, such as CdWO$_4$, requires several different steps, that we are going to explain.

The first requirement is the knowledge of the isotopic composition of the elements of which the crystals are made, the exposure time and the cooling time. 
The isotopic composition of natural and enriched cadmium is already indicated in table \ref{tab_106cd_composition}. Other materials, namely tungsten and oxygen, are in their natural composition, that can be easily found in literature.

The exposure and cooling time is known for each detector. The enriched scintillator has been kept at ground level for only seven days, and it has been stored underground for 120. This crystal is particularly meaningful for the experiment, so particular care has been kept to avoid any type of contamination.
The crystals of natural material instead have been kept 180 days at ground level, and 120 days deep underground. We then expect a higher level of contamination in these objects with respect to the levels inside the enriched crystal.

Once we have these data, we use them inside the two programs that calculate the cosmogenic activation. The program is able to evaluate the activation inside one element for each run, so for every crystal, three different runs must be performed.

The activity the softwares calculate is given in counts kg$^{-1}$ day$^{-1}$. To have a realistic evaluation, we want to scale the results according to the relative mass abundance of the three materials inside the crystal.
At first, we calculate the mean mass of the element, according to the isotopic composition, then, we sum all the masses. At the end, we multiply each activity data for the ratio between the mass of the element from which we obtain the activity we are dealing with and the sum of the masses of the elements.

We use the activity data to perform a first selection on the isotopes we want to simulate to reconstruct the background. In particular, we select the nuclides for which the activity after the cooling time, and without any mass correction, is greater than 0.01 counts kg$^{-1}$ day$^{-1}$ on the data produced by the two softwares, without taking into account the previously introduced correction. 


For the remaining isotopes, we compile a table of nuclear decay data, that we obtain from \cite{nudat}. For each nuclide, we note the decay modes and the Q-values of the decays. We also note the half-life given by the programs. 
All these data allow us to perform a further selection. This selection is necessary to reduce the simulation power and time needed to reconstruct the background contribution.
A second cut is performed by selecting only the isotopes with a  half-life longer than 100 days. For the other contribution, we can state that they will completely disappear during the data-taking of the experiment.
The capability of the system to perform a pulse-shape discrimination analysis allows us to remove from the tables of interesting isotopes all the pure $\alpha$ emitters, whose events will be rejected easily in the first data selection.

After this rough selection is performed, we obtain a list of possible contaminants for all the crystals from both the softwares.
The results for the enriched detector are reported in table \ref{tab_activia_contaminats} for the data from Activia and COSMO1. In table \ref{tab_activia_nat} we find the results for natural crystals. As we expect, the number of radioactive contaminants is higher for the two $^{nat}$CdWO$_4$ detectors, since their exposure to cosmic rays was longer (180 days) than for the enriched detector (7 days).

\begin{table}
{\relsize{-2}
\begin{tabular}{cccccc}
\hline
Isotope&Activity (Activia) & Activity (COSMO1)& half-life [days] &Decay mode & Q-value [keV]\\
       &[counts kg$^{-1}$ day$^{-1}$] &[counts kg$^{-1}$ day$^{-1}$]                                &                  &           &              \\
\hline
V-49&- &0.0043&331&1&601.9\\
Fe-55&- &0.0041&985.5&1&231.21\\
Co-57&- &0.0037&271&1&836\\
Zn-65&-&0.0037&243.8&1&1352.1\\
Ge-68&-&0.0067&287&1&106.9\\
Se-75&0.0090&- &118.5&1&864.7\\
Rh-101&0.029&0.097 &1205.3&1&542\\
 & & & &3&157.41\\
Rh-102&0.0067&0.040&206&-1&1150\\
 & & & &1&2323\\
 & & & &3&98.7\\
 Ag-108&-&0.012&47500&-1&1650\\
 & & & &1&1922\\
 & & & &3&79.131\\
Ag-110m&0.028 &0.25&252&-1&2892.9\\
 & & & &1&889\\
 & & & &3&116.48\\
Cd-109&0.024&0.23 &453&1&214.3\\
 & & & &3&59.9\\
 & & & &3&259.5\\
 Cd-113&-&0.048&5329&-1&322\\
 & & & &3&263.7\\
Ce-139&0.0070&- &137.6&1&264.6\\
 & & & &3&754.24\\
Pm-143&0.0088&- &265&1&1041.3\\
Dy-159&0.029&- &144&1&365.6\\
 & & & &3&218\\
Lu-173&0.054&0.0092 &500&1&670.8\\
Lu-174&-&0.0056&142&1&1374.3\\
 & & & &3&126.2\\
Hf-172&0.044&0.091&683&1&350\\
Ta-179&-&0.13&665&1&105.6\\
Ta-182&0.074&0.38&115&-1&1814.5\\
 & & & &3&356.47\\
 W-178&-&0.39&121&1&91.3\\
W-181&0.23&-&121&1&188\\
Re-184&0.014&0.082&165&1&1481\\
 & & & &3&188.046\\
 \hline
 \end{tabular}
 }
\caption{Contaminants in $^{106}$CdWO$_4$ calculated by Activia and COSMO1. Decay modes are labeled as follows: -1=beta-, 1=electronic capture, 3=internal transition. The decay modes and Q-values are from \cite{nudat}. }\label{tab_activia_contaminats}
\end{table}

\newpage
  \thispagestyle{empty}

\begin{table}[H]
{\relsize{-2}
 \begin{tabular}{cccccc}
\hline
Isotope &Activity (Activia)            & Activity (COSMO1) & Half-life [days]&Decay mode & Q-value [keV]\\
        &[counts kg$^{-1}$ day$^{-1}$] &[counts kg$^{-1}$ day$^{-1}$] &      &           &              \\
\hline
Na-22&0.0039&-&950.35&2&2842.2\\
Ca-45&0.0052&-&165&-1&255.8\\
V-49&0.022&0.0032&330&1&601.9\\
Mn-54&0.021&-&312&1&1377.2\\
Fe-55&0.0054&-&986.2&1&231.21\\
Co-57&0.015&-&271&1&836\\
Zn-65&0.039&0.0038&244.1&1&1352.1\\
Ge-68&0.020&-&270.8&1&106.9\\
Se-75&0.094&-&118.5&1&864.7\\
Y-88&0.049&-&106.6&2&3622.6\\
 & & & &3&442.62\\
 & & & &3&392.87\\
Rh-101&0.44&0.28&1205.3&1&542\\
 & & & &3&157.41\\
Rh-102&0.39&0.19&206&-1&1150\\
 & & & &1&2323\\
 & & & &3&98.7\\
Ag-108&0.020&0.034&46400&-1&1650\\
 & & & &1&1922\\
 & & & &3&79.131\\
Ag-110m&1.8&0.76&252&-1&2892.9\\
 & & & &1&889\\
 & & & &3&116.48\\
Cd-109&1.6&0.63&453&1&214.3\\
 & & & &3&59.9\\
 & & & &3&259.5\\
Cd-113&0.093&0.16&5113&-1&322\\
 & & & &3&263.7\\
Ca-45&0.014&0.011&165&-1&255.8\\
V-49&0.025&0.013&330&1&601.9\\
Mn-54&0.022&0.014&312&1&1377.2\\
Fe-55&0.0061&-&986.2&1&231.21\\
Co-57&0.012&0.011&271&1&836\\
Zn-65&0.030&0.022&244.1&1&1352.1\\
Ge-68&0.013&0.0093&270.8&1&106.9\\
Se-75&0.038&0.025&118.5&1&864.7\\
Y-88&0.011&0.028&106.6&2&3622.6\\
 & & & &3&442.62\\
 & & & &3&392.87\\
Rh-101&0.014&0.011&1205.3&1&542\\
 & & & &3&157.41\\
Rh-102&0.015&0.031&206&-1&1150\\
 & & & &1&2323\\
 & & & &3&98.7\\
Cd-109&0.033&0.011&453&1&214.3\\
 & & & &3&59.9\\
 & & & &3&259.5\\
Sn-113&0.047&0.025&115&1&1036.6\\
 & & & &3&77\\
Te-121&0.068&0.0060&154&1&1054\\
 & & & &3&212.189\\
Ba-133&0.0080&-&3906&1&517.5\\
 & & & &3&288\\
Ce-139&0.12&-&137.6&1&264.6\\
 & & & &3&754.24\\
Pm-143&0.18&0.077&265&1&1041.3\\
Pm-144&0.020&0.041&363&1&2331.7\\
Sm-145&0.027&0.057&340&1&620\\
Gd-153&0.0055&0.014&242&1&483.6\\
 & & & &3&93.34\\
Dy-159&0.51&0.031&144&1&365.6\\
 & & & &3&218\\
Tm-170&0.0058&0.033&129&1&314.4\\
 & & & &-1&968\\
Lu-173&1.2&0.21&500&1&670.8\\
Lu-174&0.0052&0.027&1204&1&1374.3\\
 & & & &3&126.2\\
Hf-172&1.0&2.2&683&1&350\\
Hf-178&0.0053&-&11300&4&2445.79\\
Ta-179&3.8&3.0&665&1&105.6\\
Ta-182&1.2&6.2&115&-1&1814.5\\
 & & & &3&356.47\\
W-181&3.8&6.4&121&1&188\\
Re-184&0.26&1.5&165&1&1481\\
 & & & &3&188.046\\
 \hline
 \end{tabular}
}
 \caption{Contaminants in natural CdWO$_4$ calculated by Activia and COSMO1. Decay modes are labeled as follows: -1=beta-, 1=electronic capture, 3=internal transition. The decay modes and Q-values are from \cite{nudat}.}\label{tab_activia_nat}

\end{table}

\newpage
\section{Monte Carlo simulation results}

 \subsection{Simulation of 2$\nu\epsilon\beta^+$ effect}
The aim of this work is to use the data produced by cosmogenic evaluation software to identify the most dangerous sources of background produced inside the crystals. 

We need at first a spectrum of the effect we are interested to observe. From this first result, we can start to search for the most dangerous contaminations among the isotopes listed in the table \ref{tab_activia_contaminats} for the $^{106}$CdWO$_4$ detector. For the $^{nat}$CdWO$_4$ crystals, we notice in table \ref{tab_activia_nat} that some contaminants result particularly active with respect to the others. So we will focus on them.

The first simulation performed using the GEANT4 code was the 2$\nu\epsilon\beta^+$ decay of $^{106}$Cd. 
The $2\nu2\epsilon$ channel has an expected half-life lower than $2\nu\epsilon\beta^+$ channel, but the search for this kind of decay in our crystal is not possible. A contamination of $^{113m}$Cd, whose Q-value is 586 keV, with an activity  of $116\pm4$ Bq Kg$^{-1}$\cite{first_enriched} covers every signal in this range. Its effect inside our experiment is presented in figure \ref{fig_113activity}. \cite{first_enriched}

The initial conditions of the decays were produced using DECAY0. The simulation took the events from the file text produced as output from this code and used the data to generate the primary event. Then, the interaction and propagation was ruled by the Monte Carlo code implemented inside GEANT4. The output of the code was contained inside a TTree structure in a ROOT file. 
 

\begin{figure}[H]
 \centering
 \includegraphics[width=\textwidth]{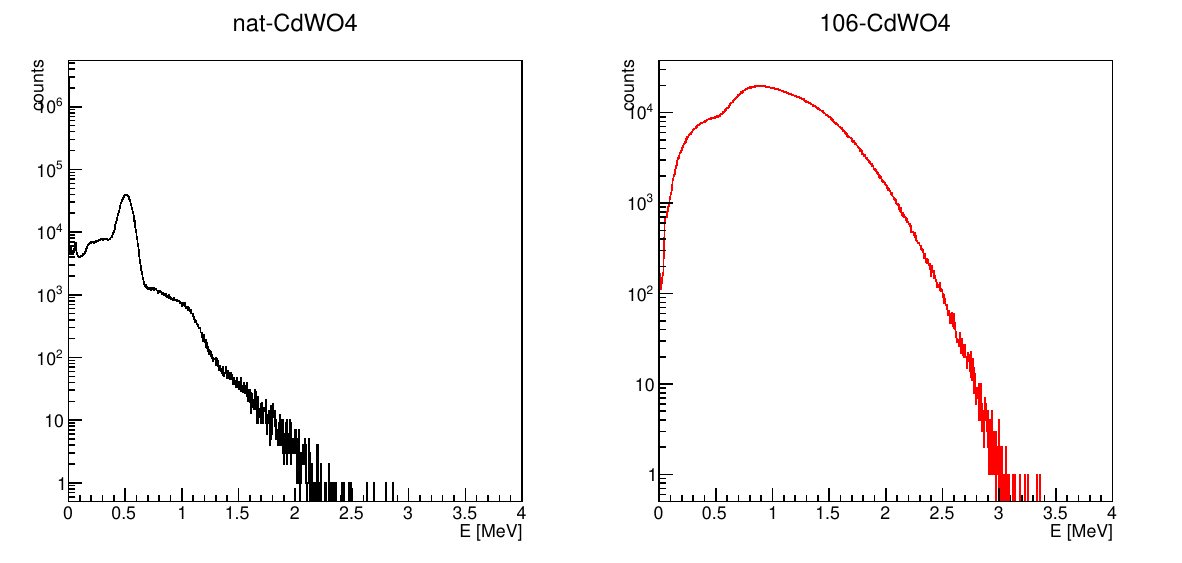}
 \caption{Energy deposited inside $^{nat}$CdWO$_4$ crystals and $^{106}$CdWO$_4$ crystal from $2\nu\epsilon\beta^+$ decay of $^{106}$Cd inside the enriched crystal. }\label{fig_spectra}
\end{figure}

 In figure \ref{fig_spectra} spectra collected from the detectors are reported. Since the energy spectrum from the two enriched crystals is practically equal, only one is shown. 
 In the spectrum from enriched crystal, on the right side of figure \ref{fig_spectra}, in red, we notice the shape we expect for a $\beta$ decay. We do not find any low energy peak, that should be produced by de-excitations of the nucleus after the electronic capture, but it is probable that such kind of signal is covered by low energy spectrum of $\beta^+$
 
 The spectrum in black in figure \ref{fig_spectra} is deposited inside one of the natural crystals. Its shape  presents the same main features of the energy spectrum deposited inside the $^{106}$CdWO$_4$. It is clearly visible the peak from annihilation $\gamma$ quanta at 511 keV, and the large number of events in the previous sections of the spectra is due to Compton scattering of these events.
 
 The Compton background extends also up to $\sim 1$ MeV, and it is due to the cases in which both the $\gamma$ quanta from positron annihilation interact with the same crystal. The number of these events is too low to observe a proper peak, but it is clear that, after this limit value, the background reduces significantly in both the spectra.
 
 The low number of counts that we observe at higher energies can be produced by high energy positrons produced in the decay, that escape the enriched crystal and interact with the natural crystals. This can be related to the events in the cases in which the decay happens on the enriched detector surfaces.

\begin{figure}[H]
 \centering
 \includegraphics[width=0.7\textwidth]{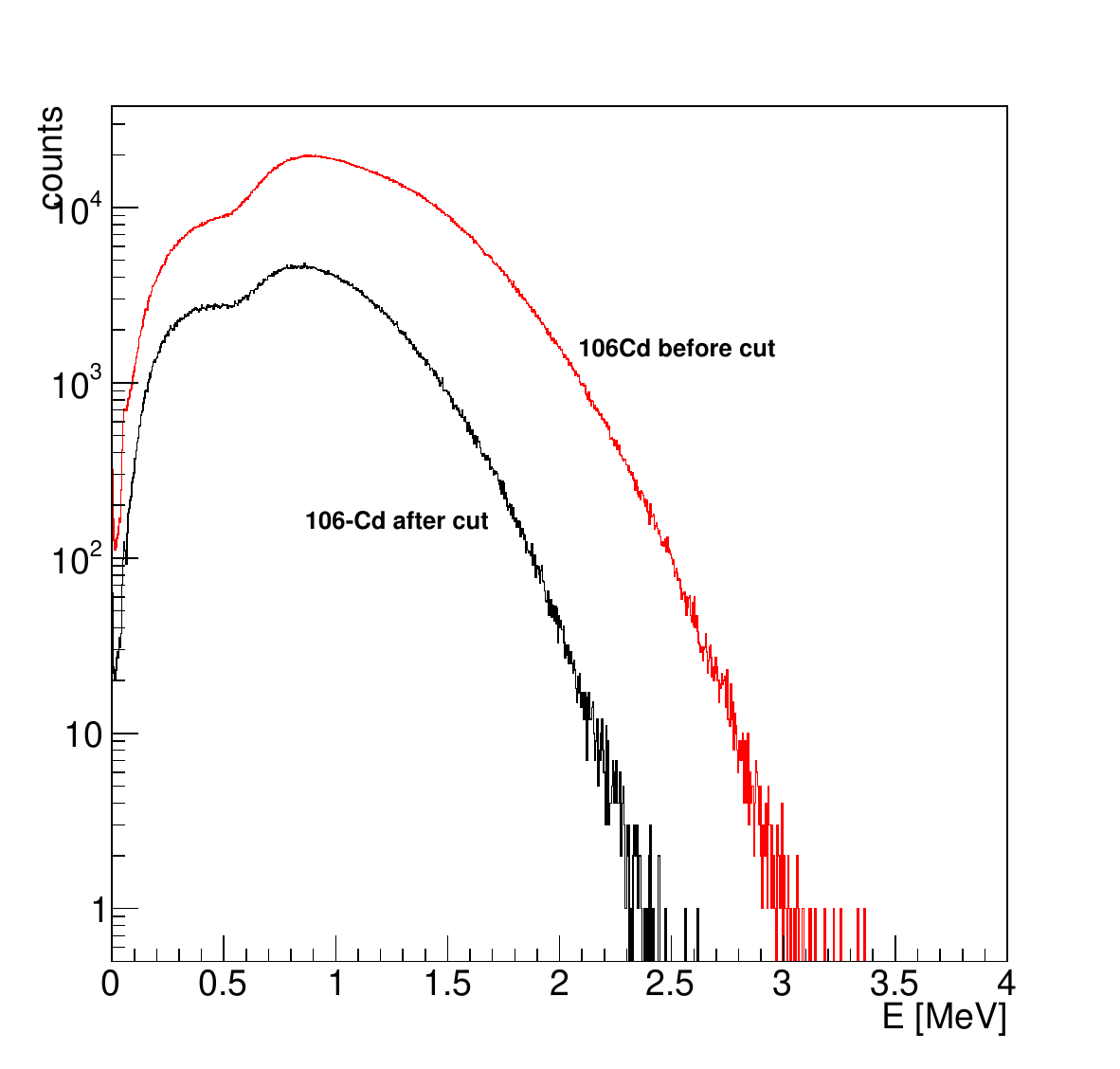}
 \caption{Comparison between the spectrum from enriched crystal with no selection and spectrum in coincidence with an energy deposition of 511 keV inside one of the two natural crystals.}\label{fig_coincidence}
\end{figure}

The first selection we want to perform is to connect the energy deposited inside the natural crystals to the energy deposition inside the enriched crystal.
We want to select for our analysis all the events inside the enriched crystal that are related to an energy deposition inside the natural crystal that can be considered inside the peak at 511 keV, generated by the $\gamma$ quanta from annihilation of positrons.

We select for analysis only the events for which the energy deposition inside one of the two natural detectors is a fixed range from the annihilation energy of 511 keV. The range used in this test is 4$\sigma$. 
The value of $\sigma$ at 511 keV is known from the resolution data shown in figure \ref{fig_calibration}. 
The result of the coincidence selection is reported in black in figure \ref{fig_coincidence}. In the same figure, the spectrum of events inside enriched crystal without any selection is plotted in red for comparison.

We immediately notice a reduction of the total number of counts. The events that survive this cut are 1022973, from which we obtain an efficiency of 20.5\% after the coincidence cut.

\subsection{Simulation of $^{106}$CdWO$_4$ contaminants}

\begin{table}[b]
\centering
{\relsize{-1}
 \begin{tabular}{lcccc}
 \hline
 Nuclide & Half-life  & Q-value & Activity by COSMO1 & Activity by Activia \\
         & [days]     & [keV]   & [counts kg$^{-1}$ day$^{-1}$] & [counts kg$^{-1}$ day$^{-1}$] \\
 \hline
 $^{102m}$Rh & 1365 $\pm$ 4 & 2323$\pm$ 5 & 0.064 & 0.0067\\
 $^{184}$Re & 169 $\pm$ 8 & 1481 $\pm$ 4 & 0.082 & 0.014\\
 $^{65}$Zn  & 243.93 $\pm$ 0.09  & 1352.1 $\pm$ 0.3  & 0.0037 &  - \\
 $^{108}$Ag & 47500 $\pm$ 100 & 1922 $\pm$ 5 & 0.012 & - \\
 $^{174}$Lu & 142 $\pm$ 2 & 1374.3 $\pm$ 1.6  & 0.0056 & - \\
 $^{110m}$Ag& 249.83 $\pm$ 0.04 & 2892.9 $\pm$ 1.5 & 0.25 & 0.028\\
 $^{182}$Ta & 114.74 $\pm$ 0.12 & 1814.5 $\pm$ 1.7 & 0.39 & 0.074\\
 \hline
 \end{tabular}
 }
\caption{Selected contaminants inside the enriched crystal and their properties.}\label{tab_selected_enr}
\end{table}

 In the section dedicated to the cosmogenic activation, we present the first stage of selection performed on the tables produced by COSMO1 and Activia softwares. We suppose that the crystal has been exposed to cosmogenic activation for 7 days, and it has been kept underground for 120 days.
 
 To be able to perform a more detailed analysis of the contaminations, a further selection must be performed in order to reduce the number of the candidates.
  Since all the contaminants suggested by the calculations have very low activities, and none of them exceeds the others significantly, we do not take into account this parameter in our selection at first. As we will see later, for contaminations of non-enriched crystals we are going to focus on the most active contaminants instead.


 Inside the tables, nuclides that decay by electronic capture (EC) and have the higher Q-value are selected, since they are able to produce a decay spectrum comparable with the one produced by $2\nu\epsilon\beta^+$ decay of $^{106}$Cd.
 Among the isotopes that decay by EC, we choose to start our work from the simulation of those for which $\beta^+$ decay is also allowed. The list of these candidates is quite short, and we are able to test all of them with our simulation. They are:  $^{65}$Zn, $^{102}$Rh, $^{174}$Lu. We also simulate $^{108}$Ag and $^{184}$Re because of the high Q-value of their decay.
 
 Then, we consider also two $\beta^-$ emitters, $^{110m}$Ag and $^{182}$Ta, because of their high Q-value. We suppose they produce bremsstrahlung irradiation that produces a signal in the two natural detectors inside the coincidence window. 
 A summary of the selected isotopes and their properties is reported in table \ref{tab_selected_enr}.
 
 We start now to present the results of the simulations. Each simulation consists in 5$\times $ 10$^6$ decays.
 
 \subsubsection{$^{102}$Rh and $^{102m}$Rh}

 $^{102}$Rh appears in both COSMO1 and Activia calculations with different values of activity (see table \ref{tab_activia_contaminats}). 
 This nuclide decays by $\beta^-$ and EC processes. Also a $\beta^+$ decay is possible, in competition with EC decay. The Q-value of this last reaction is 2323$\pm$5 keV \cite{nudat}, and the intensity we expect from $\beta^+$ decay is 14.7\%\cite{nudat}. In figure \ref{fig_rh102_decay} the decay scheme for the electronic capture or $\beta^+$ decay is shown. It can also be produced in a metastable state whose energy is 140.75 keV and the half-life is 3.742 yr.
 
Because of all these characteristics, $^{102}$Rh and its metastable state $^{102m}$Rh can be an important background contaminant in our experiment. Since it is not possible to simulate its decay using DECAY0 libraries, that are able to produce initial conditions only for a small number of usual contaminants, we use the GEANT4 event generator. This means that we have to use a different simulation, for which the physics and geometry is exactly the same we used for the simulation of double-$\beta$ process, only the classes that rule the input options are different.
 From the simulation we expect a lower rate of events after selection, since only a fraction of events emits positron. 
 
 \begin{figure}[H]
  \centering
  \includegraphics[width=\textwidth]{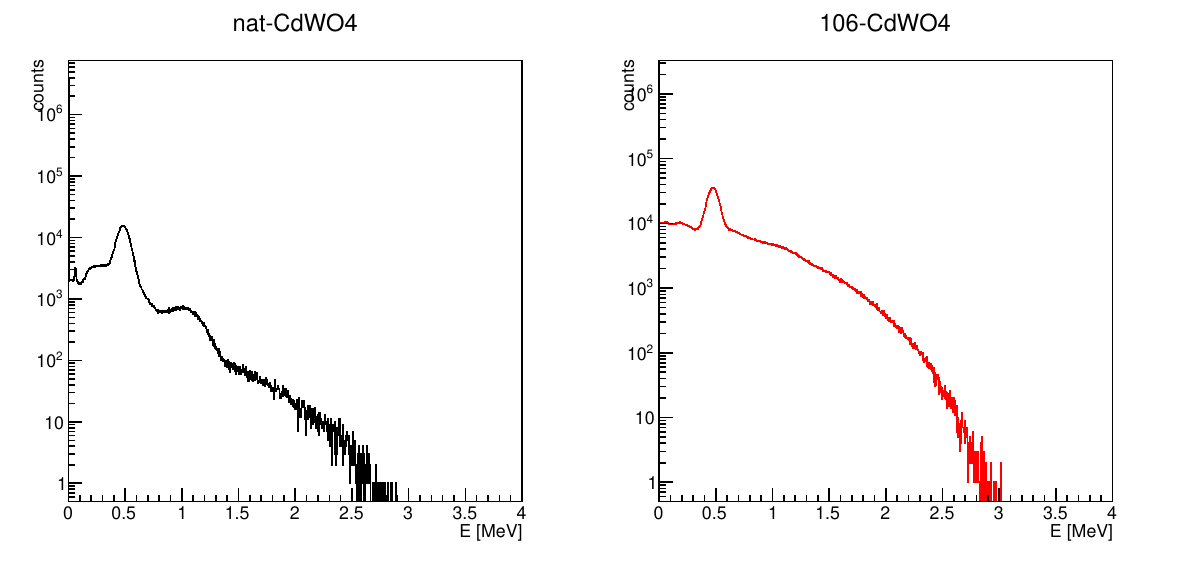}
  \caption{Spectra of $^{102}$Rh and $^{102m}$Rh produced inside the detectors. Since the energy spectrum from the two natural crystals is practically equal, only one is reported.}\label{fig_rh102_spectra}
 \end{figure}

 In figure \ref{fig_rh102_spectra} the energy distributions deposited inside the three detectors are shown. The natural crystal spectra are really close to those we have already discussed for $^{106}$Cd $2\nu\epsilon\beta^+$ decay. 
 
 They show a clear peak at 511 keV, due to the annihilation $\gamma$ quanta, and a second peak at $\sim1$ MeV, for those events in which both the $\gamma$ rays have been collected by the same detector. There is also a tail of counts, up to 3 MeV, due to the events in the high energy range that deposit their energy inside the $^{nat}$CdWO$_4$ crystals.
 
 The energy distribution inside the enriched detector is continuous and decreases to the end-point that in this case  results higher than the Q-value of the reaction. The difference can be attributed to the metastable state, that enhances the energy involved in the decay. We notice that a peak at 511 keV is superimposed on this structure in the collected energy spectrum. The peak is due to $\gamma$ quanta from positron annihilation that do not escape the $^{106}$CdWO$_4$ detector.
 
 Also for these data, a selection is performed in order to extract only the events in the enriched crystal related to an energy deposition inside the natural crystals between 4$\sigma$ from the 511 keV peak. This selection will be applied to all the collected data.
 
 \begin{figure}[H]
  \centering
  \includegraphics[width=0.7\textwidth]{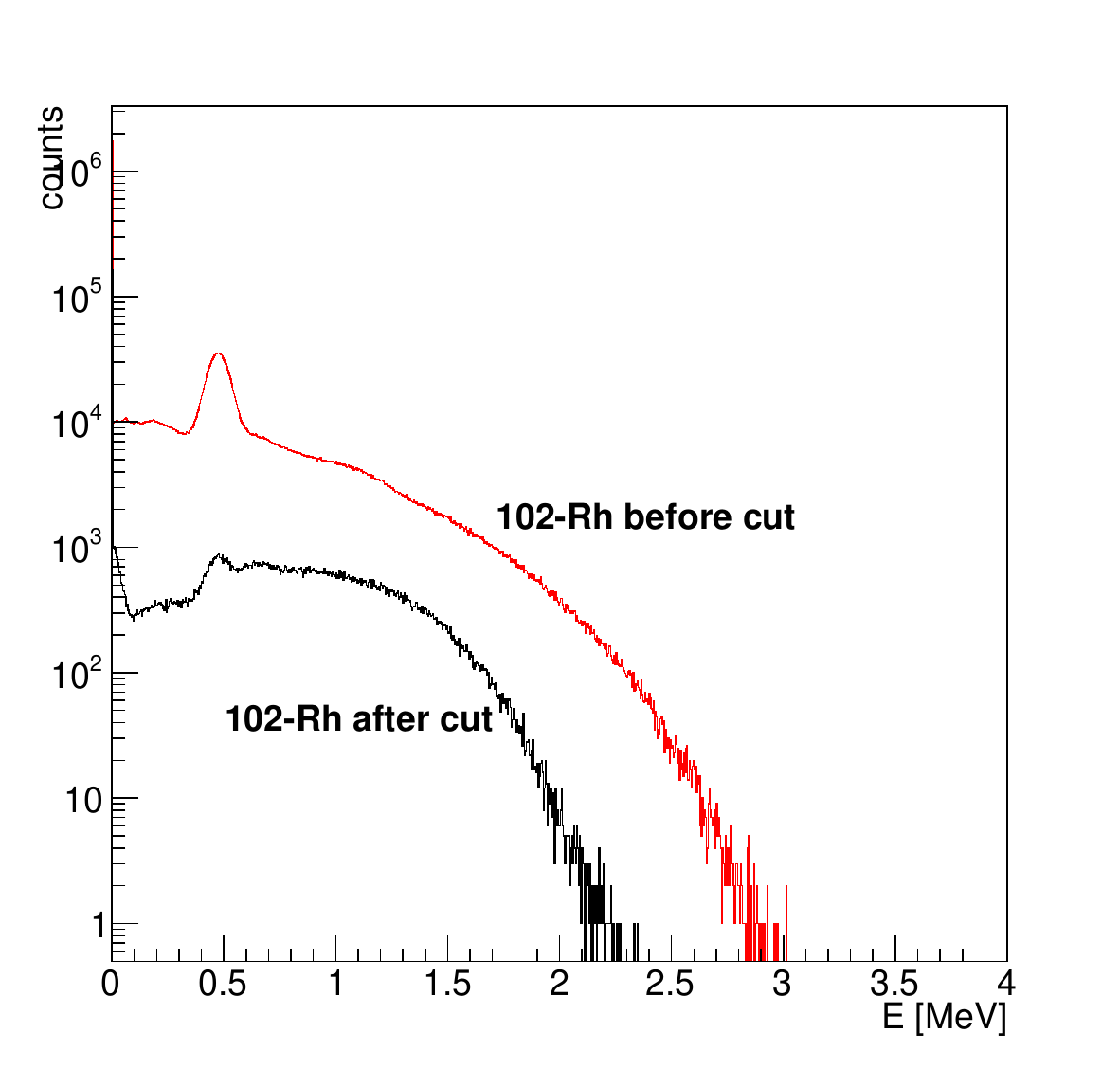}
  \caption{Comparison between the energy deposited inside the crystal by $^{102}$Rh before and after the selection of events related to 511 keV energy deposition inside natural crystals.}\label{fig_rh102_coincidence}
 \end{figure}
 
 The result of this selection is shown in figure \ref{fig_rh102_coincidence}. After the selection, 370540 events survive. This means that the efficiency of the cut is 7.4\%. The efficiency is lower than the one we observe for the $2\nu\epsilon\beta^+$ decay of $^{106}$Cd(20.5\%), but still significant.

 From this cut we observe that the 511 keV peak that we can appreciate inside the enriched detector spectrum is reduced. Also the end-point of energy distribution results lower after the cut (it changes from $\sim3$ MeV to $\sim2$ MeV), as we noticed for the previous selection performed on data about $2\nu\epsilon\beta^+$ decay of $^{106}$Cd.
 
 
\begin{figure}[H]
  \centering
  \includegraphics[width=0.7\textwidth]{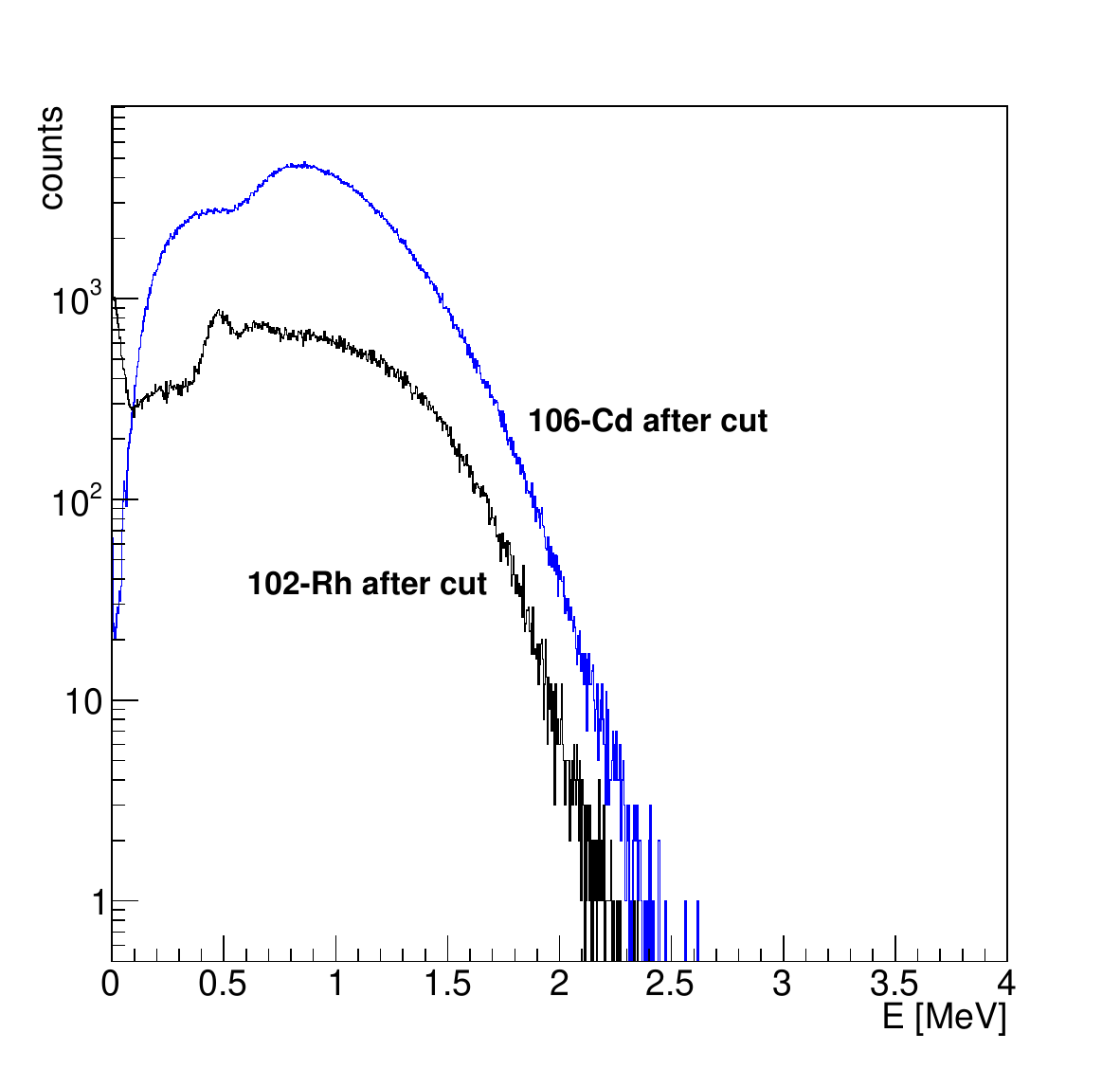}
  \caption{Comparison between the spectrum from $2\nu\epsilon\beta^+$ decay of $^{106}$Cd and $^{102}$Rh spectrum.}\label{fig_rh102_comparison}
 \end{figure}

To understand whether this effect is actually able to mimic the one we are interested in, we produce a new figure, in which we superimpose the histograms after the selection cuts for the $^{102}$Rh and for the $2\nu\epsilon\beta^+$ decay of $^{106}$Cd.
 
From figure \ref{fig_rh102_comparison}, where we can see in blue the spectrum from $2\nu\epsilon\beta^+$ decay of $^{106}$Cd and in black the $^{102}$Rh deposited energy, we can appreciate that the two spectra have some similarities. The shape has some differences but the range is the same, and the positron emission allows it to survive to coincidence selection.
 So, because of the similar Q-value and the $\beta^+$ emission, $^{102}$Rh can be an important background source.
 
 \subsubsection{$^{184}$Re}
 
 
$^{184}$Re Q-value is slightly lower than the one of $^{102}$Rh, only 1481 keV \cite{nudat}, but we have not found any other nuclide that decays by electronic capture with a Q-value higher than 2 MeV. This one has a Q-value high enough to be taken into account as background source in our calculations. It decays only by electronic capture \cite{nudat}, no other channels are in competition. The decay scheme is shown in figure \ref{fig_re184_decay}.
 
 
 
 \begin{figure}[H]
  \centering
  \includegraphics[width=\textwidth]{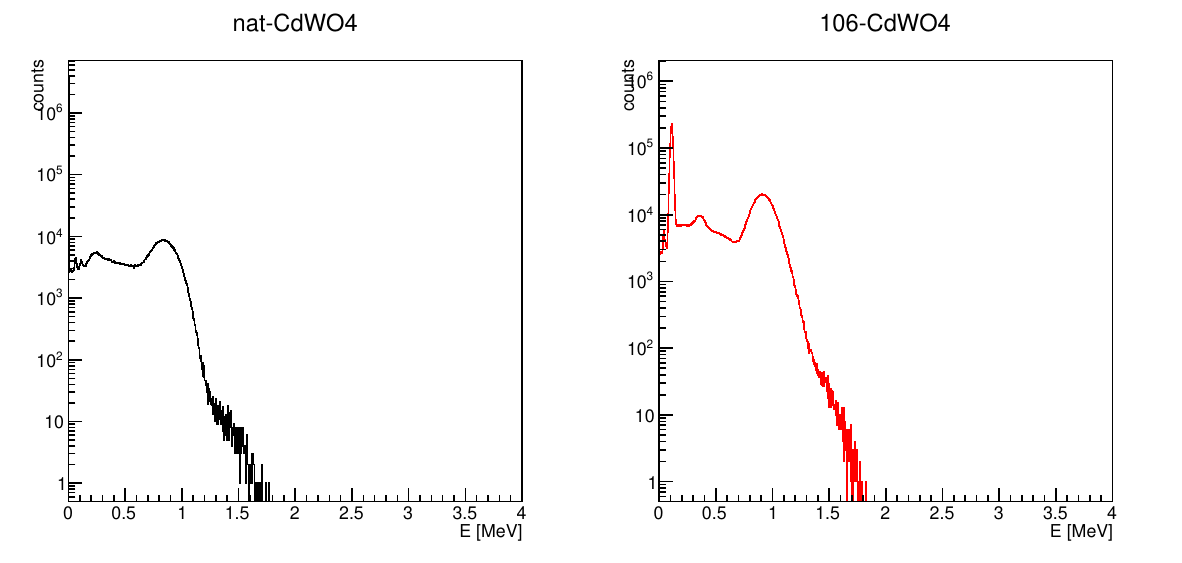}
  \caption{Energy deposited inside the three detectors by $^{184}$Re. Since the energy spectrum from the two natural crystals is practically equal, only one is reported.}\label{fig_re184_spectra}
 \end{figure}
 
 In figure \ref{fig_re184_spectra} we show the energy deposition inside the detectors. Since the spectra inside the two $^{nat}$CdWO$_4$ are very close, we present only two spectra for the three detectors.  We immediately notice that the energy distributions are widely different from the previous case we presented. In the spectra from natural crystals, in black, there is no track of the peak at 511 keV caused by positron annihilation. We therefore do not expect to find it, since in the documentation $\beta^+$ decay is not suggested as possible.
 
 We therefore notice in all the three spectra a peak around 1.0 MeV. Moreover, at $\sim 110$ keV in the spectrum from $^{106}$CdWO$_4$ crystal we have a sharp peak probably due to some de-excitation processes that follows the electronic capture.

 \begin{figure}[H]
  \centering
  \includegraphics[width=0.6\textwidth]{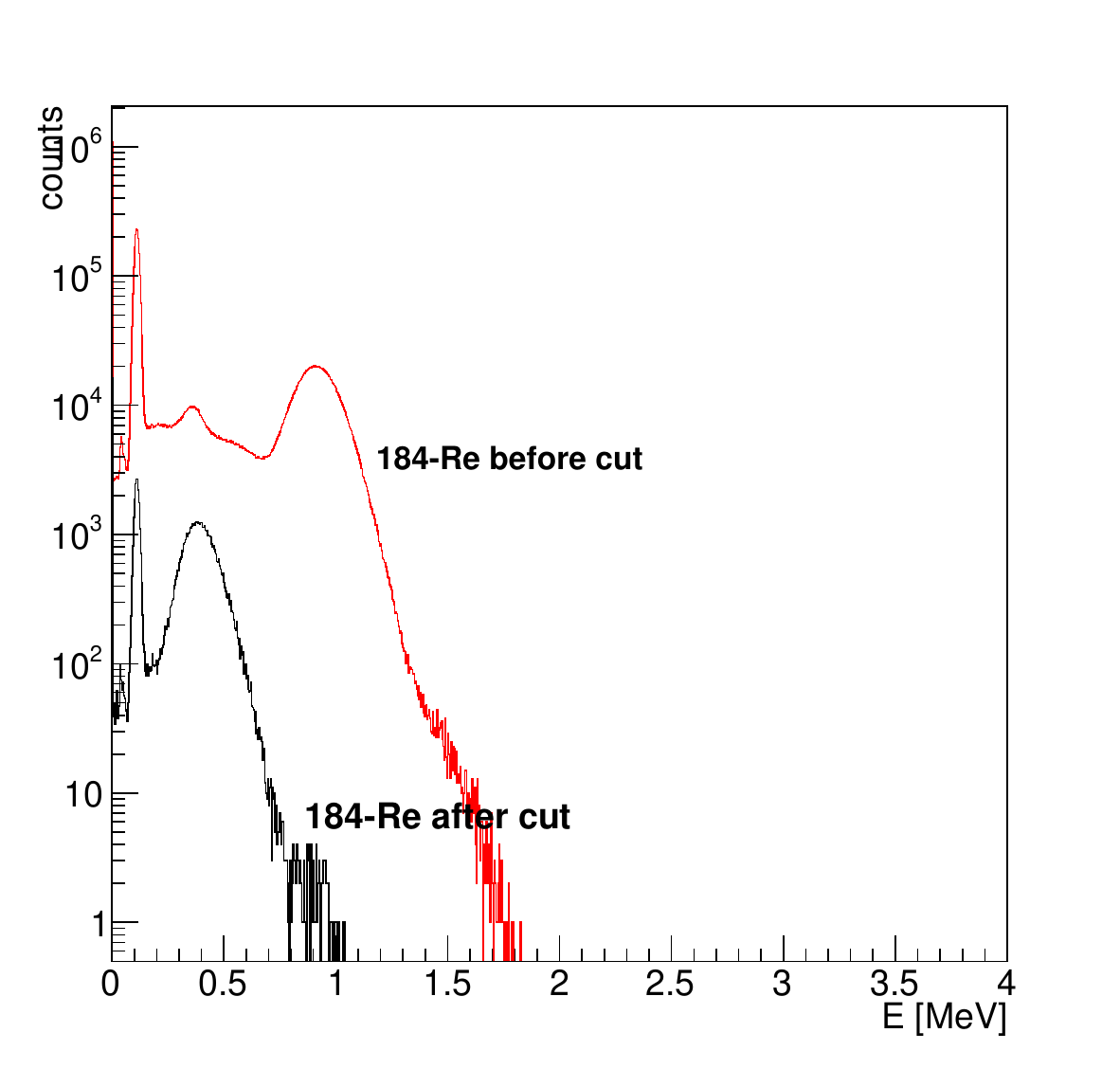}
  \caption{Comparison of spectra deposited in enriched crystal by $^{184}$Re decay before and after the coincidence selection.}\label{fig_re184_coincidence}
 \end{figure}

 \begin{figure}[H]
   \centering
  \includegraphics[width=0.7\textwidth]{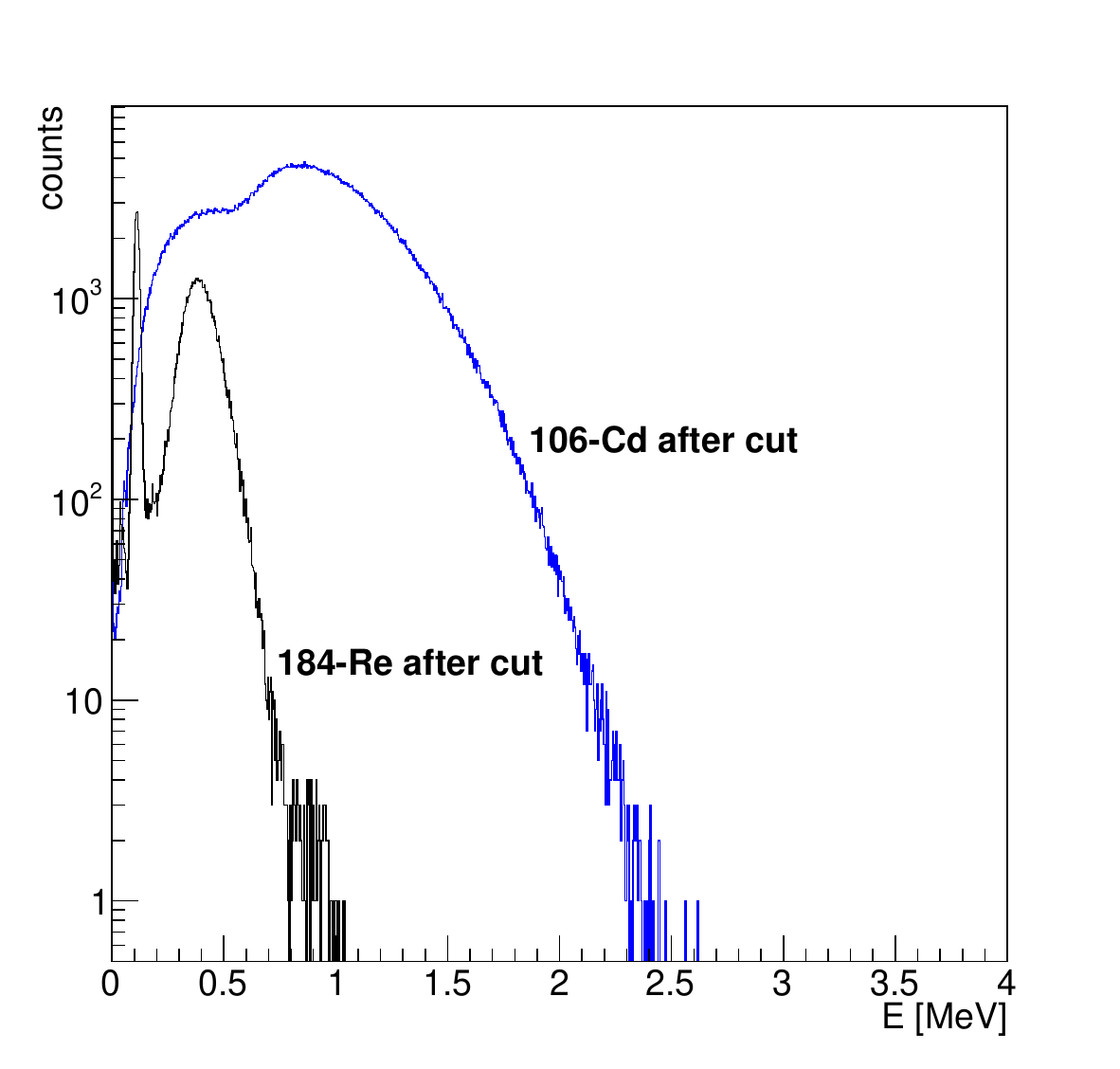}
  \caption{Comparison between the spectrum from $2\nu\epsilon\beta^+$ decay of $^{106}$Cd and $^{184}$Re spectrum.}\label{fig_re184_comparison}
 \end{figure}
 
 The analysis procedure we perform to select the interesting events is the same previously reported for $^{102}$Rh: we select inside the Ntuple structure that contains the data of all the events in which there is an energy deposition inside a 4$\sigma$ interval from 511 keV.
 
In this case the coincidence selection modifies widely the shape of the energy spectrum, as we can notice in figure \ref{fig_re184_coincidence}. The number of counts that survive this selection is 97648. The efficiency of the cut is then 2.0\%.  The shape that results is widely different from the one that the effect we are searching for  produces, as shown in figure \ref{fig_re184_comparison}.  

The high number of counts in the low energy range makes this nuclide not completely negligible. Its presence inside the enriched crystal could be a background source in the study of the low energy region of the spectrum.

\subsubsection{$^{182}$Ta}

 $^{182}$Ta is reported as a possible contaminant for the enriched crystal by both the activation softwares (see table \ref{tab_activia_contaminats}). 
 The Q-value of the $\beta^-$ decay is 1814.5 keV \cite{nudat}. 
 In this case, the daughter nuclide $^{182}$W is stable.
 
 \begin{figure}[H]
  \centering
 \includegraphics[width=\textwidth]{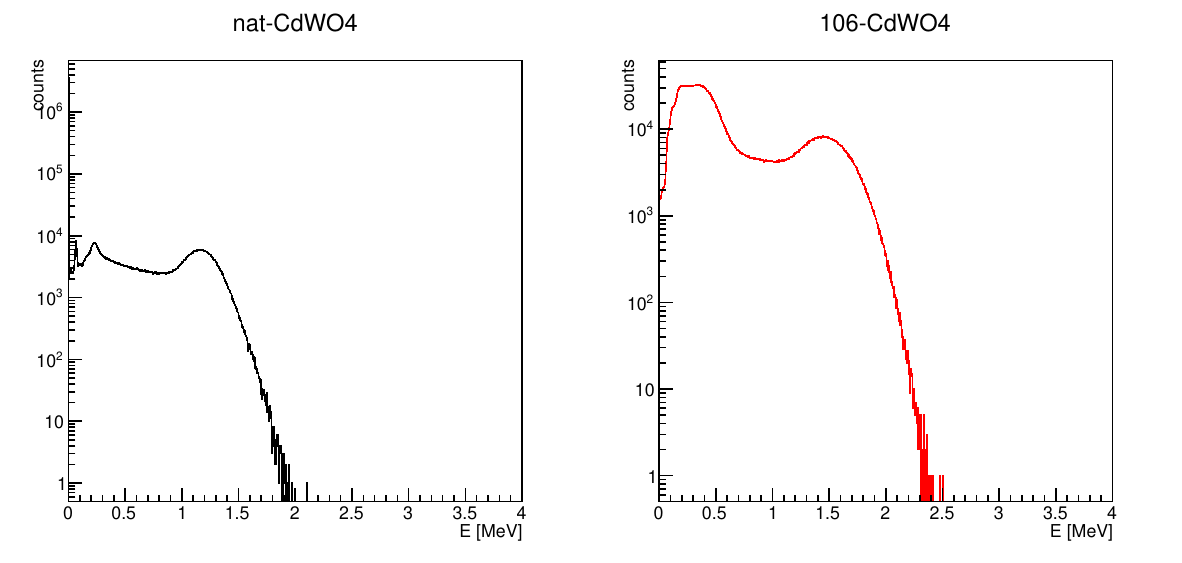}
 \caption{Energy deposited inside the three detectors by $^{182}$Ta decays. Since the energy spectrum from the two natural crystals is practically equal, only one is reported.}\label{fig_ta182_1_spectra}
 \end{figure}
 
 The simulated responses of $^{106}$CdWO$_4$ and $^{nat}$CdWO$_4$ detectors to decays of $^{182}$Ta in the natural detectors are presented in figure \ref{fig_ta182_1_spectra}. The energy deposited inside the $^{106}$CdWO$_4$ detector presents two peaks at $\sim0.2$ MeV and $\sim2$ MeV. Inside the $^{nat}$CdWO$_4$ detectors the shape of the energy distribution is quite different. There are small sharp peaks in the low energy range (under 0.5 MeV), and a continuous distribution that finishes with a large peak-like structure at the end point, near 1.5 MeV.
   
 \begin{figure}[H]
 \centering
 \includegraphics[width=0.7\textwidth]{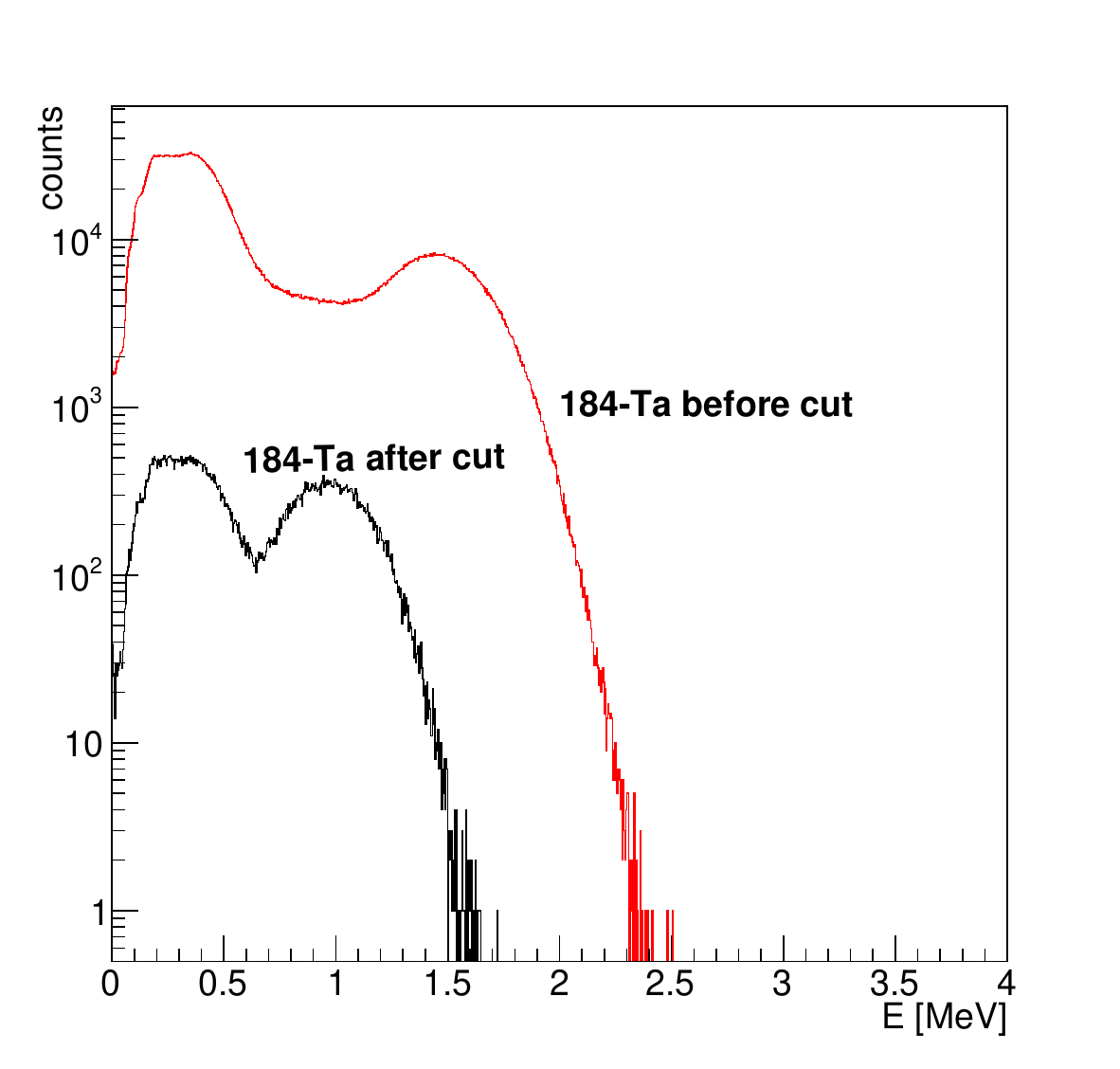}
 \caption{Comparison of spectra deposited in enriched crystal by $^{182}$Ta decay before and after the coincidence selection.}\label{fig_ta182_1_coincidence}
\end{figure}

A significant number of events, 91597, survives the coincidence selection. The efficiency is 1.8\%. The shape of the spectrum is shown in figure \ref{fig_ta182_1_coincidence}; we can appreciate two large peaks at $\sim0.4$ MeV and at $\sim1$ MeV.

 \begin{figure}[H]
   \centering
  \includegraphics[width=0.7\textwidth]{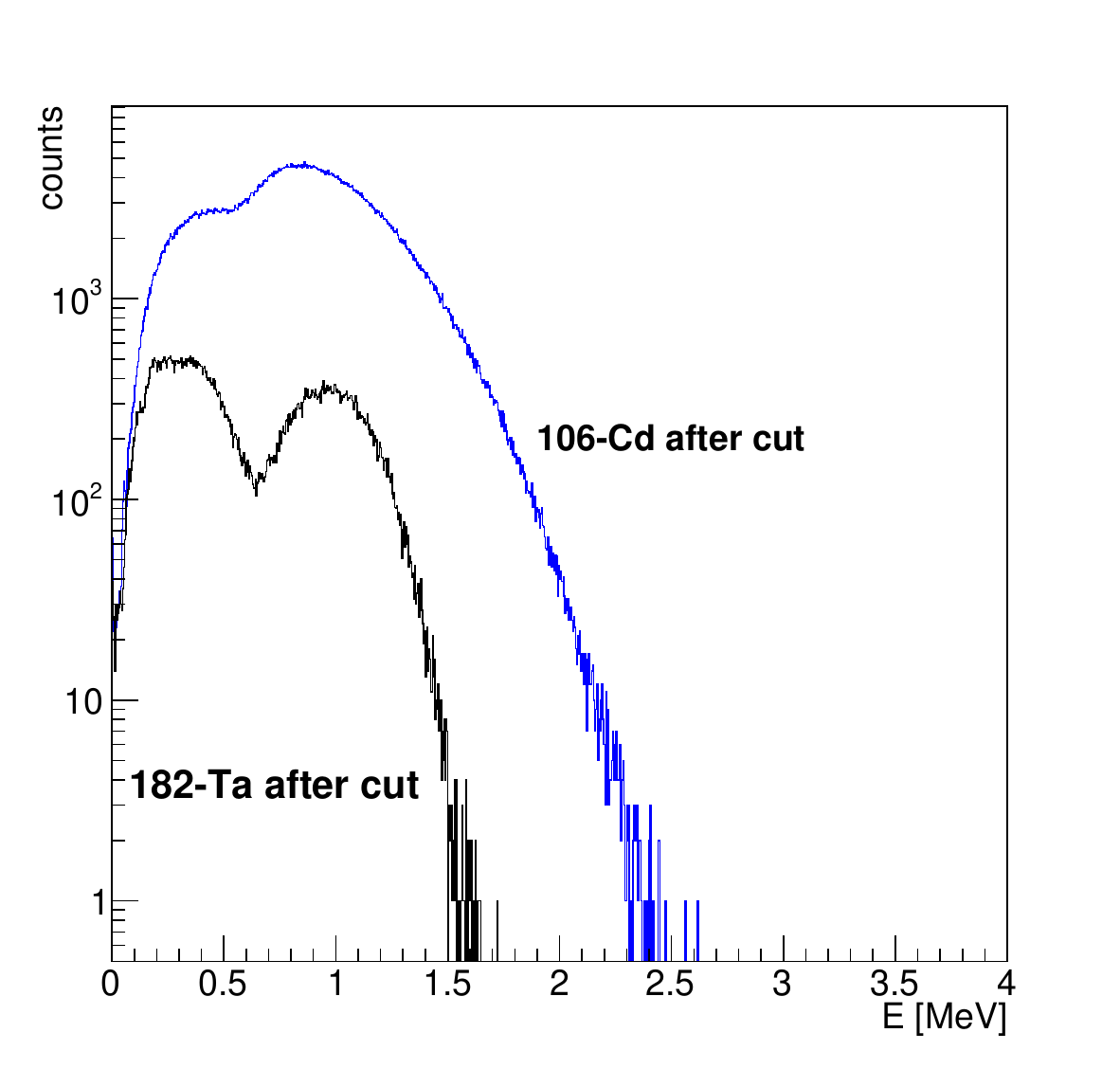}
  \caption{Comparison between the spectrum from $2\nu\epsilon\beta^+$ decay of $^{106}$Cd and $^{182}$Ta spectrum.}\label{fig_ta182_1_comparison}
 \end{figure}
 
 In figure \ref{fig_ta182_1_comparison} the energy distribution generated by $^{182}$Ta decay is shown and in comparison to the spectrum obtained for $2\nu\epsilon\beta^+$ decay of $^{106}$Cd. Because of the shape of the spectrum, this contaminant should be taken into account in the background reconstruction.
 
\subsubsection{$^{65}$Zn}

 $^{65}$Zn is a radioactive isotope that could contaminate the enriched crystal according to COSMO1 calculation. 
 We choose to run a simulation of this isotope for the high Q-value (1352.1 keV \cite{nudat}) and because in NuDat archives it is shown to decay also with positron emission in some cases. The expected intensity of this emission is low, about 1\% \cite{nudat} of the decays, but this kind of signals has a certain importance, as in the case of $^{102}$Rh, because it is expected to survive to the coincidence selection cut. Its decay scheme is reported in figure \ref{fig_zn65_decay}.
 
 
 \begin{figure}[H]
  \centering
 \includegraphics[width=\textwidth]{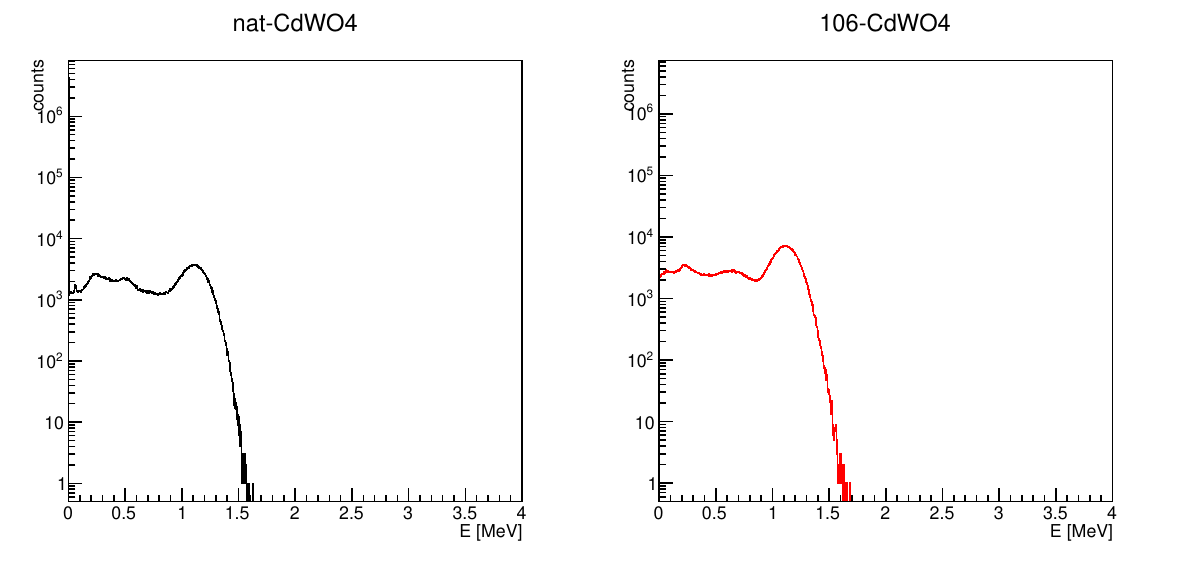}
 \caption{Energy deposited inside the three detectors by $^{65}$Zn decays. Since the energy spectrum from the two natural crystals is practically equal, only one is reported.}\label{fig_zn65_spectra}
 \end{figure}

 As usual, we report the energy deposited in all the three crystals in figure \ref{fig_zn65_spectra}. We notice that in this case the energy deposited inside the three detectors has almost the same shape.
 
 Starting from the higher energies, we find a first peak at $\sim1.3$MeV. Then, we find a continuous distribution with some superimposed structures. The little variation they cause is too low to allow an exact identification of the peak position by fit. 
 
 On the collected data we perform the usual cut to select the events in coincidence. The two black spectra inside figure \ref{fig_zn65_spectra}, connected to the natural detectors, show no structures inside the natural crystals at 511 keV, so we expect to not have a large number of events that survive the coincidence cut.
 
 \begin{figure}[H]
 \centering
 \includegraphics[width=0.6\textwidth]{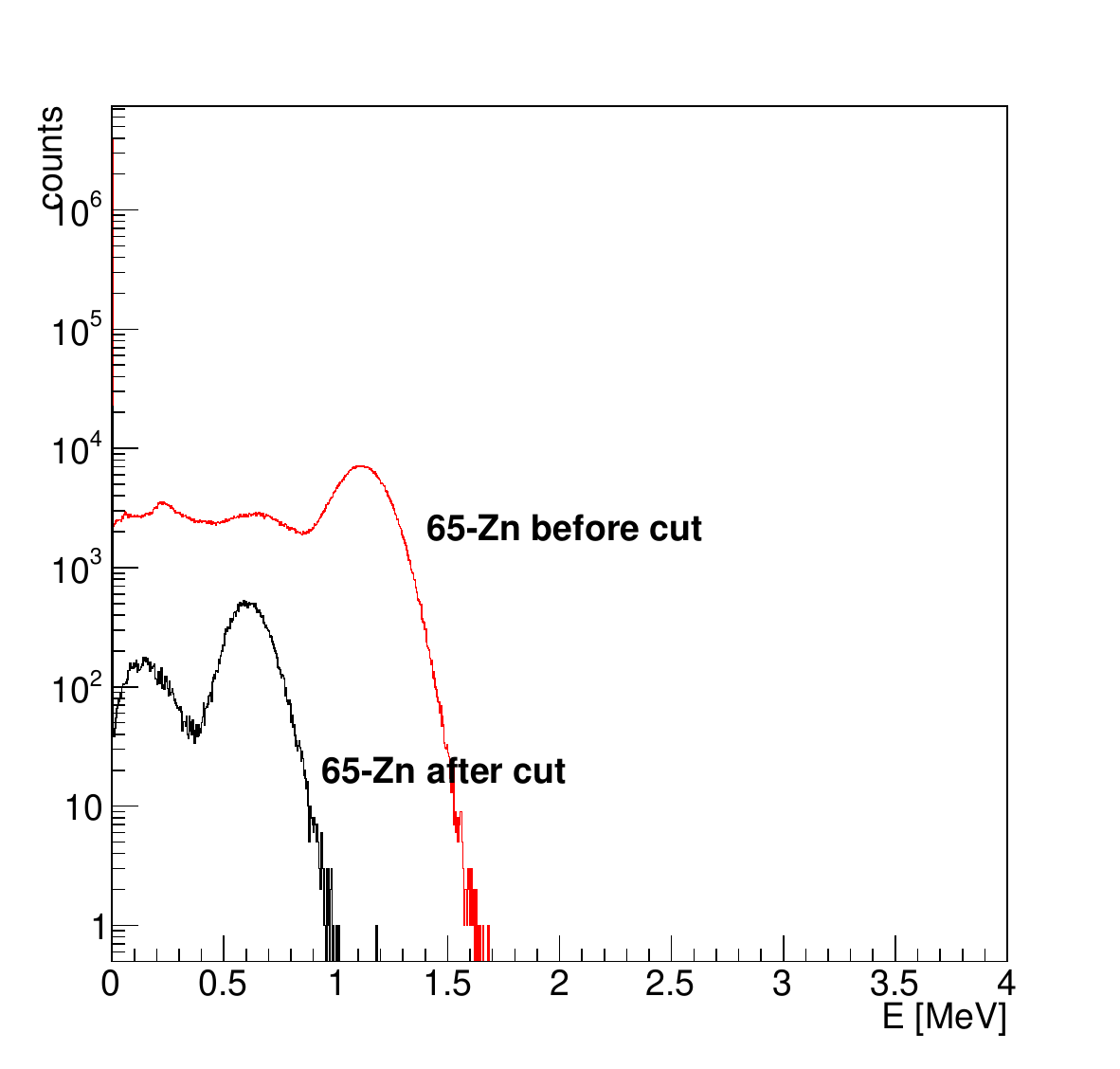}
 \caption{Comparison of spectra deposited in enriched crystal by $^{65}$Zn decay before and after the coincidence selection.}\label{fig_zn65_coincidence}
\end{figure}

 \begin{figure}[H]
  \centering
  \includegraphics[width=0.7\textwidth]{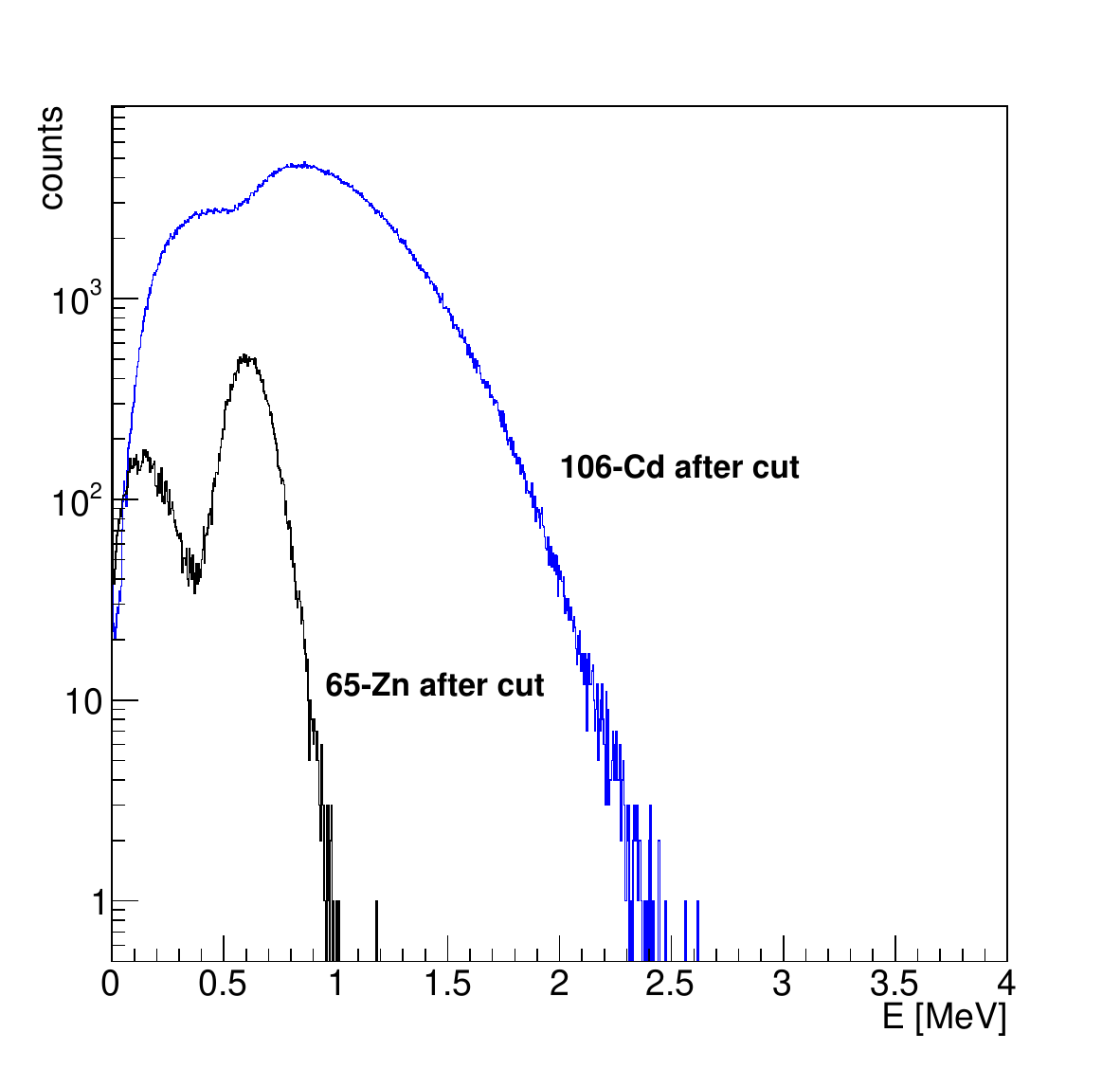}
  \caption{Comparison between the spectrum from $2\nu\epsilon\beta^+$ decay of $^{106}$Cd and $^{65}$Zn spectrum.}\label{fig_zn65_comparison}
 \end{figure}

The effect of the selection on the data, performed according to the same procedure used for the other contaminants and already presented, is shown in figure \ref{fig_zn65_coincidence}, whereas in figure \ref{fig_zn65_comparison} the result is compared to the $2\nu\epsilon\beta^+$ decay of $^{106}$Cd. The efficiency of our detector after this cut is 1.2\%. The energy deposited by $^{65}$Zn covers the lower range of the spectrum, with a significant number of counts. 

The shape of the deposited energy is instead different from the effect we are searching for, but this contamination could be significant in the lower energy range of the energy spectrum deposited inside the enriched crystal.

\subsubsection{$^{108}$Ag}

Another isotope selected for this analysis is $^{108}$Ag. It has a Q-value of 1922 keV \cite{nudat} for EC decay, but it can also undergo to $\beta^-$ decay with a Q-value of 1650 keV \cite{nudat}, according to NuDat archives. The Q-value for EC is higher than $^{184}$Re and $^{65}$Zn, but its presence is reported only by COSMO1 calculations (see table \ref{tab_activia_contaminats}). 
The decay scheme is reported in figure \ref{fig_ag108_decay}.


\begin{figure}[H]
 \centering
 \includegraphics[width=\textwidth]{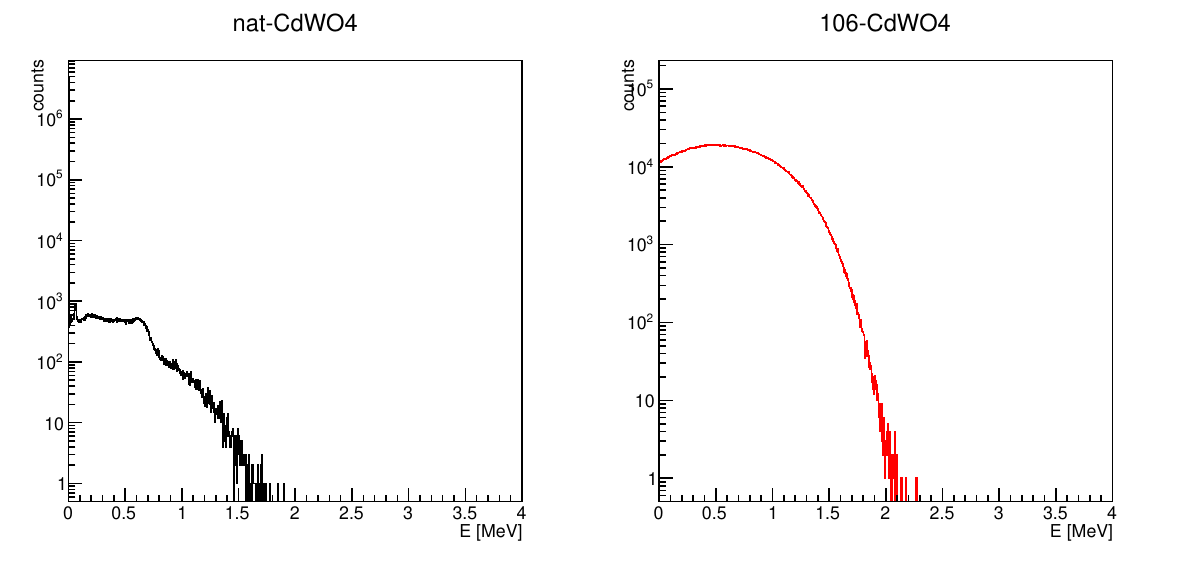}
 \caption{Energy deposited inside the three detectors by $^{108}$Ag. Since the energy spectrum from the two natural crystals is practically equal, only one is reported.}\label{fig_ag108_spectra}
\end{figure}

The result of the simulation is reported in figure \ref{fig_ag108_spectra}. In this case there is no trace of structures inside the energy spectra deposited inside the natural crystals. The shape of the energy distribution is exactly the one we expect for $\beta$ decay, that results to be the main decay channel of this isotope, with a continuous distribution that decreases to the end point corresponding to the Q-value of the reaction. 

\begin{figure}[H]
 \centering
 \includegraphics[width=0.6\textwidth]{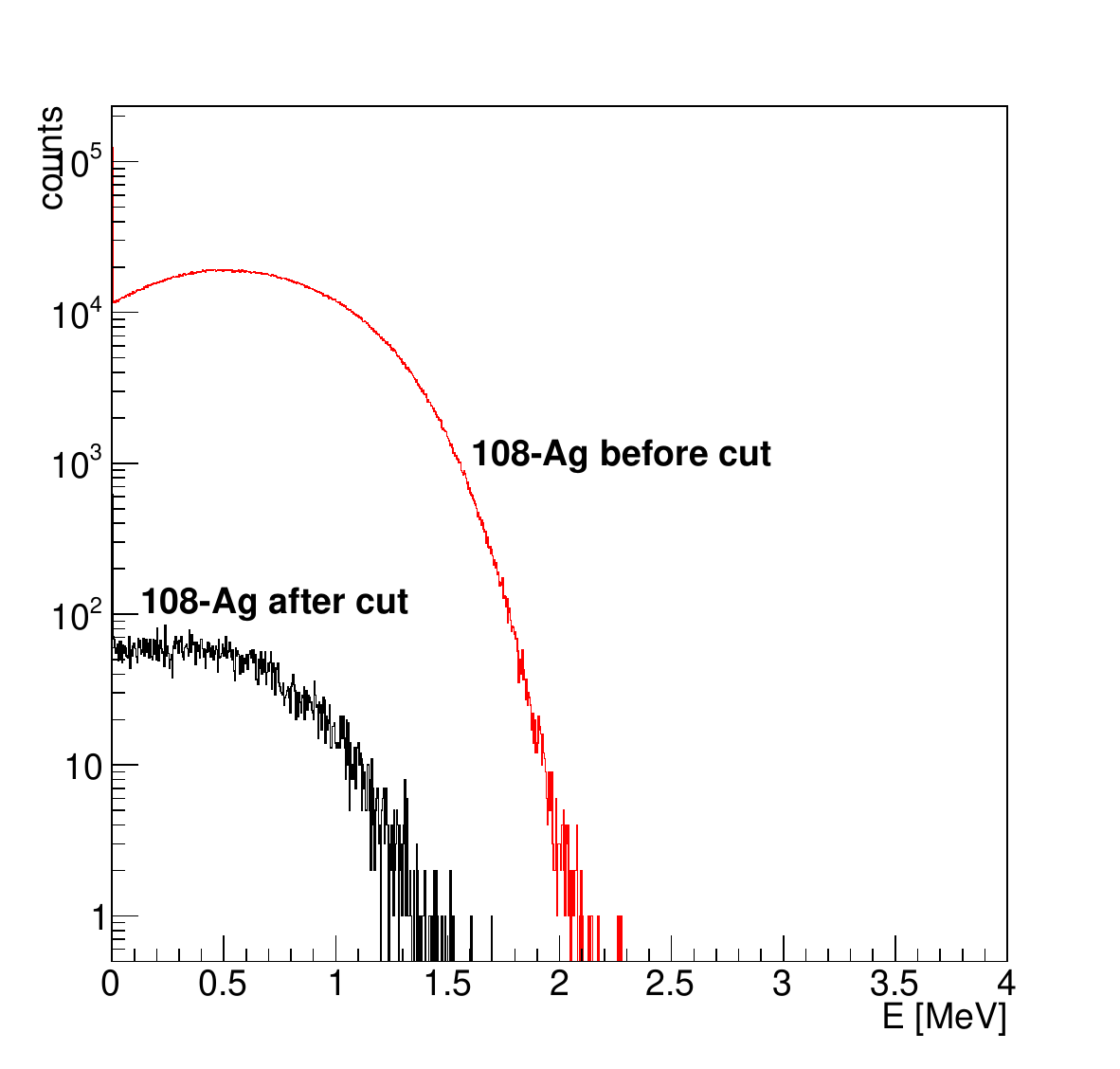}
 \caption{Comparison of spectra deposited in enriched crystal by $^{108}$Ag decay before and after the coincidence selection.}\label{fig_ag108_coincidence}
\end{figure}

Inside the two lateral detectors, we find the tails of the energy deposition. From this direct observation, we can expect that  the coincidence cut should remove a large number of events from the final distribution. The energy distribution resulting after the selection is reported in figure \ref{fig_ag108_coincidence}.

Only few events remain after this selection, namely 13165, and only at low energies. The efficiency is then 0.26\%. 

 \begin{figure}[H]
  \centering
  \includegraphics[width=0.7\textwidth]{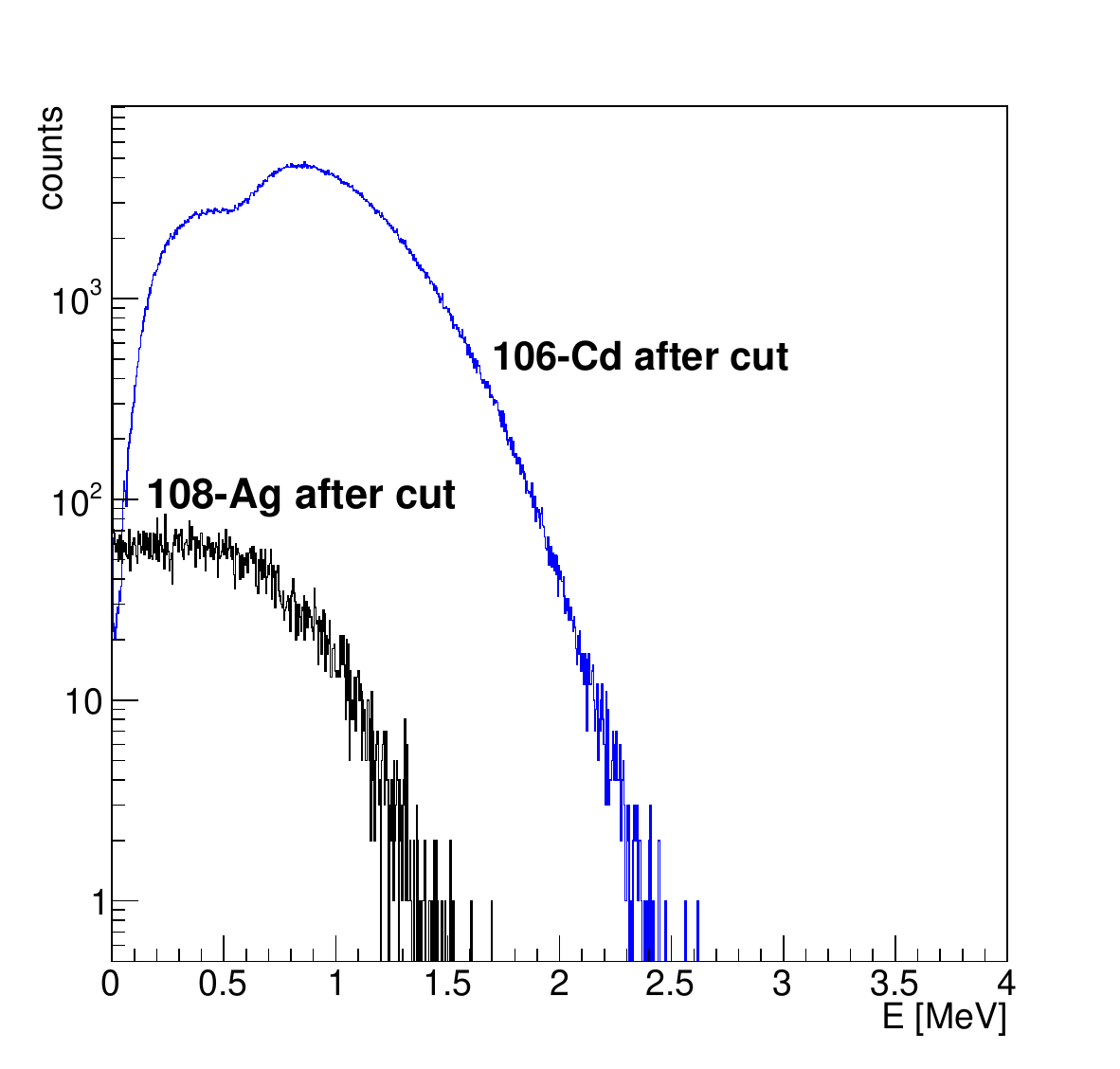}
  \caption{Comparison between the spectrum from $2\nu\epsilon\beta^+$ decay of $^{106}$Cd and $^{108}$Ag spectrum.}\label{fig_ag108_comparison}
 \end{figure}

In figure \ref{fig_ag108_comparison} we report the comparison between the energy deposited inside the $^{106}$CdWO$_4$ crystal by $^{108}$Ag decay and by $2\nu\epsilon\beta^+$ decay of $^{106}$Cd after the coincidence selection. 

\subsubsection{$^{110m}$Ag}
 
The nuclide has different decay modes. The most probable is $\beta^-$ decay (99.7\% \cite{nudat}), but also EC is possible (0.3\% \cite{nudat}). The Q-value for the $\beta^-$ events is 2892.9 keV \cite{nudat}, and for EC 889 keV \cite{nudat}. The metastable state at 117.59 keV is the main source of decays from this contamination, since its half-life is 250 days long. The half life of the most probable $\beta^-$ decay is only 24.6 s. In figure \ref{fig_ag110_decay} the decay scheme for the $\beta^-$ process is reported, since it is the most probable according to NuDat archive.
 
 
 \begin{figure}[H]
 \centering
 \includegraphics[width=\textwidth]{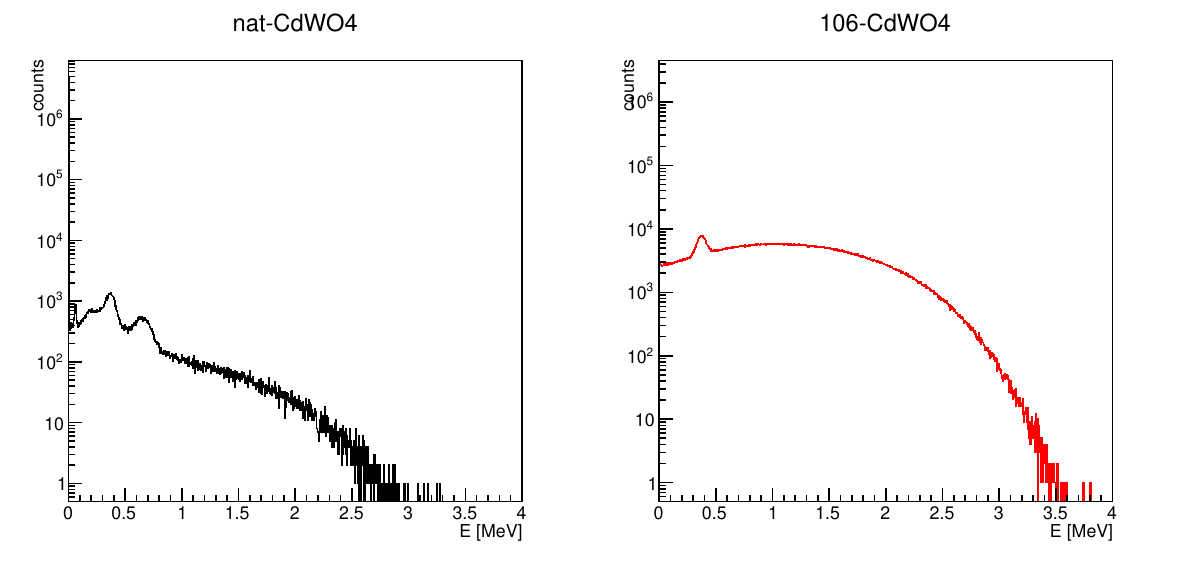}
 \caption{Energy deposited inside the three detectors by $^{110m}$Ag. Since the energy spectrum from the two natural crystals is practically equal, only one is reported.}\label{fig_ag110_1_spectra}
\end{figure}

In figure \ref{fig_ag110_1_spectra} the spectra of the energy deposited inside the three detectors are shown. Inside the enriched crystal there is a continuous distribution up to 3.5 MeV, with a superimposed peak at $\sim 0.5$ MeV. The shape of the spectra inside the two $^{nat}$CdWO$_4$ crystals is more complex. There are several irregular peaks inside the lower energy range of the spectrum, up to 1 MeV. They are superimposed on a continuous distribution that slowly decreases to the end-point at $\sim 3.5$ MeV.

 \begin{figure}[H]
 \centering
 \includegraphics[width=0.7\textwidth]{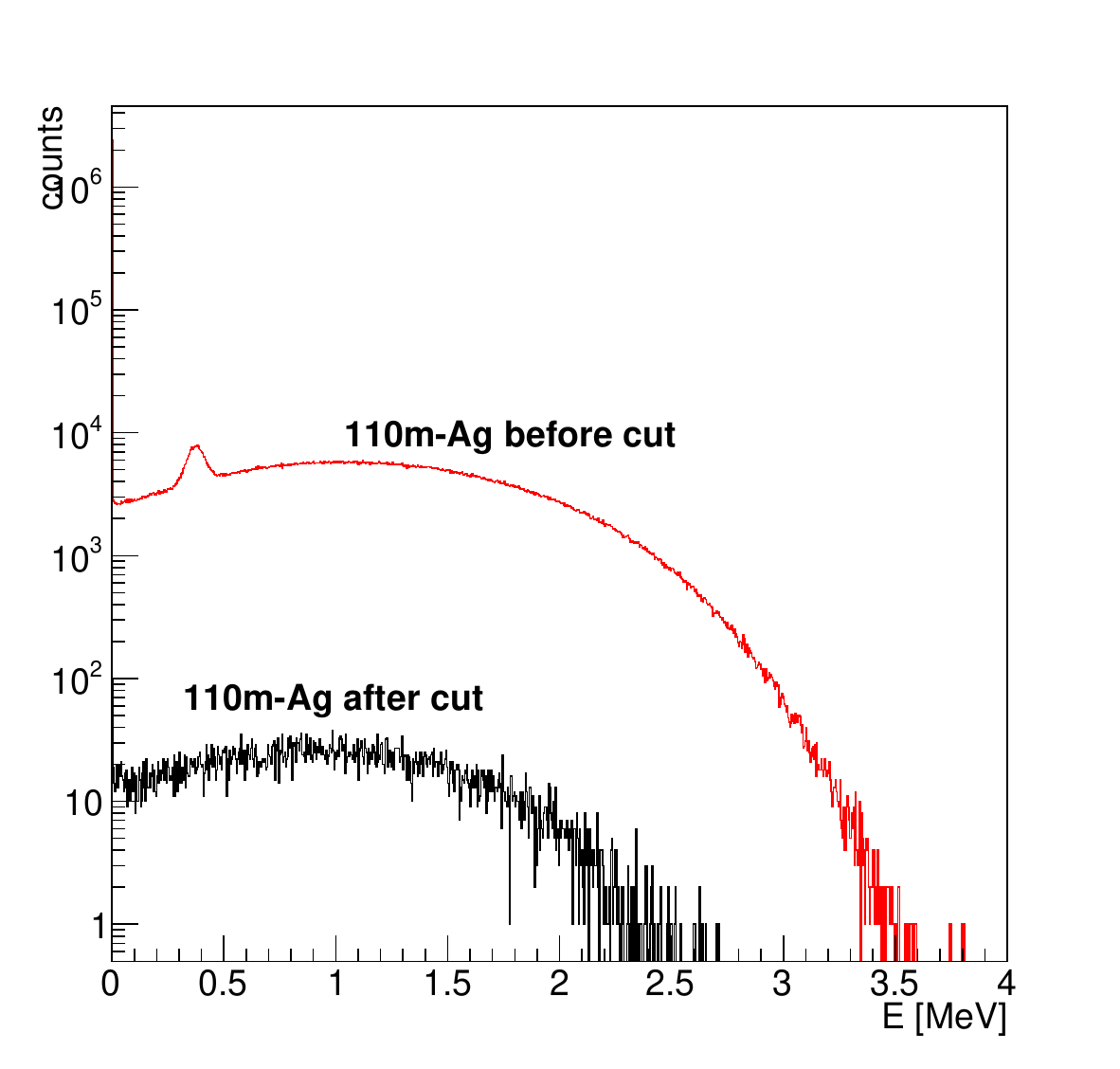}
 \caption{Comparison of spectra deposited in enriched crystal by $^{110}$Ag decay before and after the coincidence selection.}\label{fig_ag110_1_coincidence}
\end{figure}

In figure \ref{fig_ag110_1_coincidence} the result of the usual coincidence selection is shown. The surviving events are 10061, so the efficiency is 0.20\%. 

 \begin{figure}[H]
  \centering
  \includegraphics[width=0.7\textwidth]{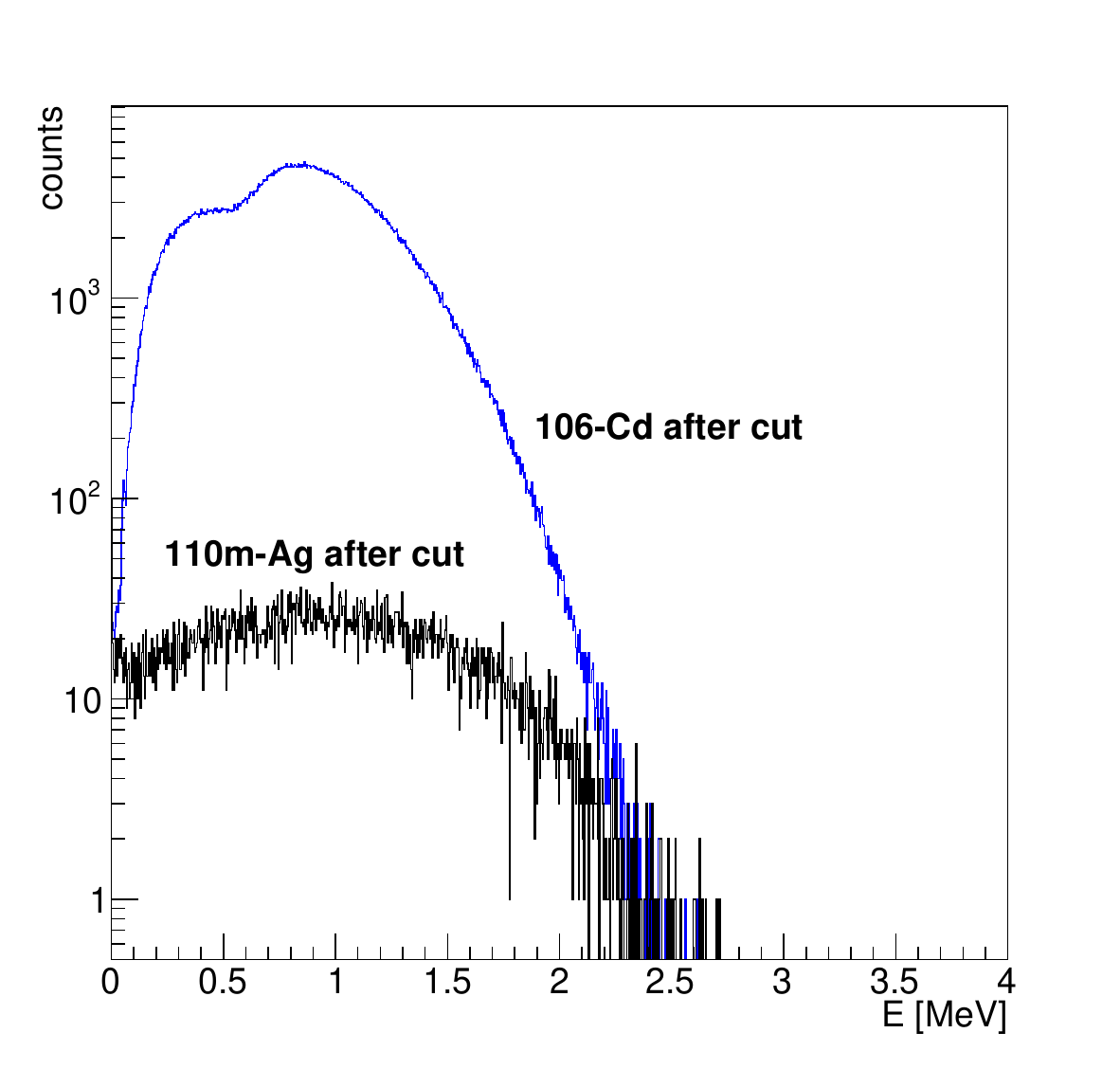}
  \caption{Comparison between the spectrum from $2\nu\epsilon\beta^+$ decay of $^{106}$Cd and $^{110m}$Ag spectrum.}\label{fig_ag110_1_comparison}
 \end{figure}

In figure \ref{fig_ag110_1_comparison} the spectrum obtained inside the enriched crystal after the coincidence selection is superimposed on  the one obtained for $2\nu\epsilon\beta^+$ decay of $^{106}$Cd. The range of the two processes is quite the same, but the coincidence selection strongly reduces the number of events caused by $^{110m}$Ag. 


\subsubsection{$^{174}$Lu}

Another nuclide that COSMO1 indicates as a possible contaminant of the enriched crystal after cosmogenical activation is $^{174}$Lu. 
We choose to run a simulation of its decay because this isotope, that usually decays by electronic capture with a Q-value of 1374.3 keV \cite{nudat}, has a small probability of decaying by positron emission, and so it is possible that its decay produces 511 keV $\gamma$ quanta from positron annihilation.


\begin{figure}[H]
 \centering
 \includegraphics[width=\textwidth]{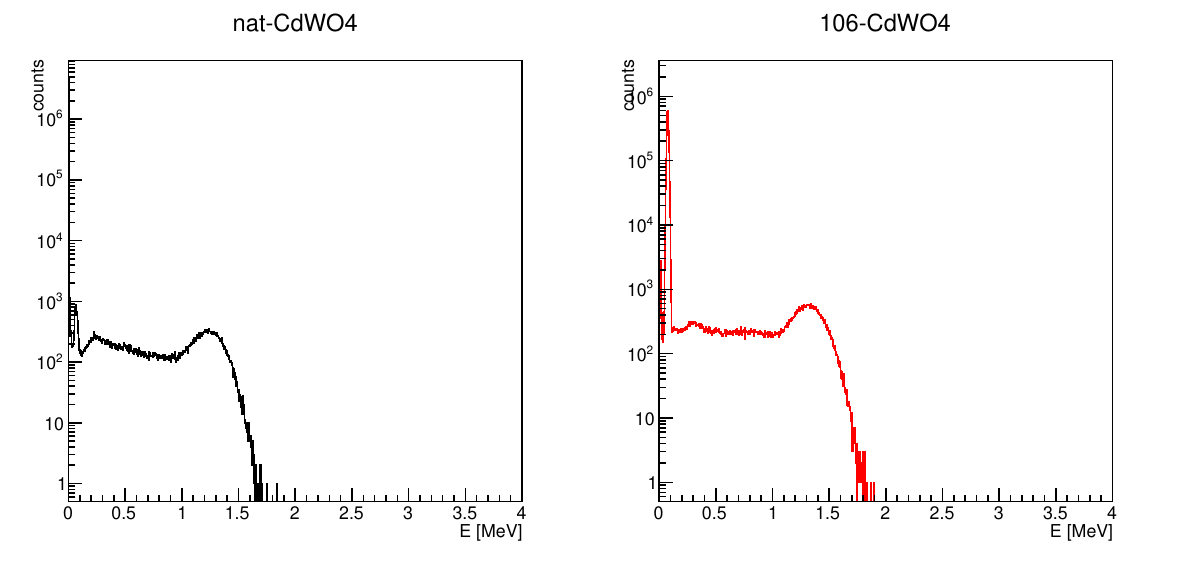}
 \caption{Energy deposited inside the three detectors by $^{174}$Lu. Since the energy spectrum from the two natural crystals is practically equal, only one is reported.}\label{fig_lu174_spectra}
\end{figure}

In this case, as we noticed before for $^{65}$Zn, the energy distribution deposited inside the three detectors is similar, as we can see in figure \ref{fig_lu174_spectra}. We notice the same main structures.

In the low energy range there is a sharp peak at $\sim$100 keV, that is mainly visible in the energy distribution deposited in the enriched detector, where the decay takes place. Because of the low energy of this emission, we do not expect to find it in the other detectors, and in fact only a little signal is present there.

We have then a continuous energy distribution, with the edge at 1.5 MeV. At this final energy, a structure similar to a peak is visible.

 \begin{figure}[H]
 \centering
 \includegraphics[width=0.6\textwidth]{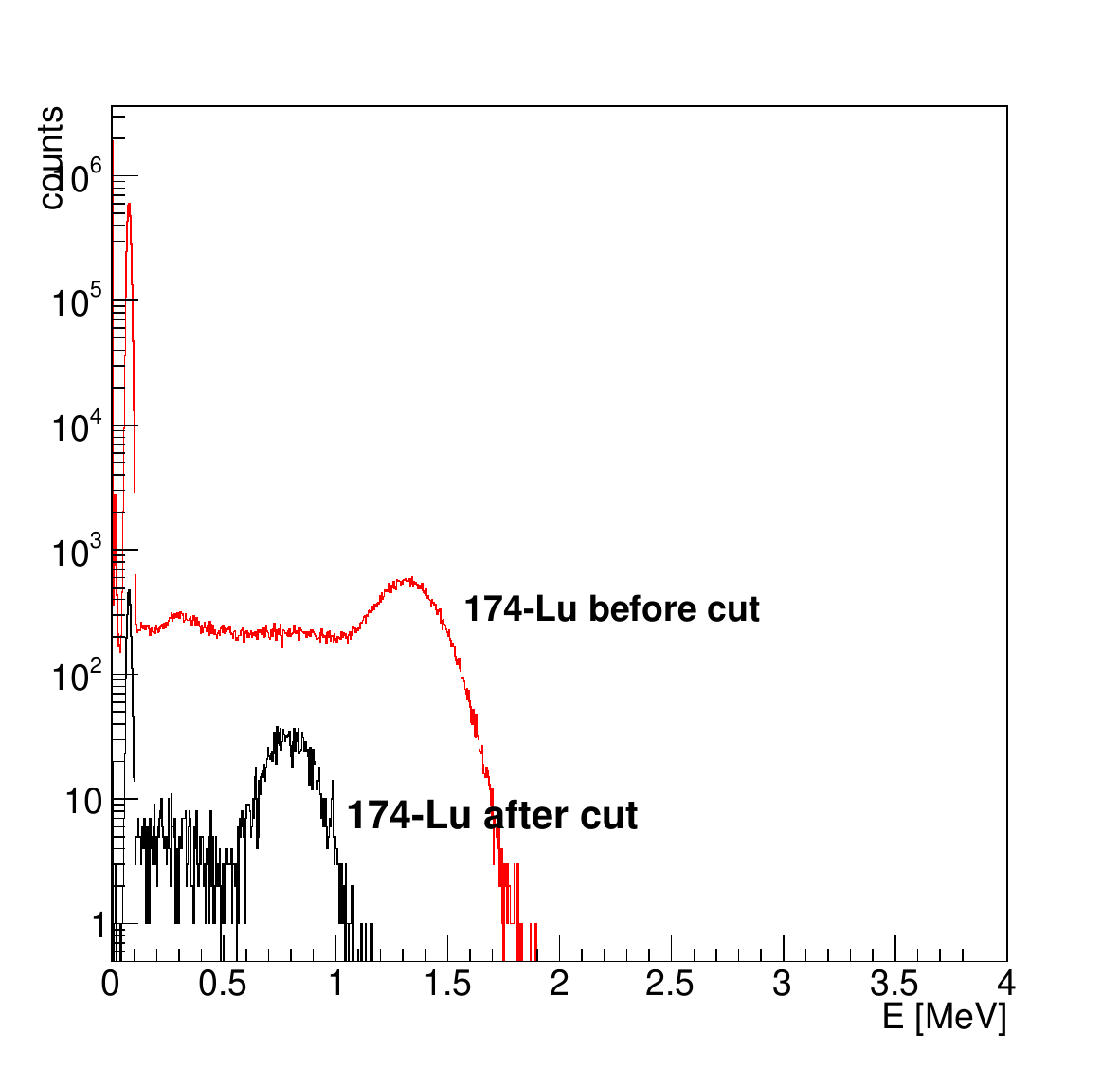}
 \caption{Comparison of spectra deposited in enriched crystal by $^{174}$Lu decay before and after the coincidence selection.}\label{fig_lu174_coincidence}
\end{figure}

 \begin{figure}[H]
  \centering
  \includegraphics[width=0.7\textwidth]{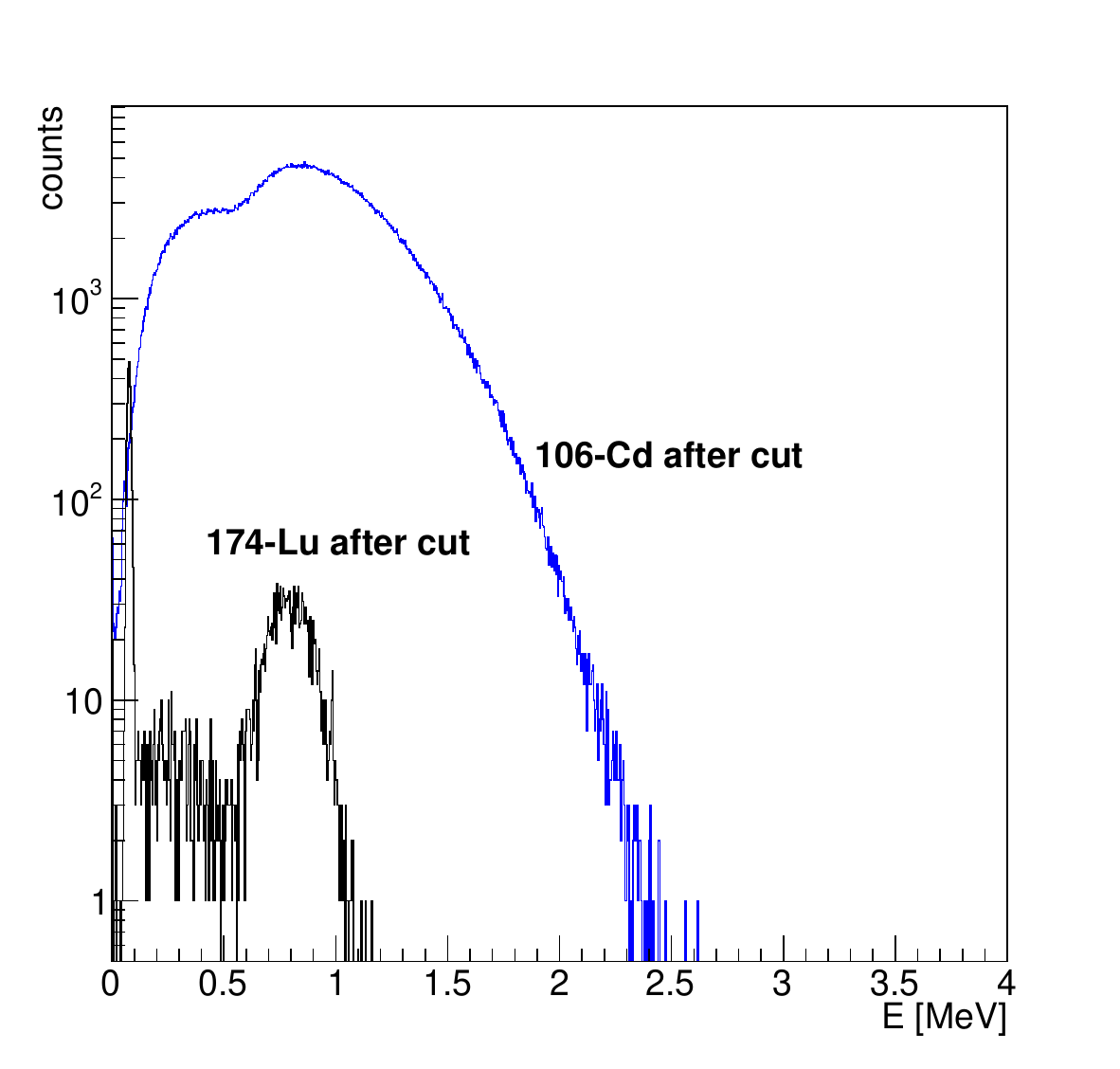}
  \caption{Comparison between the spectrum from $2\nu\epsilon\beta^+$ decay of $^{106}$Cd and $^{174}$Lu spectrum.}\label{fig_lu174_comparison}
 \end{figure}

For this isotope we performed the usual coincidence cut, whose result is shown in figure \ref{fig_lu174_coincidence}. 

The shape of the spectrum changes drastically after this selection, and there is a shift of the maximum emission to the lower energies. The number of events that survives this cut is only 4830, so the efficiency is 0.097\%. This is the lowest effect we have met yet. If all the other nuclides we present in this analysis are actually inside the enriched crystal, this one should not affect significantly our measurement, since it survives the cuts only in a fraction of events smaller than the others.

 \subsection{Simulation of $^{nat}$CdWO$_4$ contaminants}
 
 \begin{table}[hb]
 \centering
 {\relsize{-2}
 \begin{tabular}{lcccc}
 \hline
 Nuclide & Half-life  & Q-value & Activity by COSMO1 & Activity by Activia \\
         & [days]     & [keV]   & [counts kg$^{-1}$ day$^{-1}$] & [counts kg$^{-1}$ day$^{-1}$] \\
 \hline
 $^{172}$Hf & 683 $\pm$ 11 & 350 $\pm$ 50 & 2.2& 1.0\\
 $^{182}$Ta & 114.74 $\pm$ 0.12 & 1814.5 $\pm$ 1.7 & 6.2 &1.2\\
 $^{184}$Re & 169 $\pm$ 8 & 1481 $\pm$ 4& 1.5 & 0.26\\
 $^{110m}$Ag& 249.83 $\pm$ 0.04 & 2892.9 $\pm$ 1.5 &0.76 &1.8 \\
 $^{173}$Lu & 500 $\pm$ 4 & 670.8 $\pm$ 1.7 & 0.21 & 1.2\\
 \hline
 \end{tabular}
 }
\caption{Selected contaminants inside the natural crystal and their properties.}\label{tab_selected_nat}
\end{table}
 
 From the nuclide selection of possible radioactive contaminants inside the $^{nat}$CdWO$_4$ natural crystals calculated by COSMO1 and Activia softwares, we select the most active to be simulated, in order to understand their effect in our measurements.
 
 The criteria differs from the ones we applied to enriched crystal. The main reason is that there are orders of magnitude of difference between the nuclide activity values in this case, whereas in the main crystal all the isotopes had similar activities, and they were lower than that reported in natural crystals.
 The differences are due to the longer exposure to cosmogenical activation.
 
 We apply to these contributions to the background the same coincidence analysis previously introduced for the enriched crystal contaminants.
 
 This study is performed using a third version of the simulation, in which the decay source is selected randomly inside the $^{nat}$CdWO$_4$ crystals. The simulation uses the GEANT4 event generator to build primary particles. As we previously pointed out, all the simulations have the same geometry, physics and acquisition structures, whereas the input mechanism and the places where the decay takes place change.
 
 We start our analysis from the nuclides that present a huge activity in the calculations from both the softwares, COSMO1 and Activia. The nuclides are $^{172}$Hf, $^{179}$Ta, $^{182}$Ta, $^{181}$W. 
 
 We immediately notice that $^{179}$Ta and $^{181}$W have an extremely low Q-value (105.6 keV \cite{nudat} for  $^{182}$Ta and 188 keV \cite{nudat} for $^{181}$W). Moreover, the daughter nuclides are stable, so it is not possible for their decay to produce any signal that can survive the selection cuts, because the energy deposited inside the two natural crystals will always be largely under 511 keV.
 
 If we take into account also the isotopes that result to be particularly abundant only in one of the two calculation results, we find three more nuclides from Activia, $^{110m}$Ag, $^{109}$Cd and $^{173}$Lu, and $^{184}$Re from COSMO1.
 
 Again, $^{109}$Cd has a Q-value too low (214.3 keV \cite{nudat}) to be interesting for this analysis, and its daughter nuclide is stable. 
 
 We have then 5 isotopes to focus on. For each of them, we run a simulation consisting in 5$\times$10$^6$ events.
 
 \subsubsection{$^{172}$Hf}
 
 $^{172}$Hf is a radioactive nuclide that should be produced by cosmogenical activation inside the two natural crystals according to the calculation from both COSMO1 and Activia (see table \ref{tab_activia_nat}). 
 The decay scheme is reported in figure \ref{fig_hf172_decay}.
 The Q-value of $^{172}$Hf EC decay is 350 keV \cite{nudat}. It is a quite low value, but the daughter nuclide, $^{172}$Lu, is radioactive. It decays by EC with a half-life of 6 days \cite{nudat}, and the Q-value of the reaction is 2519.3 keV \cite{nudat}. 
 
 
 \begin{figure}[H]
  \centering
  \includegraphics[width=\textwidth]{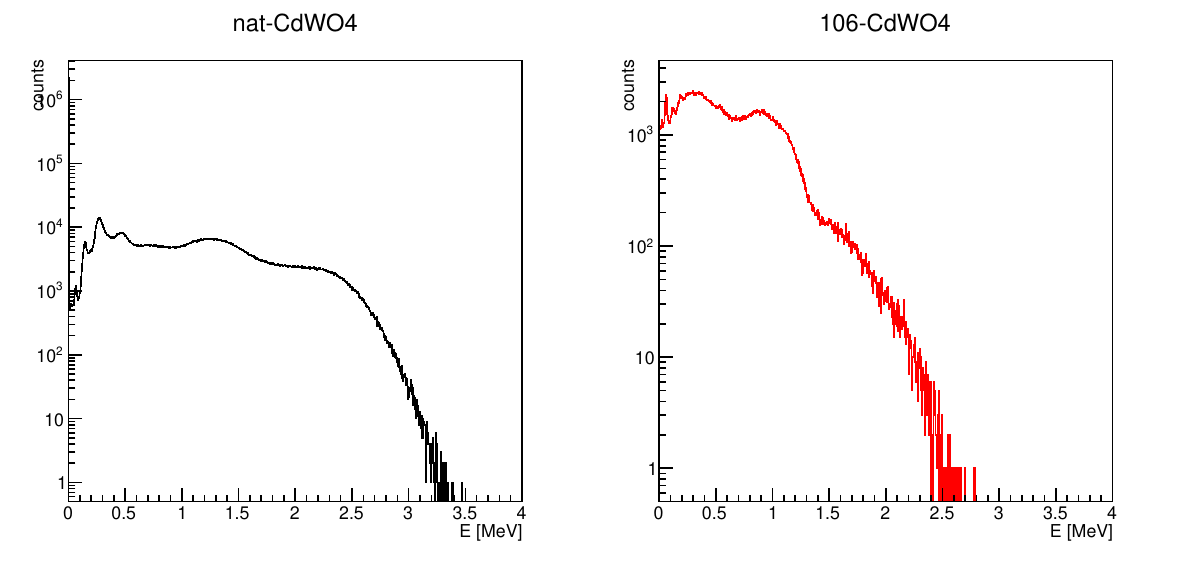}
  \caption{Energy distribution due to $^{172}$Hf and $^{172}$Lu produced by cosmogenic activation inside the  $^{nat}$CdWO$_4$ crystals. Since the energy spectrum from the two natural crystals is practically equal, only one is reported.}\label{fig_hf172_spectra}
 \end{figure}
 
 In figure \ref{fig_hf172_spectra} the energy distributions deposited inside the three detectors are reported. We notice in the spectra collected by $^{nat}$CdWO$_4$ crystals some structures at low energies, that derive from the de-excitation of the nuclei after the EC processes. Then, we have a continuous structure, with a small local maximum at $\sim$1.5 MeV. Then, the energy starts to decrease until $\sim3.5$ MeV. 
 
 Inside the $^{106}$CdWO$_4$ crystal we cannot resolve the structures at low energies, and the maximum on the spectrum is reached at lower energy, near $\sim1.0$ MeV. In this spectra, the energy distribution stops at $\sim$2.5 MeV.
 
  \begin{figure}[H]
  \centering
  \includegraphics[width=0.6\textwidth]{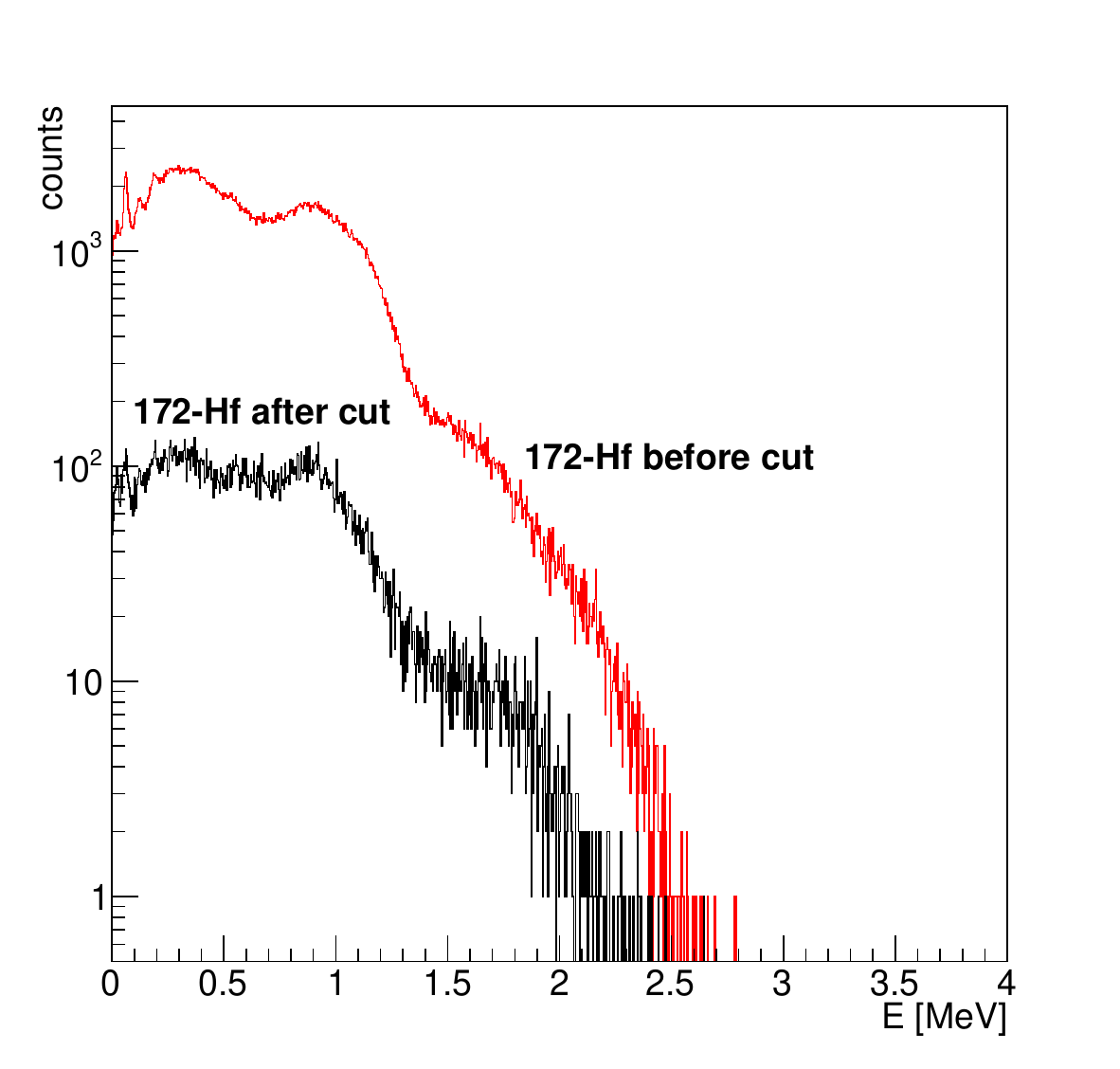}
  \caption{Comparison between the energy deposited inside the enriched crystal by $^{172}$Hf and $^{172}$Lu before and after the selection of events related to 511 keV energy deposition inside natural crystals.}\label{fig_hf172_coincidence}
 \end{figure}
 
 The usual coincidence cut is applied to the data collected from this simulation, and the result is shown in figure \ref{fig_hf172_coincidence}. We know from figure \ref{fig_hf172_spectra} that the energy region we use to select is widely populated, since the decay takes place inside that crystal. And since the distribution of the energy deposited in the enriched crystal shows a large number of events, the result we obtain is not unexpected. 
 
 
 
 
 For the efficiency calculation, we use the previously presented method, and we select the events in the enriched crystal that are in coincidence with energy deposition inside the two natural detectors inside a 4$\sigma$ interval from 511 keV. At the same time, we care that there is energy deposition inside the enriched crystal, because we focus our analysis about the characteristics of double-$\beta$ decay of $^{106}$Cd on the shape of the spectrum released inside this detector. 
 
 The number of surviving events is 28439, and the efficiency is then 0.57\%. 
 
 \begin{figure}[H]
  \centering
  \includegraphics[width=0.7\textwidth]{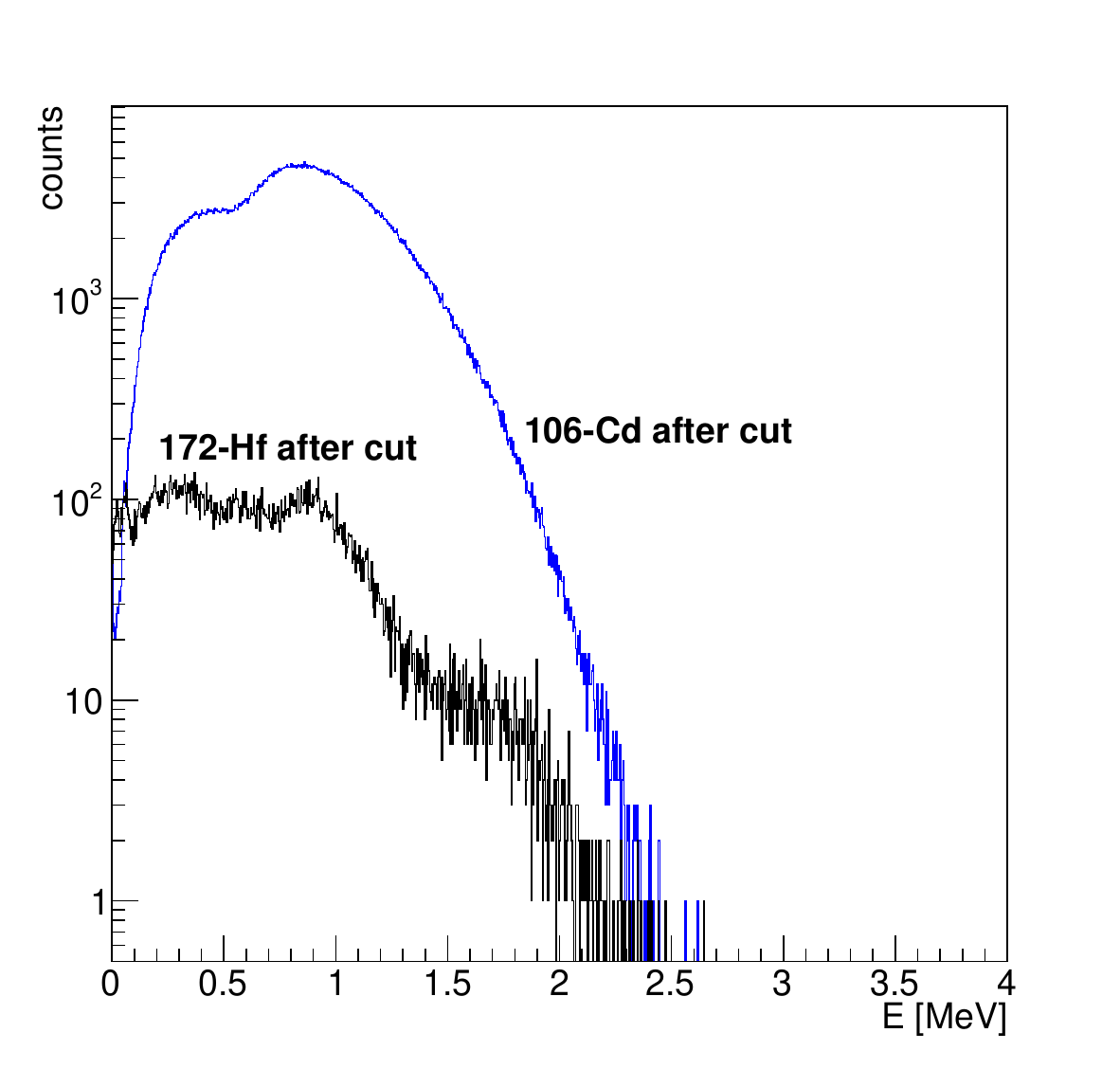}
  \caption{Comparison between the spectrum from $2\nu\epsilon\beta^+$ decay of $^{106}$Cd and $^{172}$Hf spectrum.}\label{fig_hf172_comparison}
 \end{figure}
 
 In figure \ref{fig_hf172_comparison} we present the usual comparison between the spectrum obtained for $2\nu\epsilon\beta^+$ decay of $^{106}$Cd  and the spectrum from $^{172}$Hf and $^{172}$Lu after the coincidence cut. 
 
 We can see that the effect of this contamination is significant inside all the range where we expect to find the energy distribution from $^{106}$Cd decay. The fact that this contamination is inside the natural crystals makes the coincidence cut ineffective on it, and the effective presence of the contaminant should be carefully checked, mainly because of the high activity foreseen by the theoretical calculations.
 
 \subsubsection{$^{182}$Ta}
%
%
 $^{182}$Ta is a radioactive nuclide that should be generated by cosmogenical activation inside the two $^{nat}$CdWO$_4$ crystals, according to both COSMO1 and Activia calculations (see table \ref{tab_activia_nat}). 
 The main properties of this isotope have been discussed previously, since it is also taken into account as a contaminant for enriched crystal. The decay scheme is reported in figure \ref{fig_ta182_decay}.
 
 \begin{figure}[H]
  \centering
  \includegraphics[width=\textwidth]{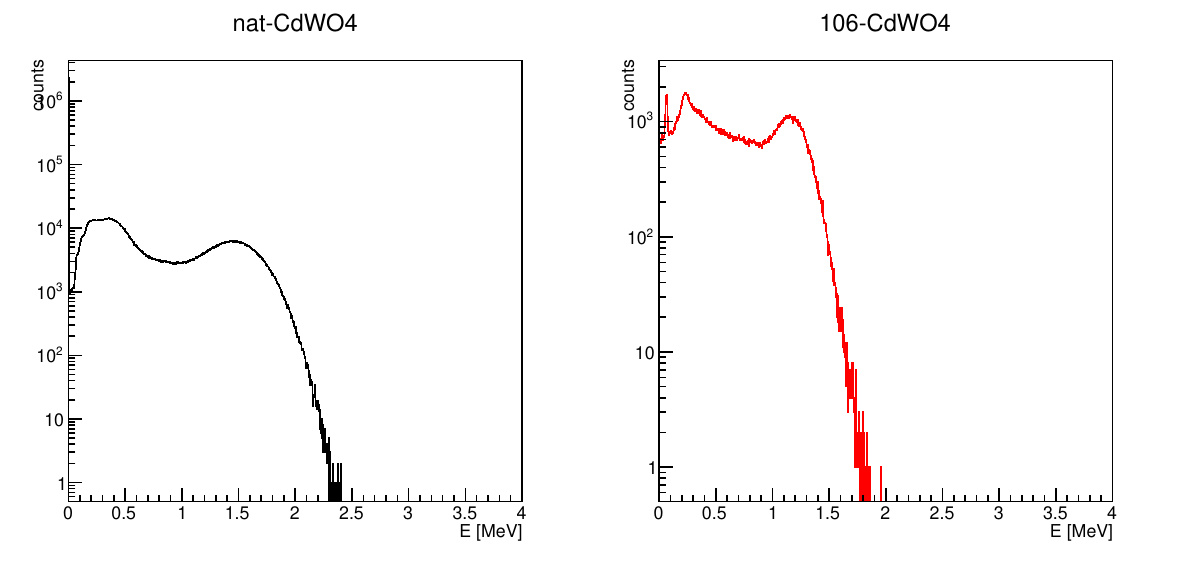}
  \caption{Energy distribution due to $^{182}$Ta produced by cosmogenic activation inside the  $^{nat}$CdWO$_4$ crystals. Since the energy spectrum from the two natural crystals is practically equal, only one is reported.}\label{fig_ta182_spectra}
 \end{figure}
 
 The simulated response of $^{106}$CdWO$_4$ and $^{nat}$CdWO$_4$ detectors to decays of $^{182}$Ta in the natural detectors are presented in figure \ref{fig_ta182_spectra}.
 Inside the two natural crystal we find practically the same energy spectrum, whereas inside the enriched crystal a degradation of the deposited energy modifies the shape and reduces the end-point of the distribution to a lower value.
 
   \begin{figure}[H]
  \centering
  \includegraphics[width=0.6\textwidth]{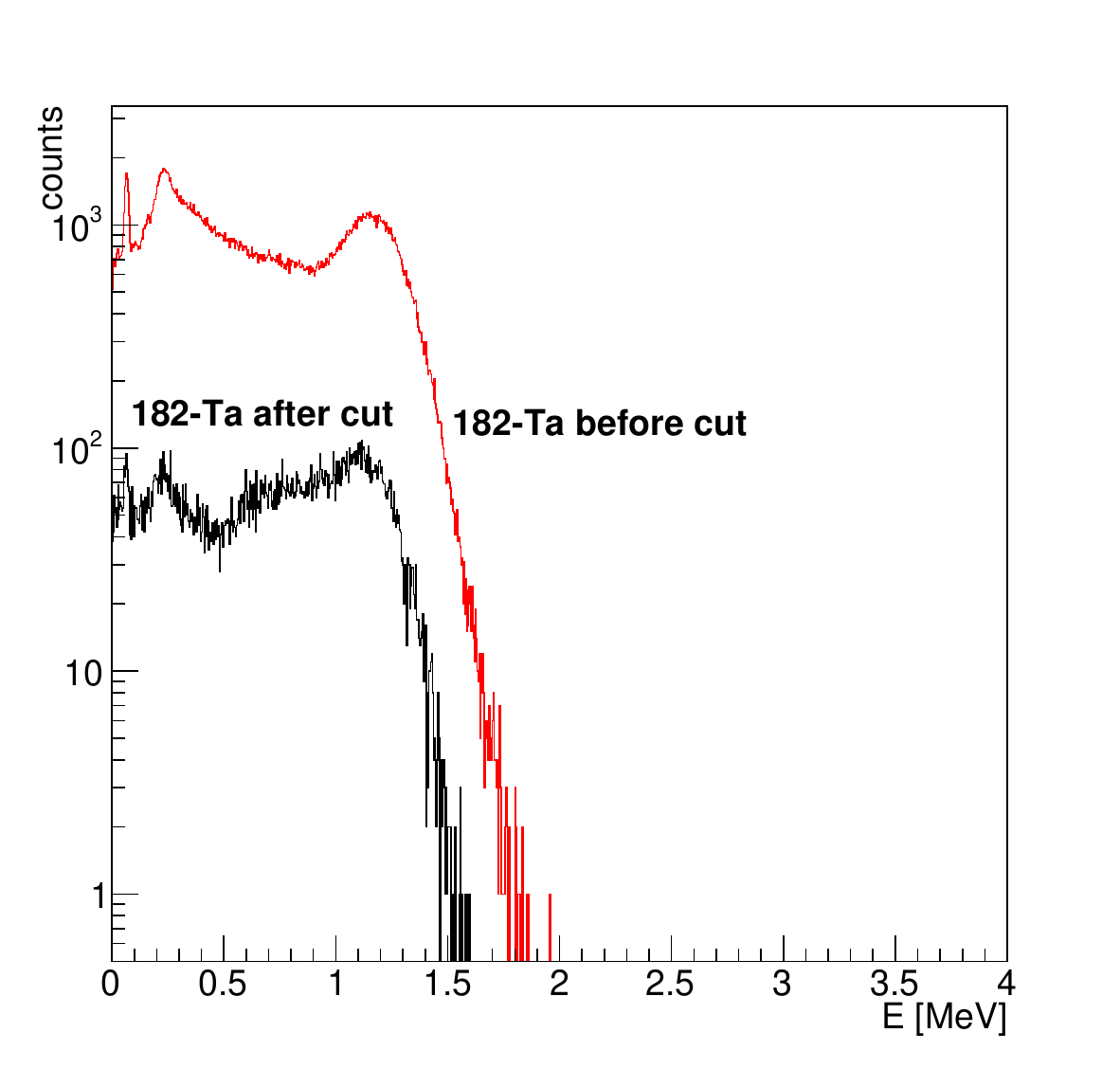}
  \caption{Comparison between the energy deposited inside the enriched crystal by $^{182}$Ta before and after the selection of events related to 511 keV energy deposition inside natural crystals.}\label{fig_ta182_coincidence}
 \end{figure}
 
 The usual coincidence selection is applied to the data collected from the simulation. The effect of this analysis on the energy deposited inside the $^{106}$CdWO$_4$ crystal is reported in figure \ref{fig_ta182_coincidence}.
 
 
 After the selection previously presented, we obtain 21243 events inside the enriched crystal in coincidence with events in the selected energy range in natural detectors. Only 0.42\% of $^{182}$Ta decays provide events in the enriched detector after the coincidence cut application.
  
 \begin{figure}[H]
  \centering
  \includegraphics[width=0.7\textwidth]{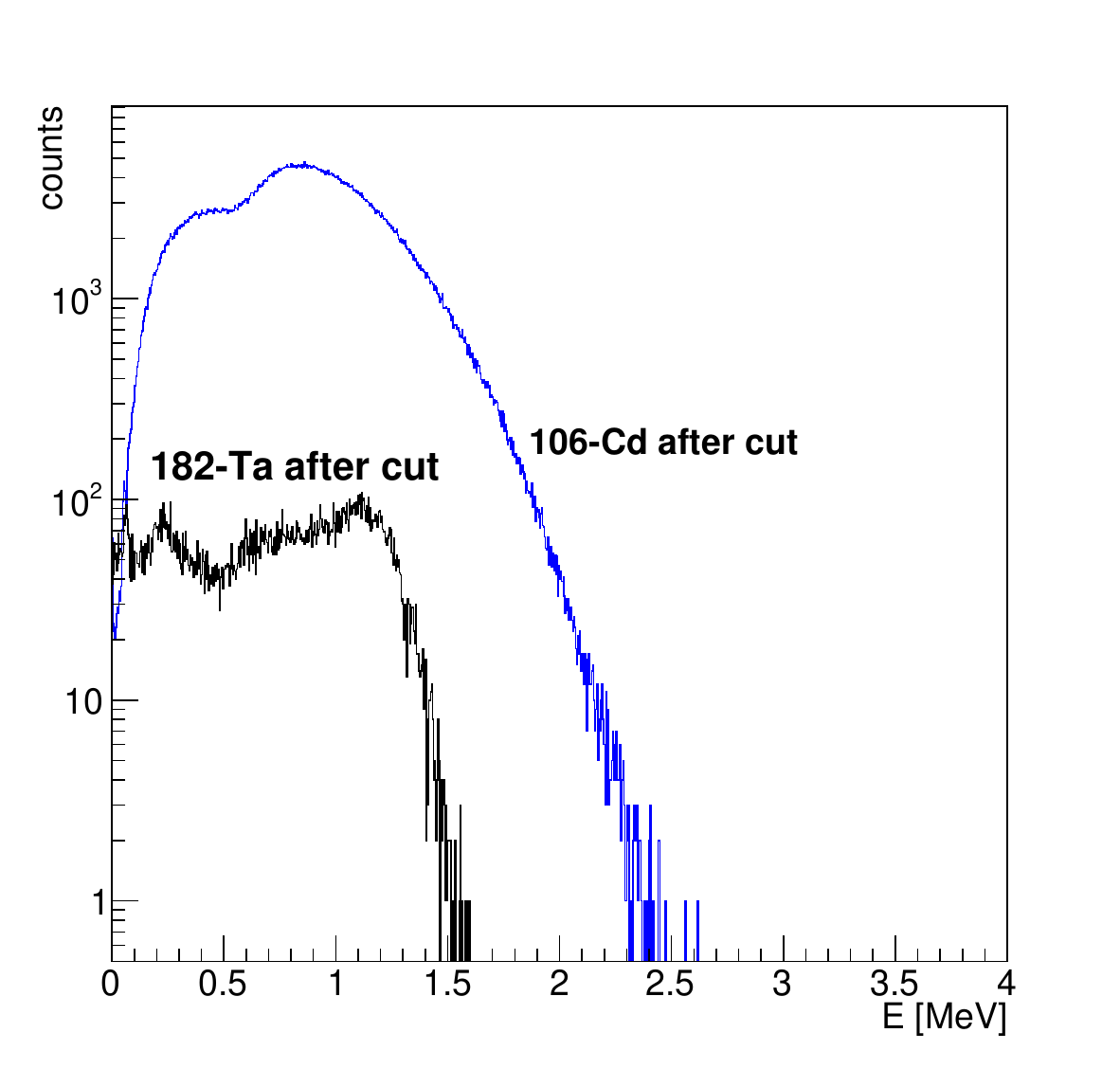}
  \caption{Comparison between the spectrum from $2\nu\epsilon\beta^+$ decay of $^{106}$Cd and $^{182}$Ta spectrum.}\label{fig_ta182_comparison}
 \end{figure}
 
 In figure \ref{fig_ta182_comparison} we report, as usual, the overlap between the spectrum deposited in the enriched detector by $^{182}$Ta decay and by $2\nu\epsilon\beta^+$ decay of $^{106}$Cd. The effect is again remarkable, even if it covers only a part of the range interested by double-$\beta$ effect.
 
 \subsubsection{$^{184}$Re}

 We already discussed the main features of $^{184}$Re, since it is reported to be one of the contaminants inside the enriched crystal we analyzed. 
 The decay scheme is reported in figure \ref{fig_re184_decay}. 
 
 
  \begin{figure}[H]
  \centering
  \includegraphics[width=\textwidth]{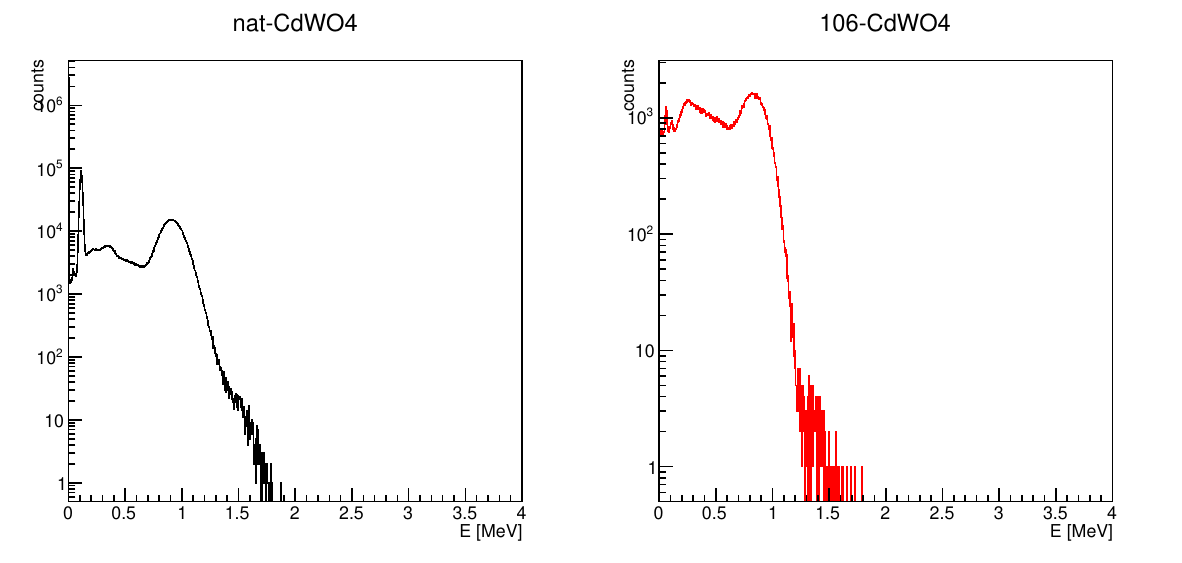}
  \caption{Energy distribution due to $^{184}$Re produced by cosmogenic activation inside the  $^{nat}$CdWO$_4$ crystals. Since the energy spectrum from the two natural crystals is practically equal, only one is reported.}\label{fig_re184_2_spectra}
 \end{figure}
 
 In figure \ref{fig_re184_2_spectra} we show the energy distributions inside the detectors. In the spectra inside the enriched crystals we can see an X-ray peak at $\sim$13 keV, and another peak at $\sim$1 MeV. All these features completely disappear in the enriched crystal spectrum, where only a continuous energy distribution is shown.
 
 \begin{figure}[H]
  \centering
  \includegraphics[width=0.7\textwidth]{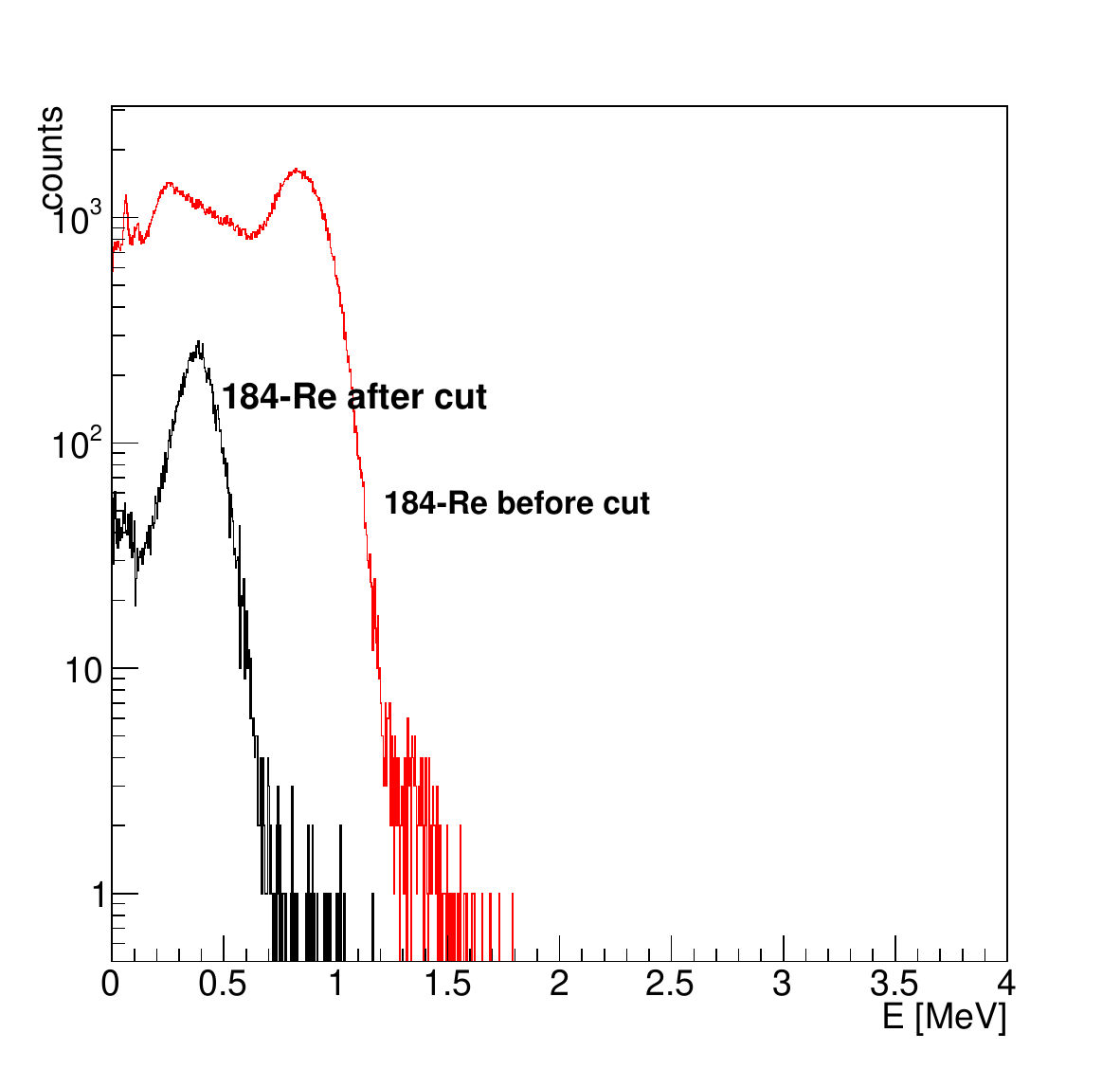}
  \caption{Comparison between the energy deposited inside the enriched crystal by $^{184}$Re before and after the selection of events related to 511 keV energy deposition in natural crystals.}\label{fig_re184_2_coincidence}
 \end{figure}
 
 Figure \ref{fig_re184_2_coincidence} shows the result of the coincidence cut on the collected data. The shape we obtain from this simulation is really close to the one from the same isotope located in the enriched crystal. 
 
 After the cut, the number of surviving events is 15251, and the efficiency is 0.31\%. 
 
   \begin{figure}[H]
  \centering
  \includegraphics[width=0.7\textwidth]{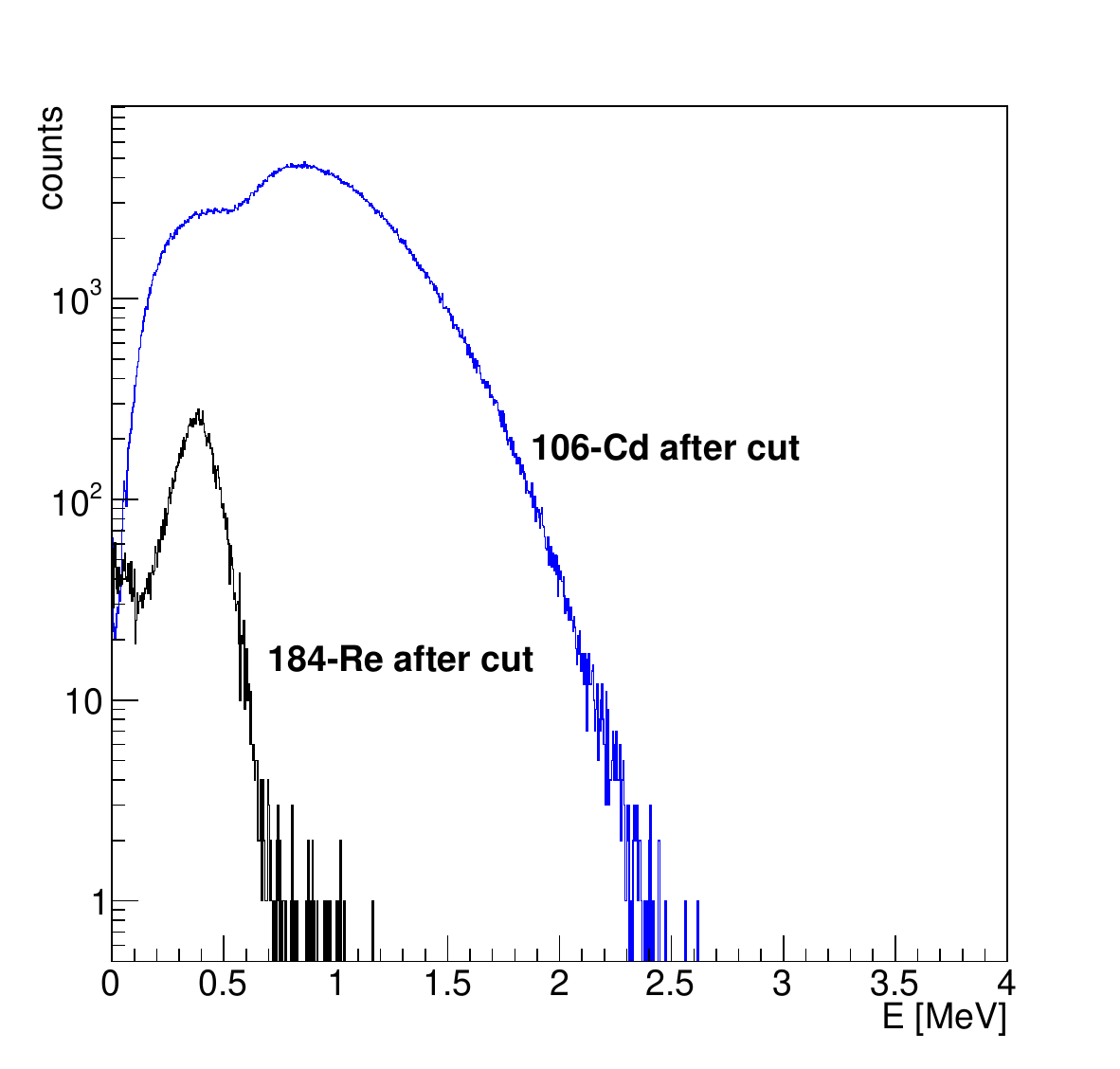}
  \caption{Comparison between the spectrum from $2\nu\epsilon\beta^+$ decay of $^{106}$Cd and $^{184}$Re spectrum.}\label{fig_re184_2_comparison}
 \end{figure}

 The effect can be still relevant, as we can appreciate in figure \ref{fig_re184_2_comparison}, where the spectrum obtained from $^{184}$Re and $2\nu\epsilon\beta^+$ decay of $^{106}$Cd are shown together.
 
 This contamination should be taken into account in further studies, since it seems to resist the coincidence cut and cover a large range of the energy region of interest.

 \subsubsection{$^{110m}$Ag}
 
 
 $^{110m}$Ag decay scheme (figure \ref{fig_ag110_decay}) and the features of its decay modes have been reported previously, since this isotope is predicted to be a contamination for the enriched crystal.
   
 \begin{figure}[H]
  \centering
  \includegraphics[width=\textwidth]{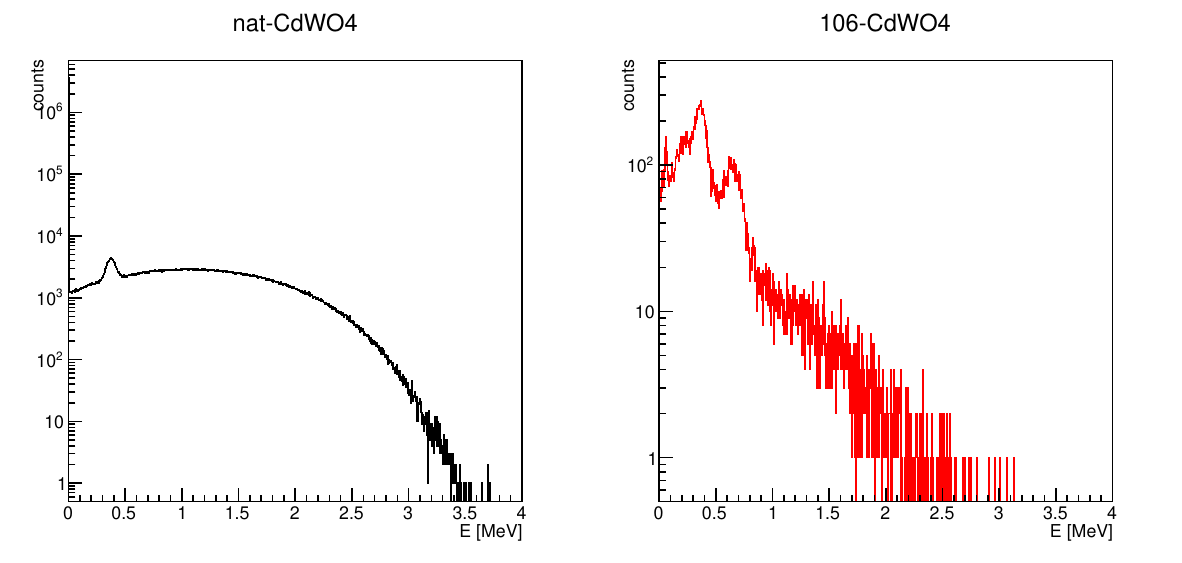}
  \caption{Energy distribution due to $^{110m}$Ag produced by cosmogenic activation inside the  $^{nat}$CdWO$_4$ crystals. Since the energy spectrum from the two natural crystals is practically equal, only one is reported.}\label{fig_ag110_spectra}
 \end{figure}
 
 In figure \ref{fig_ag110_spectra} the spectra from the energy deposited inside the detectors is shown. We immediately notice that the effects inside the enriched detector is low. This difference could be attributed to the fact that electrons emitted by $\beta$ decay are usually stopped after few millimeters inside the surrounding matter, so only the events happening near the emitting crystal surface and secondary particles can deposit energy inside the other crystal.
 
 A peak with energy $\sim$0.4 MeV is clearly visible over the continuous emission inside the $^{nat}$CdWO$_4$ detectors spectra. 
 
 Inside the $^{106}$CdWO$_4$ detector energy distribution, two peaks are distinguishable. 
 
   \begin{figure}[H]
  \centering
  \includegraphics[width=0.6\textwidth]{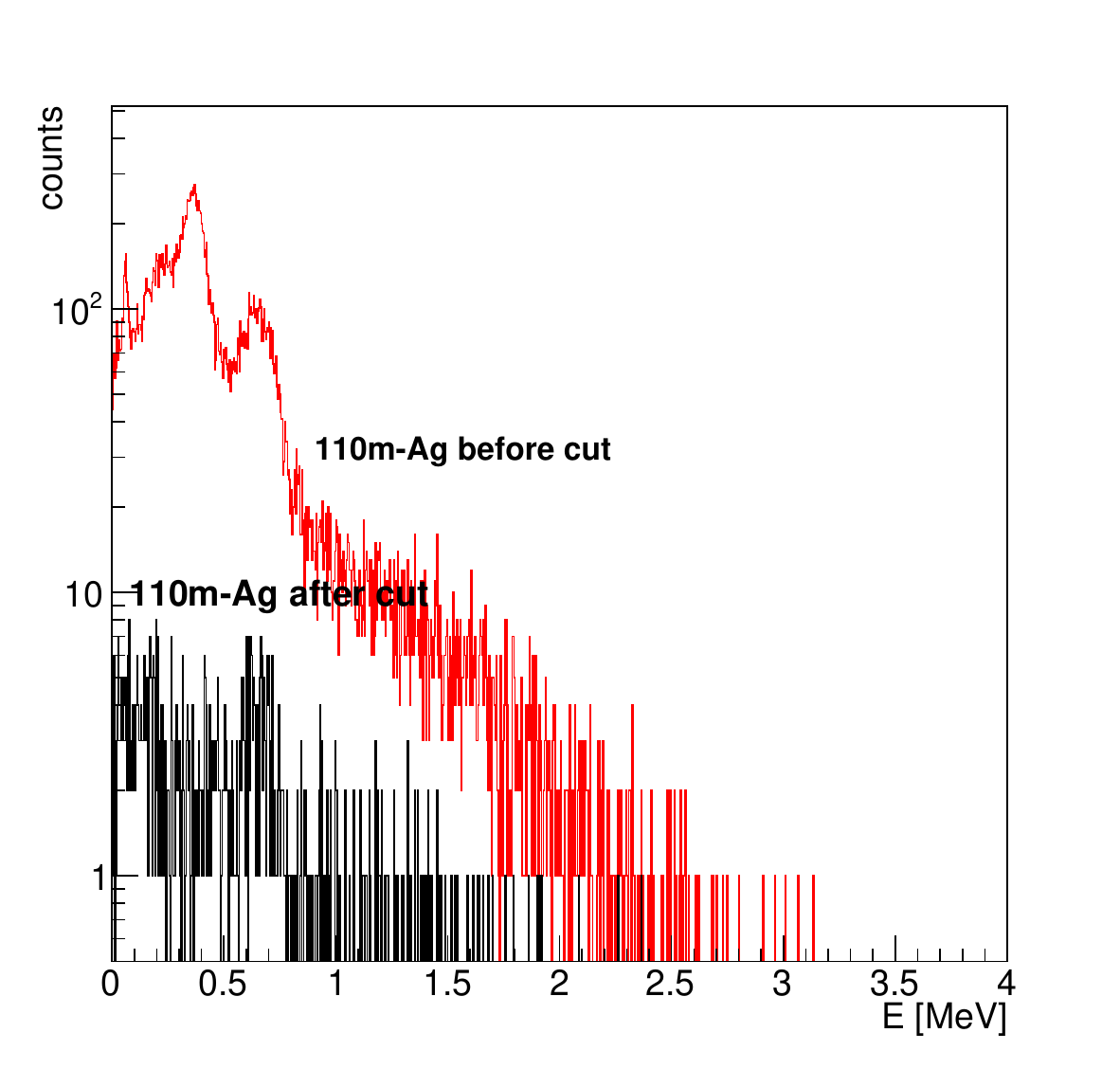}
  \caption{Comparison between the energy deposited inside the enriched crystal by $^{110m}$Ag before and after the selection of events related to 511 keV energy deposition inside natural crystals.}\label{fig_ag110_coincidence}
 \end{figure}
 
 In figure \ref{fig_ag110_coincidence} the results of coincidence selection on the spectrum deposited inside the enriched detector is reported. The low number of counts that interests this crystal is further reduced by the cut. The total number of surviving events is only 667, from which an efficiency $\leq0.02$\% can be deduced. 
 
\begin{figure}[H]
  \centering
  \includegraphics[width=0.7\textwidth]{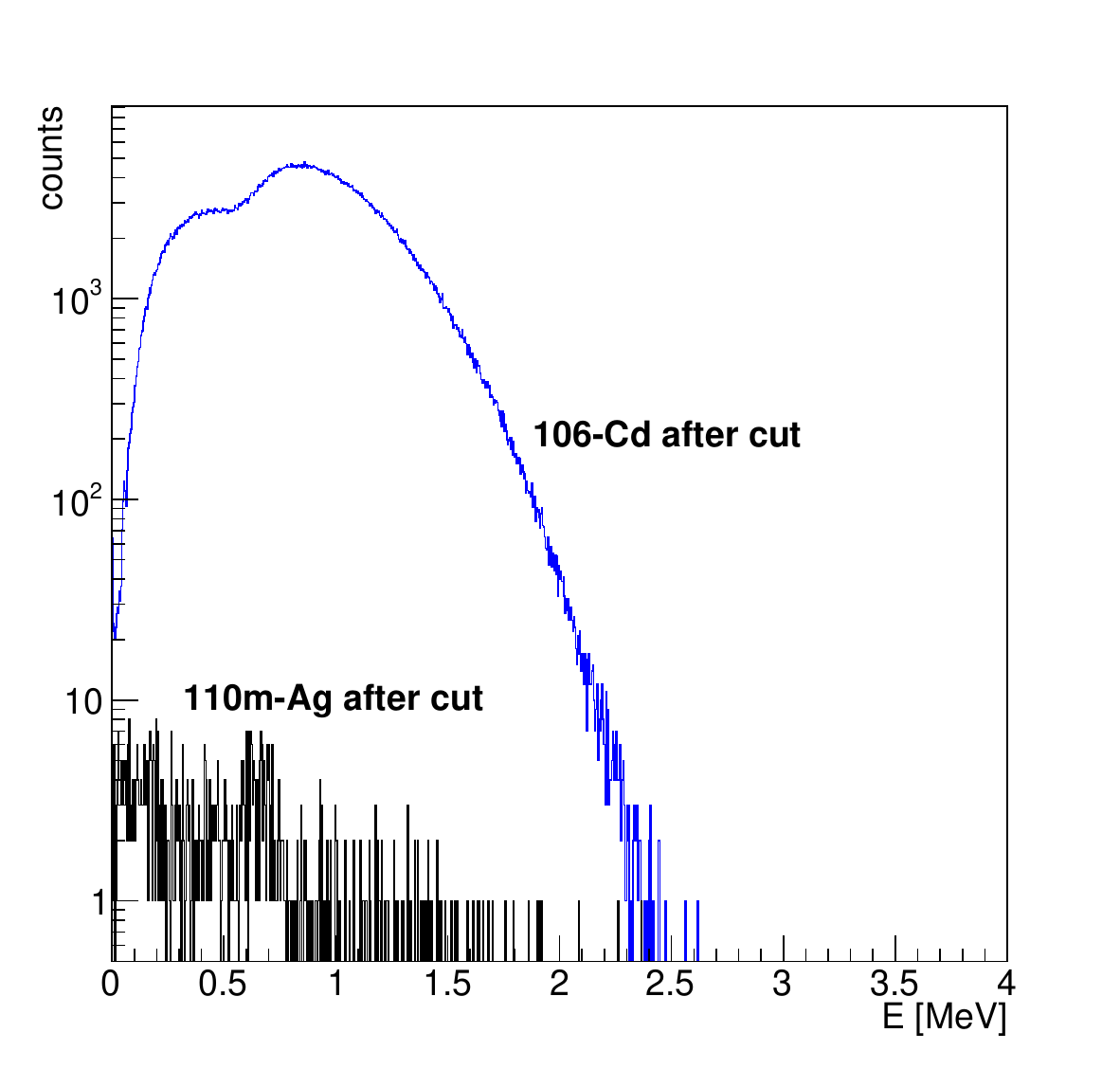}
  \caption{Comparison between the spectrum from $2\nu\epsilon\beta^+$ decay of $^{106}$Cd and $^{110m}$Ag spectrum.}\label{fig_ag110_comparison}
 \end{figure}
 
 The comparison between the effect of coincidence selection on energy deposited by $2\nu\epsilon\beta^+$ decay of $^{106}$Cd and $^{110}$Ag in the enriched crystal is reported in figure \ref{fig_ag110_comparison}. The effect of this isotope contamination is really low.
 From the exposed results, we expect a little significance of $^{110}$Ag contamination.
 
 \subsubsection{$^{173}$Lu}
 
 
 $^{173}$Lu decays by EC process, and the Q-value of the reaction is 670.8 keV \cite{nudat}. The decay scheme is reported in figure \ref{fig_lu173_decay}. The daughter nuclide, $^{173}$Yb, is stable. 
 
 
 \begin{figure}[H]
  \centering
  \includegraphics[width=\textwidth]{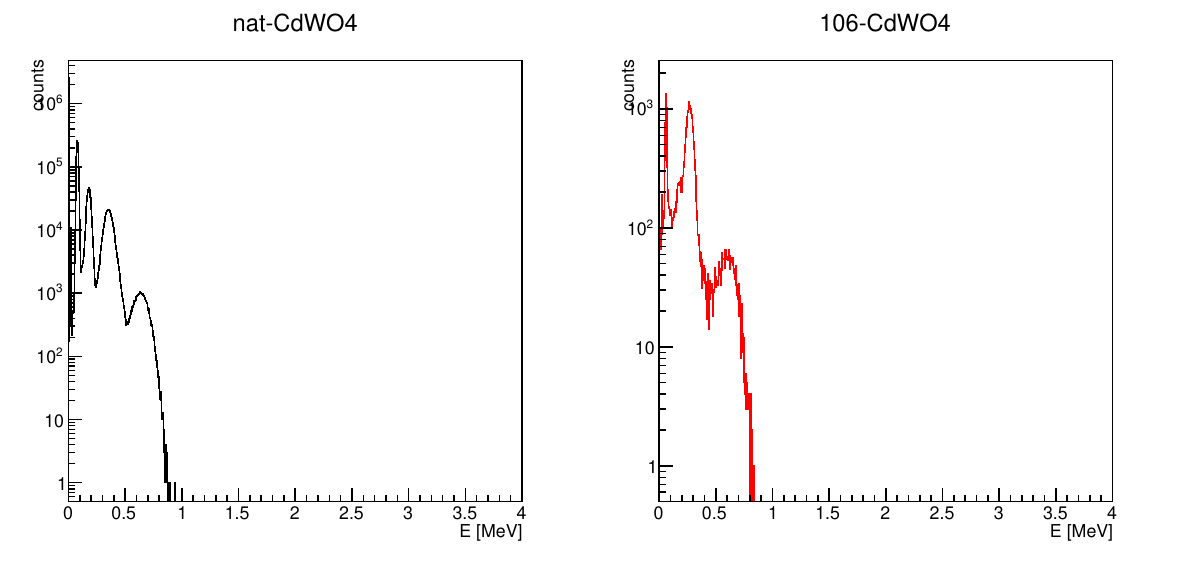}
  \caption{Energy distribution due to $^{173}$Lu produced by cosmogenic activation inside the  $^{nat}$CdWO$_4$ crystals. Since the energy spectrum from the two natural crystals is practically equal, only one is reported.}\label{fig_lu173_spectra}
 \end{figure}
 
 The three spectra in figure \ref{fig_lu173_spectra} show the energy deposition inside the detectors. Four different gamma peaks are clearly distinguishable inside the spectra deposited inside the $^{nat}$CdWO$_4$ crystal, at $\sim$80 keV, $\sim$180 keV, $\sim$350 keV and $\sim$600 keV.
 There is also an X-ray peak at $\sim$10 keV, that is not clearly visible inside the reported figure.
 
 The structures are not visible in the energy distribution deposited inside the enriched crystal, where we can see only two peaks.
 
  \begin{figure}[H]
  \centering
  \includegraphics[width=0.7\textwidth]{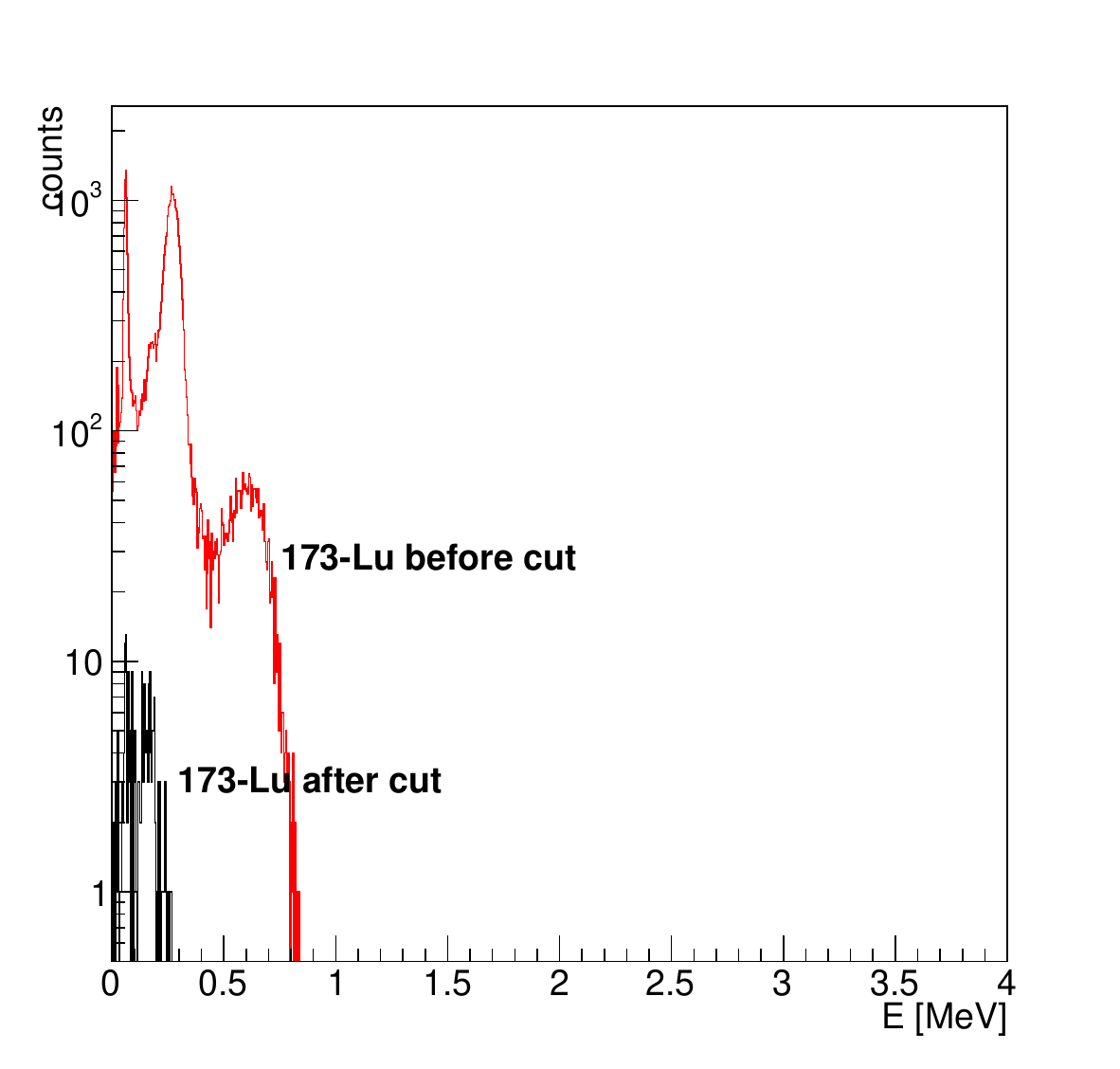}
  \caption{Comparison between the energy deposited inside the enriched crystal by $^{173}$Lu before and after the selection of events related to 511 keV energy deposition inside natural crystals.}\label{fig_lu173_coincidence}
 \end{figure}
 
 In figure \ref{fig_lu173_coincidence}, the result of the usual coincidence cut is reported. The number of events surviving the analysis is 217, and the efficiency is $\leq$0.01\%. 
 
 We can state that the effect of this isotope is not significant in our measurements. 
 
  \begin{figure}[H]
  \centering
  \includegraphics[width=0.7\textwidth]{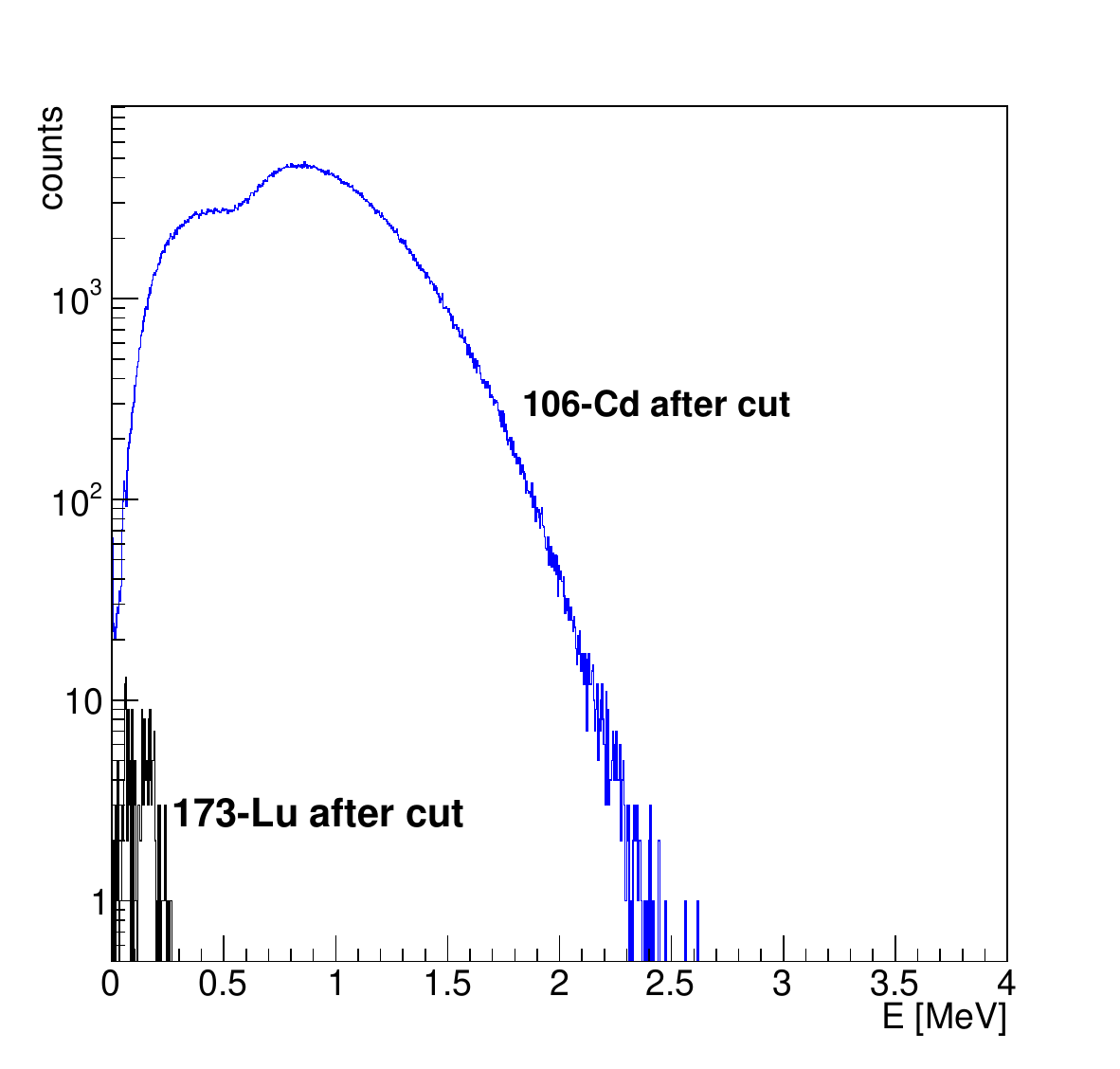}
  \caption{Comparison between the spectrum from $2\nu\epsilon\beta^+$ decay of $^{106}$Cd and $^{173}$Lu spectrum.}\label{fig_lu173_comparison}
 \end{figure}

In figure \ref{fig_lu173_comparison} the low significance of the contamination is shown by comparing the spectrum obtained from $^{173}$Lu after selection with the one obtained for $2\nu\epsilon\beta^+$ decay of $^{106}$Cd. 

Moreover, the effects of contamination appear only at low energy.
 
 \subsection{Final results}
 
 To understand the real significance of the background contribution due to the isotope we chose to simulate, we need now to calculate the activity of the $^{106}$Cd channel we want to study. To perform this calculation, we assume a half-life of 10$^{22}$ yr.
 
 The activity $a$ can be written as \cite{corvisiero}
 
 $$ a=\frac{\ln 2 \times m \times N_A}{T_{1/2}\times A}, $$
 
 where $m$ is the mass of radioactive isotope in grams, $N_A$ is the Avogadro number, $T_{1/2}$ is the half life of the nuclide and $A$ is the atomic mass.
 
 We need to know the mass of $^{106}$Cd contained inside the enriched crystal. We kow that the total mass of the crystal is 215.8 g. From the molecular composition, we know that Cd represents the 31.2\% of the total weight. So, the mass of Cd is 67.4 g. The relative abundance of the isotope we are studying is 66\%, as previously reported in table \ref{tab_106cd_composition}, so we have 44.47 g of $^{106}$Cd. 
 
 From all these data, we obtain:
 
 $$s= \frac{\ln 2 \times 44.47 \times N_A}{10^{22}\times 106}=17.5 \ counts \ yr^{-1}$$
 
 Another important step is to calculate the number of counts expected for each nuclide we simulated. The process is quite simple, since we have all the information we need about the radioactive isotopes and the crystals. 
 
 We have already weighted the activities from the activation software calculations for the effective presence of each material from which the radionuclides are produced inside the crystal, so we can easily calculate the total activity expected for each contaminant inside the detector by multiplying this value for the mass of the crystal itself.  We know from \cite{detector_development} that the mass of the enriched crystal is 0.215 kg, and the masses of the two natural detectors are 908.2 g and 918.6 g.
 The knowledge of all these data allow us to calculate the number of counts we expect in a year of data-taking. 
 
 The activity is defined as $a=\left|\frac{dN}{dt}\right|$ \cite{corvisiero}, we can use the formula \cite{corvisiero}:
 
 $$\Delta N(\Delta t)= \int_{\Delta T} a(t)dt=\int_{\Delta T} N(t)\lambda dt= N_0 \lambda\int_{\Delta T} e^{-\lambda t}dt=N_0(1-e^{-\lambda t}). $$
 
 We labeled with $\Delta N$ the number of expected counts, with $\Delta t$ the interval of integration, $a(t)$ is the activity as function of time, $N(t)$ is the number of decays as function of time, $\lambda=\ln 2/T_{1/2}$.
 $N_0$ can be deduced from the initial activity, that we can calculate by multiplying the values from COSMO1 and Activia calculations for the masses of the crystals, and from the half-life of the isotopes using the formula  $ N_0=\frac{a}{\lambda}$ \cite{corvisiero}.
 To be as conservative as possible, we always use the higher activity between the two calculated by Activia and COSMO1.
  
 

 In the calculation of the expected counts inside $^{106}$CdWO$_4$ detector, we must also take into account that only a small fraction of the decays inside $^{nat}$CdWO$_4$ affects the energy spectrum of that crystal. From the simulation, we can calculate the fraction of events that reach the enriched crystal, and use this factor to calculate the number of counts we expect inside $^{106}$CdWO$_4$. 
 The results obtained from the calculations are reported in the second column of table \ref{tab_efficiencies_enr} and table \ref{tab_efficiencies_nat}. 
 
 \begin{table}[h]
 \centering

 \begin{tabular}{ccc}
  \hline
  Source of decay  & Accumulated events  in 1yr  &Accumulated events  in 1yr\\
                   & in $^{106}$CdWO$_4$      &in $^{106}$CdWO$_4$      \\
                   & before selection         &after selection   \\
  \hline
  $^{106}$Cd ($2\nu\epsilon\beta^+$) & 17 & 3.5\\
  $^{102}$Rh &2.9& 0.18 \\
  $^{148}$Re &2.6 & 0.054 \\
  $^{182}$Ta &12 & 0.23 \\
  $^{65}$Zn  &0.038 & 0.0013 \\
  $^{108}$Ag &0.89 & 0.0022 \\
  $^{110m}$Ag&6.3 & 0.024\\
  $^{174}$Lu &0.13 & 0.00020\\
  \hline
  \end{tabular}
  
  \caption{Events calculated using activity data and simulations results for radioactive contaminations inside $^{106}$CdWO$_4$ detectors. For $2\nu\epsilon\beta$ decay of $^{106}$Cd we suppose a half-life of 10$^{22}$ yr. In the second and third row we give the number of expected counts inside the enriched detector  in the energy range between 0.05 MeV and 3 MeV before and after the coincidence selection. }\label{tab_efficiencies_enr}
\end{table}
\begin{table}
 \centering

 \begin{tabular}{ccc}
  \hline
  Source of decay & Accumulated events in 1yr    &Accumulated events in 1yr      \\
                  & in $^{106}$CdWO$_4$       &in $^{106}$CdWO$_4$                \\   
                  & before selection    &after selection                 \\ 
  \hline
  $^{182}$Ta & 104  &  6.9\\
  $^{172}$Hf &126  &  6.7 \\
  $^{184}$Re & 28  &  1.5 \\
  $^{110m}$Ag & 3.6& 0.095  \\
  $^{173}$Lu & 4.5 &  0.024 \\
  \hline
    \end{tabular}
    
   \caption{Events calculated using activity data and simulations results for radioactive contaminations inside $^{nat}$CdWO$_4$ detectors. In the second and third row we give the number of expected counts inside the enriched detector  in the energy range between 0.05 MeV and 3 MeV before and after the coincidence selection.}\label{tab_efficiencies_nat}
\end{table}


To produce a reconstruction of possible background due to cosmogenical activation that we should find inside our setup in a year of data-taking, the first step is to normalize the spectra obtained from every single contaminant to the number of counts expected for it in a year. 
\begin{figure}[H]
 \centering
 \includegraphics[width=0.7\textwidth]{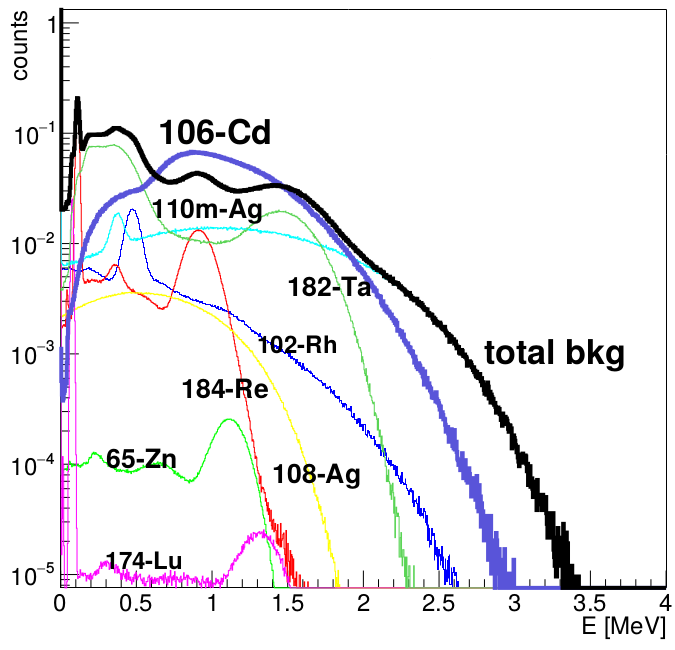}
 \caption{Spectra collected by the enriched crystal from its contaminations. }\label{fig_enr_spectra_nocut}
\end{figure}

\begin{figure}[H]
 \centering
 \includegraphics[width=0.7\textwidth]{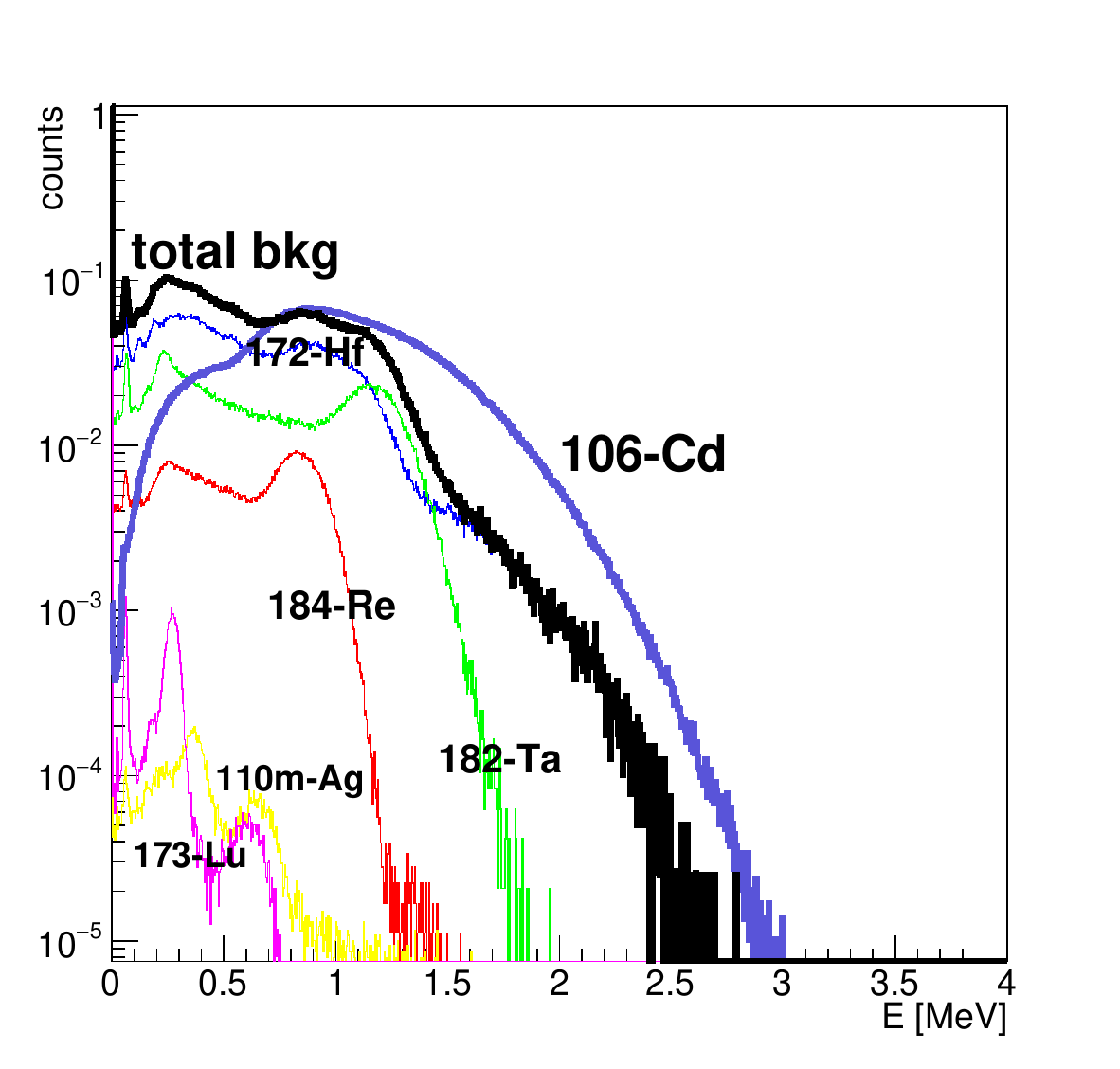}
 \caption{Spectra collected by the enriched crystal from contaminants of $^{nat}$CdWO$_4$. }\label{fig_nat_spectra_nocut}
\end{figure}

Figures \ref{fig_enr_spectra_nocut} and \ref{fig_nat_spectra_nocut} report the sum of the spectra before the coincidence selection.  We notice that double-$\beta$ effect from $^{106}$Cd is quite completely overlapped by the background produced by the cosmogenic activated nuclides. The background results to be of the same order of magnitude of the expected effect and it also mimics the shape of its spectrum. 
We want now to verify the effect the coincidence cut has on the data. 

\begin{figure}
 \centering
 \includegraphics[width=\textwidth]{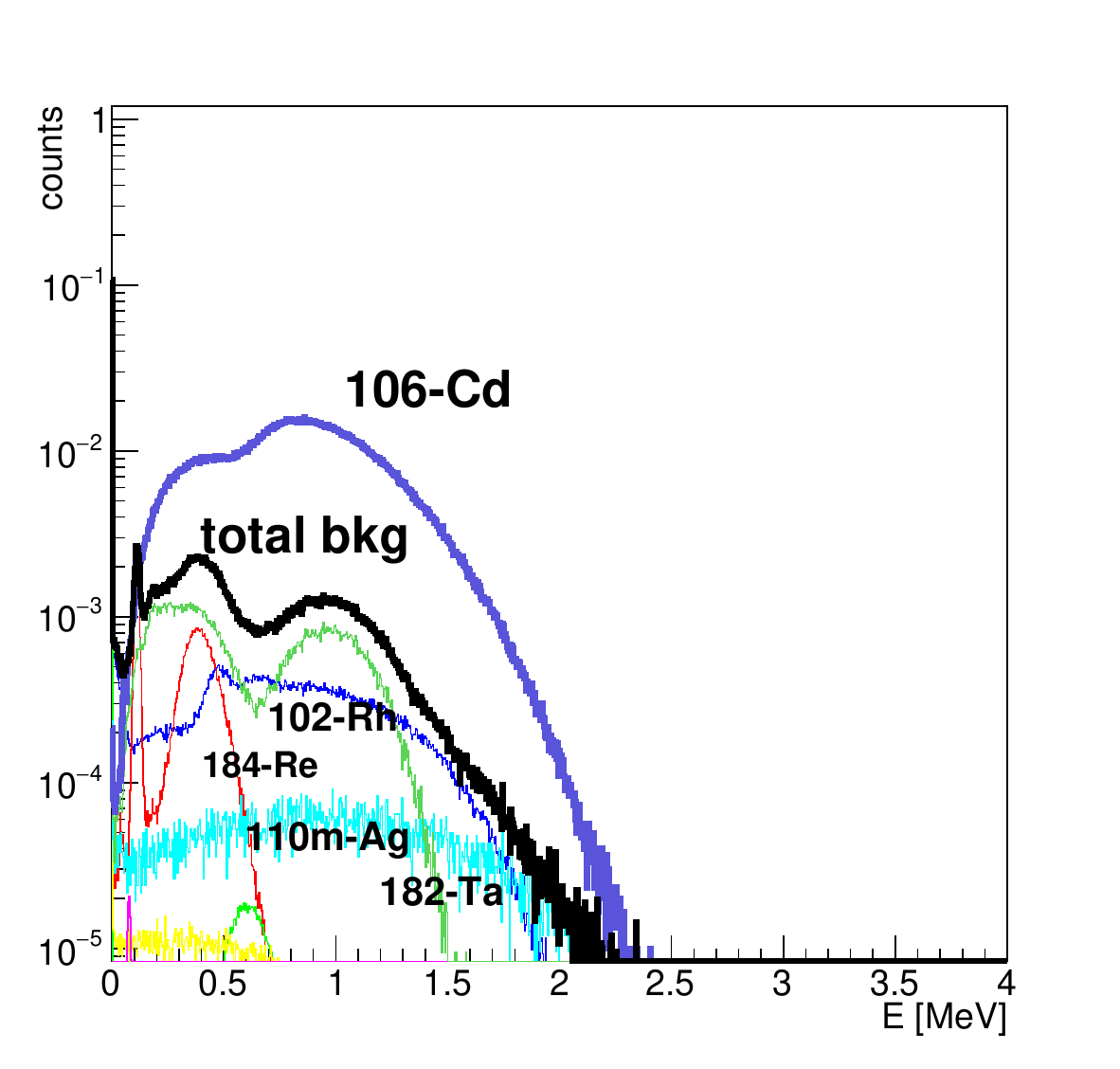}
 \caption{Spectra collected by the enriched crystal from its contaminants after the coincidence cut. }\label{fig_enr_spectra_cut}
\end{figure}

\begin{figure}
 \centering
 \includegraphics[width=\textwidth]{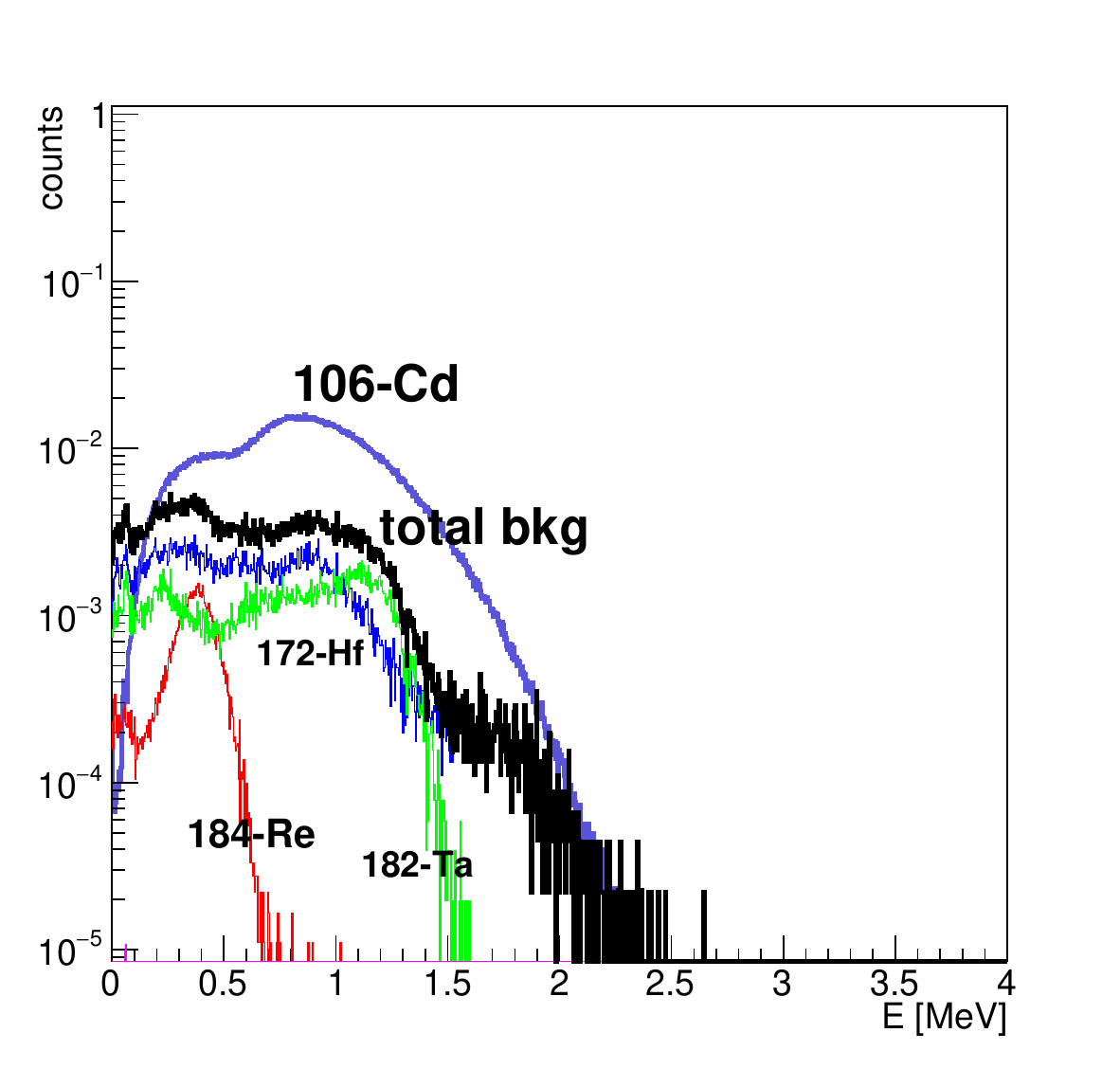}
 \caption{Spectra collected by the enriched crystal from contaminants of $^{nat}$CdWO$_4$ detectors after the coincidence cut. }\label{fig_nat_spectra_cut}
\end{figure}

Figures \ref{fig_enr_spectra_cut} and \ref{fig_nat_spectra_cut} show the effect of coincidence selection on the data.
To correctly produce these spectra, we have to take into account both the number of expected counts per year and the efficiency we deduced for the detection of each isotope. This information is reported on table \ref{tab_efficiencies_enr} and table \ref{tab_efficiencies_nat}.

It is clear that coincidence selection enhances our capability to resolve the double-$\beta$ process. We notice after this analysis that the contaminants inside the $^{nat}$CdWO$_4$ crystals give rise to a higher background than the contaminations foreseen for $^{106}$CdWO$_4$ detector. 

From figures \ref{fig_enr_spectra_cut} and \ref{fig_nat_spectra_cut} we can extract the main contributions to background. Inside the data collected and shown for cosmogenic activated contaminants in $^{106}$CdWO$_4$ we notice that there are four nuclides which contribute widely to the background. They are $^{182}$Ta and $^{184}$Re. 
From $^{nat}$CdWO$_4$ detectors we select $^{172}$Hf, $^{184}$Re and $^{182}$Ta. These two last nuclides are expected in both the detectors, and their presence emerges as one of the main contaminations for the measurement.
The next step for our analysis is to sum all these contributions, to obtain the total expected background from the nuclides we calculated in this work. 

\begin{figure}
 \centering
 \includegraphics[width=\textwidth]{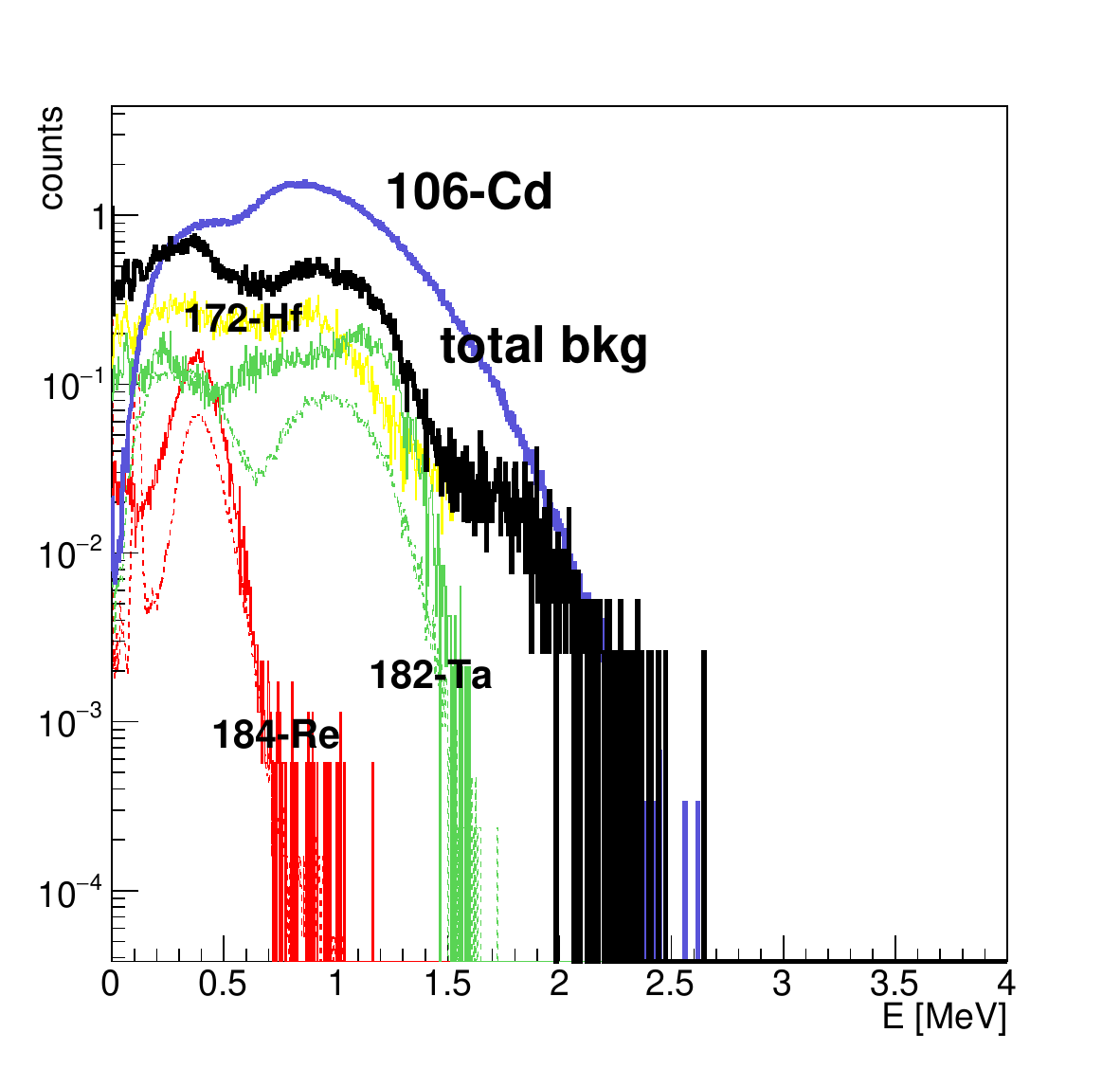}
 \caption{Sum of all the most relevant contaminant spectra compared to the searched $2\nu\epsilon\beta^+$ decay from $^{106}$Cd. Dotted lines are related to the nuclides that decay inside the enriched crystal, continued lines are related to nuclides inside the natural crystals.}\label{fig_all_contaminants}
\end{figure}

In figure \ref{fig_all_contaminants} the result of the sum is shown. 

\newpage
\section{Conclusions}

In this work, a Monte Carlo simulation of the experimental setup installed in DAMA/CRYS setup has been realized using GEANT4 software. The experiment uses a CdWO$_4$ crystal scintillator enriched in $^{106}$Cd to study the double-$\beta$ decay processes of the isotope. The detector is surrounded by two CdWO$_4$ detectors that allow a coincidence selection of the events in which a positron has been emitted using the characteristic $\gamma$ emission at 511 keV and also anticoincidence mode, in which events that involve only the main detector are selected. From previous works, e.g. \cite{gemulti}, we notice that this mode gives higher sensibility for some decay modes. 

The simulation has been used to study the effects of cosmogenic contaminants inside the three detectors used for the experiment. A list of possible contaminations for the detectors have been calculated using two softwares, COSMO1 and Activia. 
The nuclides proposed on this list have been selected taking into account the calculated activity, the half life, the decay modes and the Q-value of the decays.

We chose to simulate the decay of 12 radioactive nuclides, 7 inside the enriched crystal ($^{65}$Zn, $^{102}$Rh and $^{102m}$Rh, $^{108}$Ag, $^{110m}$Ag, $^{174}$Lu, $^{182}$Ta,  $^{184}$Re) and 5 inside the natural crystals ($^{110m}$Ag, $^{172}$Hf, $^{173}$Lu, $^{182}$Ta, $^{184}$Re). 
For each contaminant the number of expected counts in over a year of data taking have been calculated from the activation data. 

A coincidence selection is applied to the obtained data, to test the efficiency of the setup in background reduction in this mode. The algorithm selects all the events inside the enriched crystal that are in coincidence with a signal inside one of the two natural detectors with energy 511 keV.

To understand the effect of the obtained background on the measurement of $2\nu\epsilon\beta^+$ decay of $^{106}$Cd we ran a simulation of this channel using the initial conditions produced by DECAY0 software. We applied to that data the same coincidence selection used on the results for the contaminants.

To compare the results of all the simulations, we took into account the different activities foreseen for the nuclides and the efficiency calculated for the coincidence selection. The number of counts expected in a year has been calculated and can be found in tables \ref{tab_efficiencies_enr} and \ref{tab_efficiencies_nat}. Using these data, we scale the spectra to construct the background for a year long data-taking. All the contributions have been summed and we superimposed the obtained histogram on the effect expected for $2\nu\epsilon\beta^+$ decay of $^{106}$Cd with a supposed half-life of $10^{22}$ yr. 
The result is presented in figure \ref{fig_all_contaminants}. Since some of the contaminants give a contribution that is several orders of magnitude lower than the most active, their effect have been ignored in the construction of the total background. 

From this last picture, we notice that the background from cosmogenic nuclide cannot be neglected in data analysis, at least in the first year of measurements, since it is able to mimic the effect of the channel we choose to study also after the coincidence selection.

We also notice that the natural crystals are the main source of this background. The longer exposure time to cosmogenical activation produces a number of isotopes whose activity can strongly reduce the sensitivity of our setup to the double-$\beta$ decay of $^{106}$Cd. We remark at the end that almost all the contaminants have a half-life of some hundreds of days. We can expect a reduction of the contributions of these isotopes during the data-taking of the experiment.

The simulation could be used in future study of the setup to evaluate the other decay channels of $^{106}$Cd and the contaminations due to natural radioactivity from all the materials, not only from the crystals. In particular, copper and teflon contaminants should be taken into account if their exposure to cosmic rays has been of the same order of magnitude or longer than the exposure reported for natural detectors.

\newpage

\appendix
\section{Decay schemes for radioactive contaminants}
 \begin{figure}[hb]
 \centering
 \includegraphics[width=0.8\textwidth]{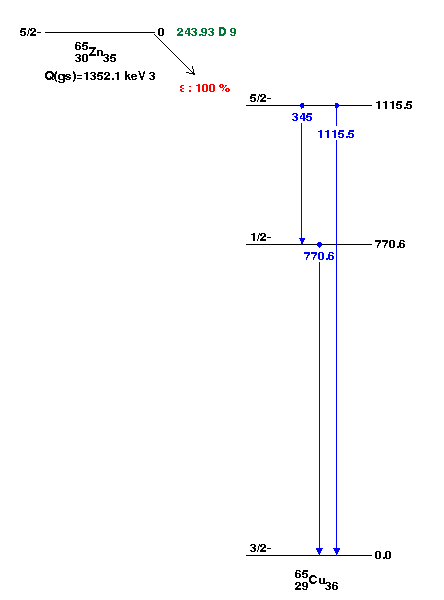}
 \caption{$^{65}$Zn decay scheme. The informations about the decay are from \cite{nudat}.}\label{fig_zn65_decay}
\end{figure}

\begin{figure}
  \includegraphics[width=\textwidth]{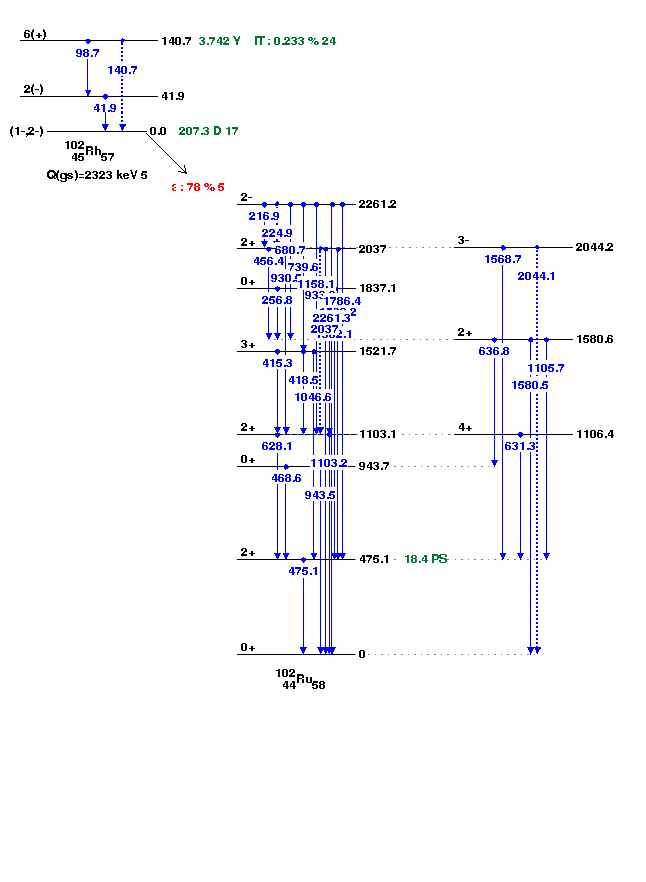}
  \caption{$^{102}$Rh and $^{102m}$Rh decay scheme. The informations about the decay are from \cite{nudat}.}\label{fig_rh102_decay}
 \end{figure}
 
\begin{figure}
 \centering
 \includegraphics[width=\textwidth]{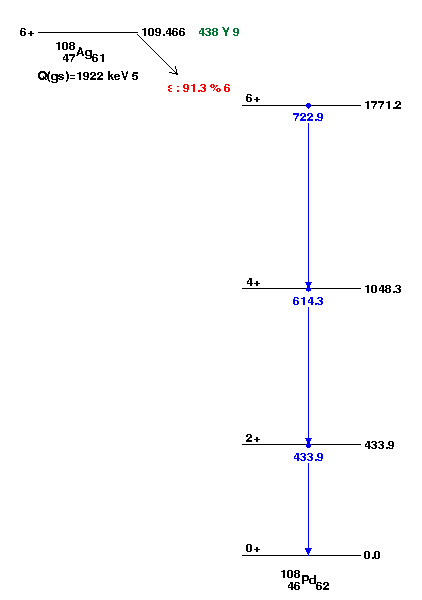}
 \caption{$^{108}$Ag decay scheme. The informations about the decay are from \cite{nudat}.}\label{fig_ag108_decay}
\end{figure}

 \begin{figure}
  \centering
  \includegraphics[width=1\textwidth]{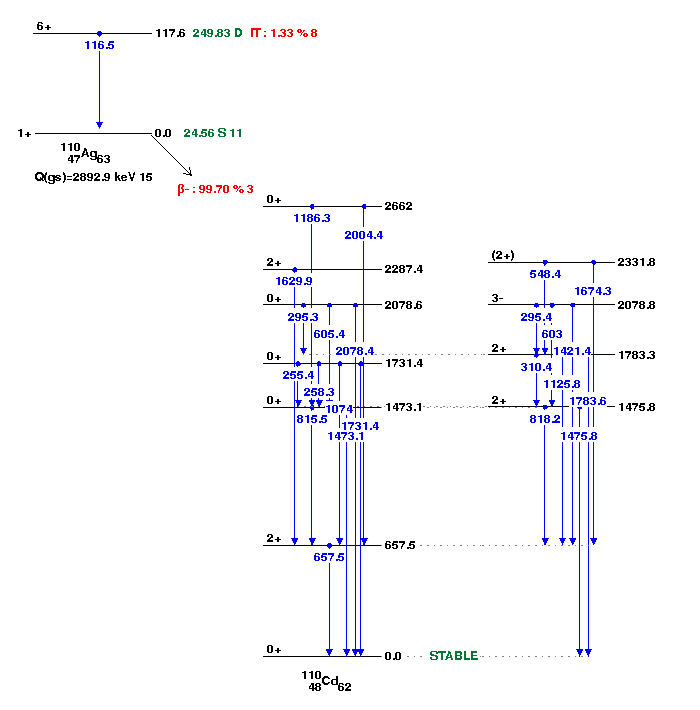}
  \caption{$^{110}$Ag and $^{110m}$Ag decay scheme.  The informations about the decay are from \cite{nudat}.}\label{fig_ag110_decay}
 \end{figure}
 
 \begin{figure}
  \centering
  \includegraphics[width=\textwidth]{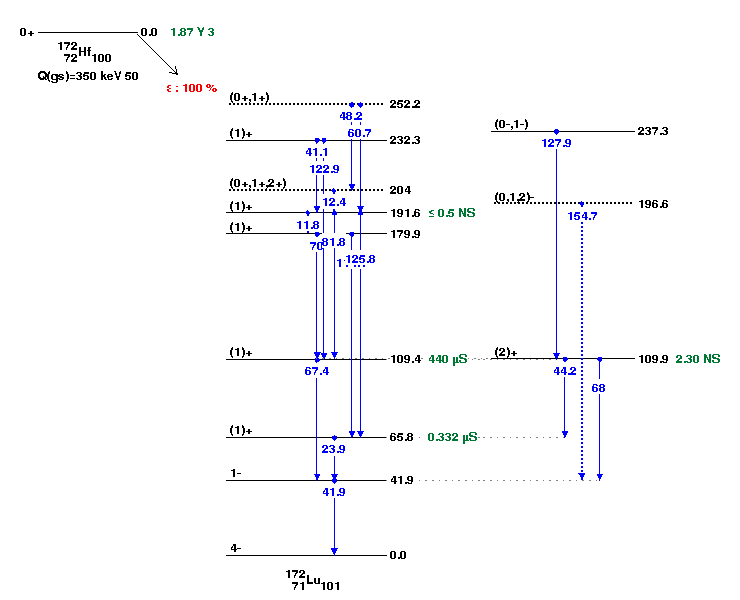}
  \caption{$^{172}$Hf decay scheme. The informations about the decay are from \cite{nudat}.}\label{fig_hf172_decay}
 \end{figure}
 \newpage
   \thispagestyle{empty}

 \begin{figure}
  \centering
  \includegraphics[angle=90,width=0.8\textwidth]{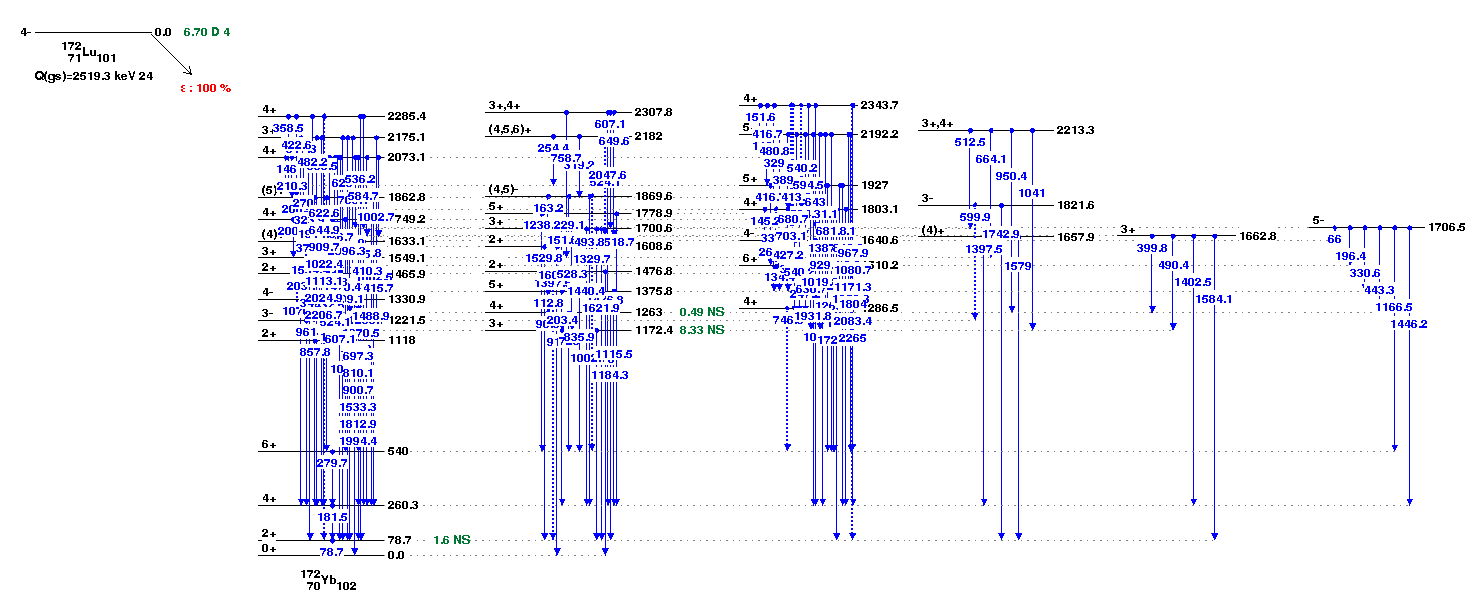}
  \caption{$^{172}$Lu decay scheme. The informations about the decay are from \cite{nudat}.}\label{fig_lu172_decay}
 \end{figure}
 
 \begin{figure}
  \centering
  \includegraphics[width=\textwidth]{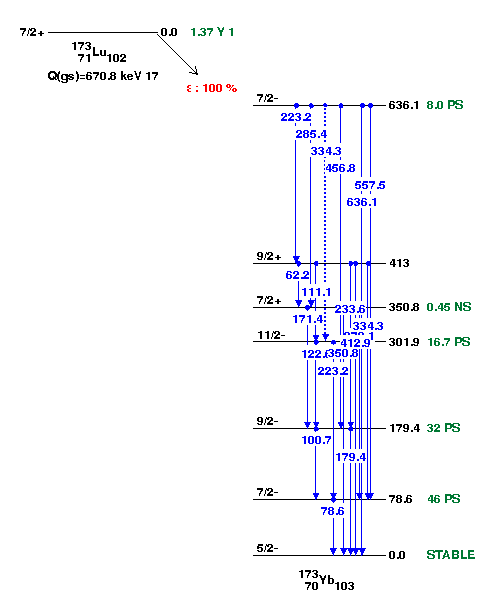}
  \caption{$^{173}$Lu decay scheme. The informations about the decay are from \cite{nudat}.}\label{fig_lu173_decay}
 \end{figure}
 
  \begin{figure}
 \centering
 \includegraphics[width=\textwidth]{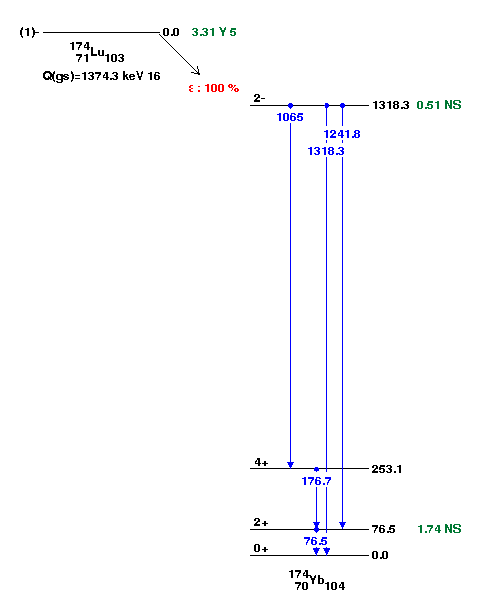}
 \caption{$^{174}$Lu decay scheme. The informations about the decay are from \cite{nudat}.}\label{fig_lu174_decay}
\end{figure}

 \begin{figure}
 \centering
 \includegraphics[width=\textwidth]{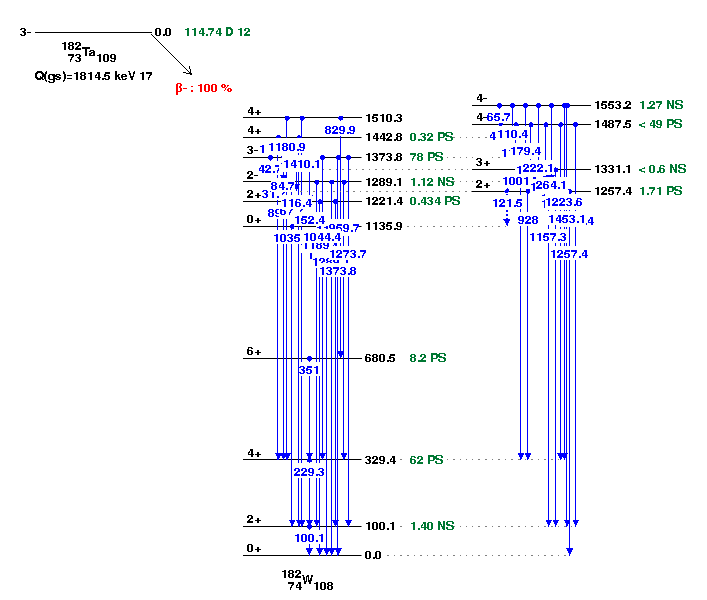}
 \caption{$^{182}$Ta decay scheme. The informations about the decay are from \cite{nudat}.}\label{fig_ta182_decay}
\end{figure}

 \begin{figure}
  \centering
  \includegraphics[width=\textwidth]{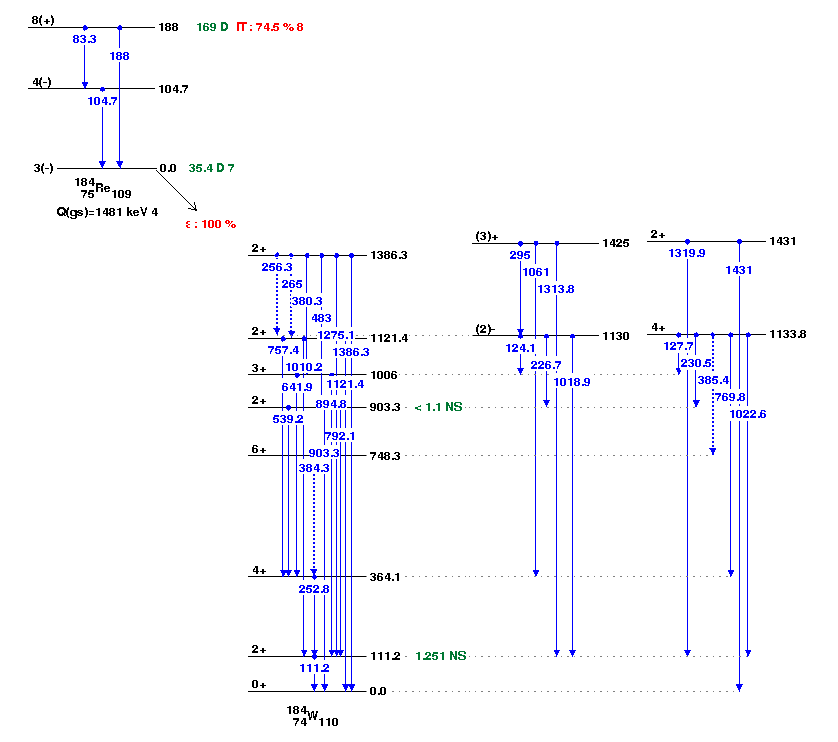}
  \caption{$^{184}$Re and $^{184m}$Re decay scheme. The informations about the decay are from \cite{nudat}.}\label{fig_re184_decay}
 \end{figure}

\newpage
\cleardoublepage

\end{document}